\documentclass[twocolumn]{aastex631}

\usepackage{comment}
\usepackage{subfigure}
\usepackage{hyperref}
\shorttitle{AASTeX v6.3.1 Sample article}
\shortauthors{Adams et al.}
\graphicspath{{./}{figures/}}

\begin{document}

\title{EPOCHS Paper II: The Ultraviolet Luminosity Function from $7.5<z<13.5$ using 180 square arcminutes of deep, blank-fields from the PEARLS Survey and Public JWST data}

\author[0000-0003-4875-6272]{Nathan J. Adams}
\affiliation{Jodrell Bank Centre for Astrophysics, University of Manchester, Oxford Road, Manchester M13 9PL, UK}

\author[0000-0003-1949-7638]{Christopher J. Conselice}
\affiliation{Jodrell Bank Centre for Astrophysics, University of Manchester, Oxford Road, Manchester M13 9PL, UK}

\author[0000-0003-0519-9445]{Duncan Austin}
\affiliation{Jodrell Bank Centre for Astrophysics, University of Manchester, Oxford Road, Manchester M13 9PL, UK}

\author[0000-0002-4130-636X]{Thomas Harvey}
\affiliation{Jodrell Bank Centre for Astrophysics, University of Manchester, Oxford Road, Manchester M13 9PL, UK}

\author[0000-0002-8919-079X]{Leonardo Ferreira}
\affiliation{Department of Physics \& Astronomy, University of Victoria, Finnerty Road, Victoria, British Columbia, V8P 1A1, Canada}

\author[0000-0002-9081-2111]{James Trussler}
\affiliation{Jodrell Bank Centre for Astrophysics, University of Manchester, Oxford Road, Manchester M13 9PL, UK}

\author[0009-0003-7423-8660]{Ignas Juodžbalis}
\affiliation{Jodrell Bank Centre for Astrophysics, University of Manchester, Oxford Road, Manchester M13 9PL, UK}

\author[0000-0002-3119-9003]{Qiong Li}
\affiliation{Jodrell Bank Centre for Astrophysics, University of Manchester, Oxford Road, Manchester M13 9PL, UK}

\author[0000-0001-8156-6281]{Rogier Windhorst}
\affiliation{School of Earth and Space Exploration, Arizona State University, Tempe, AZ 85287-1404}

\author[0000-0003-3329-1337]{Seth H. Cohen} 
\affiliation{School of Earth and Space Exploration, Arizona State University,
Tempe, AZ 85287-1404}

\author[0000-0003-1268-5230]{Rolf A. Jansen} 
\affiliation{School of Earth and Space Exploration, Arizona State University,
Tempe, AZ 85287-1404}

\author[0000-0002-7265-7920]{Jake Summers}
\affiliation{School of Earth and Space Exploration, Arizona State University,
Tempe, AZ 85287-1404}

\author{Scott Tompkins} 
\affiliation{School of Earth and Space Exploration, Arizona State University,
Tempe, AZ 85287-1404}

\author{Simon P. Driver} 
\affiliation{International Centre for Radio Astronomy Research (ICRAR) and the
International Space Centre (ISC), The University of Western Australia, M468,
35 Stirling Highway, Crawley, WA 6009, Australia}

\author{Aaron Robotham} 
\affiliation{International Centre for Radio Astronomy Research (ICRAR) and the
International Space Centre (ISC), The University of Western Australia, M468,
35 Stirling Highway, Crawley, WA 6009, Australia}

\author{Jordan C. J. D’Silva}
\affiliation{International Centre for Radio Astronomy Research (ICRAR) and the
International Space Centre (ISC), The University of Western Australia, M468,
35 Stirling Highway, Crawley, WA 6009, Australia}

\author[0000-0001-7592-7714]{Haojing Yan} 
\affiliation{Department of Physics and Astronomy, University of Missouri,
Columbia, MO 65211}

\author[0000-0001-7410-7669]{Dan Coe} 
\affiliation{AURA for the European Space Agency (ESA), Space Telescope Science
Institute, 3700 San Martin Drive, Baltimore, MD 21218, USA}

\author[0000-0003-1625-8009]{Brenda Frye} 
\affiliation{University of Arizona, Department of Astronomy/Steward
Observatory, 933 N Cherry Ave, Tucson, AZ85721}

\author[0000-0001-9440-8872]{Norman A. Grogin} 
\affiliation{Space Telescope Science Institute, 3700 San Martin Drive, Baltimore, MD 21210, USA}

\author[0000-0002-6610-2048]{Anton M. Koekemoer} 
\affiliation{Space Telescope Science Institute, 3700 San Martin Drive, Baltimore, MD 21210, USA}

\author[0000-0001-6434-7845]{Madeline A. Marshall} 
\affiliation{National Research Council of Canada, Herzberg Astronomy \&
Astrophysics Research Centre, 5071 West Saanich Road, Victoria, BC V9E 2E7,
Canada; \& ARC Centre of Excellence for All Sky Astrophysics in 3
Dimensions (ASTRO 3D), Australia}

\author[0000-0003-3382-5941]{Nor Pirzkal} 
\affiliation{Space Telescope Science Institute, 3700 San Martin Drive, Baltimore, MD 21210, USA}

\author[0000-0003-0894-1588]{Russell E. Ryan, Jr.} 
\affiliation{Space Telescope Science Institute, 3700 San Martin Drive, Baltimore, MD 21210, USA}

\author[0000-0002-2203-7889]{W. Peter Maksym} 
\affiliation{Center for Astrophysics \textbar\ Harvard  \& Smithsonian, 60 Garden St., Cambridge, MA 02138, USA}

\author[0000-0001-7016-5220]{Michael J. Rutkowski} 
\affiliation{Minnesota State University-Mankato,  Telescope Science Institute,
TN141, Mankato MN 56001, USA}

\author[0000-0001-9262-9997]{Christopher N. A. Willmer} 
\affiliation{Steward Observatory, University of Arizona, 933 N Cherry Ave,
Tucson, AZ, 85721-0009}

\author{Heidi B. Hammel}
\affiliation{Associated Universities for Research in Astronomy, Inc., 1331 Pennsylvania Avenue NW, Suite 1475, Washington, DC 20004, USA}

\author[0000-0001-6342-9662]{Mario Nonino}
\affiliation{INAF-Osservatorio Astronomico di Trieste, Via Bazzoni 2, I-34124 Trieste, Italy}

\author[0000-0003-0883-2226]{Rachana Bhatawdekar}
\affiliation{European Space Agency, ESA/ESTEC, Keplerlaan 1, 2201 AZ Noordwijk, NL}

\author[0000-0003-3903-6935]{Stephen M. Wilkins}
\affiliation{Astronomy Centre, Department of Physics and Astronomy, University of Sussex, Brighton, BN1 9QH, UK}

\author[0000-0002-7908-9284]{Larry D. Bradley}
\affiliation{Space Telescope Science Institute, 3700 San Martin Drive, Baltimore, MD 21210, USA}

\author[0000-0002-8785-8979]{Tom Broadhurst} 
\affiliation{ Department of Theoretical Physics, University of the Basque Country UPV-EHU, E-48040 Bilbao, Spain}
\affiliation{Donostia International Physics Center (DIPC), 20018 Donostia, The Basque Country, Spain}
\affiliation{IKERBASQUE, Basque Foundation for Science, Alameda Urquijo, 36-5 E-48008 Bilbao, Spain}

\author[0000-0003-0202-0534]{Cheng Cheng}
\affiliation{Chinese Academy of Sciences, National Astronomical Observatories, CAS, Beijing 100101, Peoples Republic of China}

\author[0000-0002-9767-3839]{Herv\'{e} Dole}
\affiliation{Université Paris-Saclay, CNRS, Institut d’Astrophysique Spatiale, F-91405, Orsay, France}

\author[0000-0001-6145-5090]{Nimish P. Hathi}
\affiliation{Space Telescope Science Institute, 3700 San Martin Drive, Baltimore, MD 21210, USA}

\author[0000-0002-0350-4488]{Adi Zitrin}
\affiliation{Physics Department, Ben-Gurion University of the Negev, P.O. Box 653, Beer-Sheva 8410501, Israel}



\begin{abstract}

We present an analysis of the ultraviolet luminosity function (UV LF) and star formation rate density of distant galaxies ($7.5 < z < 13.5$) in the `blank' fields of the Prime Extragalactic Areas for Reionization Science (PEARLS) survey combined with Early Release Science (ERS) data from the CEERS, GLASS, NGDEEP surveys/fields and the first data release of JADES. We use strict quality cuts on EAZY photometric redshifts to obtain a reliable selection and characterisation of high-redshift ($z>6.5$) galaxies from a consistently processed set of deep, near-infrared imaging. Within an area of 180 arcmin$^{2}$, we identify 1046 candidate galaxies at redshifts $z>6.5$ and we use this sample to study the ultraviolet luminosity function (UV LF) in four redshift bins between $7.5<z<13.5$. The measured number density of galaxies at $z=8$ and $z=9$ match those of past observations undertaken by the {\em Hubble Space Telescope} (HST). Our $z=10.5$ measurements lie between early JWST results and past HST results, indicating cosmic variance may be the cause of previous high density measurements. However, number densities of UV luminous galaxies at $z=12.5$ are high compared to predictions from simulations. When examining the star formation rate density of galaxies at this time period, our observations are still largely consistent with a constant star formation efficiency, are slightly lower than previous early estimations using JWST and support galaxy driven reionization at $z\leq8$. 

\end{abstract}

\keywords{}


\section{Introduction}\label{sec:intro}

One of the main design goals of the James Webb Space Telescope (JWST) is to expand the present redshift frontier and search for the very first galaxies to form in the Universe. Major science goals are to learn more about the physics of early galaxy and star formation, the nature of dark matter and to test our understanding of wider cosmology. A number of Guaranteed Time Observations (GTO) and Early Release Science (ERS) programmes were subsequently developed to use this new facility to conduct this search. These include, but are not limited to, a slew of early papers that have investigated potential distant galaxies \citep[e.g.,][]{Donnan2022, Adams2023, Castellano2022, Naidu2022,Finkelstein2022c,Atek2022,Yan2022,Harikane2023,Austin2023,Finkelstein2023,Leung2023,McLeod2023,Willott2023}

This search for the earliest galaxies in the Universe has a long history, dating back to the work of \citet{Hubble1931} and advancing up until today. Samples of galaxies existing within the first 10 per cent of the history of the Universe now number in the tens of thousands \citep{Duncan2014, Bouwens2015,Finkelstein2015,Harikane2021,Adams2021}, many of which were found thanks to synergies between the Hubble Space Telescope and ground-based facilities such as Subaru and VISTA.  Until the launch of JWST, the most distant galaxy candidates found had redshifts up to $z \sim 10$ \citep[e.g.][]{Bouwens2011,Bouwens2015,Finkelstein2015,McLeod2016,Oesch2018,Salmon2018,Morishita2018,Stefanon2019,Bowler2020} and claims of the detection of early galaxies stretch as far as $z\sim13$ \citep[][]{Harikane2022}.The spectroscopic confirmation of these sources has since followed, with examples at $z>7.5$ including \citet{Oesch2015,Zitrin2015,RobertsBorsani2015} and \citet{Hashimoto2018}, the highest redshift source pre-JWST was measured to exist at $z=10.957$ \citep[GNz-11:][]{Oesch2016,Jiang2020}.

Over the course of the year of 2022, JWST has been successfully launched, calibrated, and yielded its first data releases, with thanks to our colleagues working for various national and international organisations \citep[][]{Rigby2022}.   As a result, a revolution in our understanding of the early Universe is now underway and the first photometric data is now available and is being studied in some detail \citep[e.g.][Conselice et al. In Prep]{Treu2022, Finkelstein2022c}.  The first JWST data was released in July 2022, triggering rapid efforts to identify and follow up candidate galaxies at $z>11$ \citep[][]{Donnan2022, Adams2023, Castellano2022, Naidu2022,Yan2022,Finkelstein2022c,Atek2022,Harikane2023,Yan2023}. However, there are inconsistent samples of galaxies being found between the various studies published to date. There are multiple instances of some studies reporting candidate $z>11$ galaxies that either have a low-z solution in another work or are found to be insufficiently detected to be considered robust \citep{Bouwens2022c}. These results are likely due to the intricacies of reducing data from a new and complex instrument combined with the varied photometric redshift codes and approaches being undertaken. 

The NIRSpec instrument has already begun the job of confirming large numbers of high-redshift galaxies \citep{CurtisLake2022,Bunker2023,Tang2023,Wang2023}. However, it has also shown that some of these candidates are actually at lower redshifts. Recently, work by \cite[][]{arrabal2023} has shown a candidate $z=16.4$ galaxy is a galaxy at $z=4.9$ with strong emission lines. This demonstrates that it is important to be extra careful when dealing with candidate galaxies exhibiting potentially extreme properties. Beyond just redshifts, emission lines and their strengths are key to understanding fundamental galaxy properties. For example, the work of \citet{Endsley2022} has shown that emission lines can influence photometry such that 2 dex differences in stellar mass estimations were possible for one of the originally proposed massive sources presented in \citet{Labbe2022}. Further to this, CEERS NIRSpec observations analysed in \citet{Kocevski2023} have shown one such massive $z=7-9$ galaxy candidate was really a $z\sim5$ AGN.

Beyond identifying galaxies, one of the first, fundamental measurements that can be conducted with a high redshift sample of galaxies is the ultraviolet luminosity function (UV LF). This is the measure of the number density of galaxies as a function of the ultraviolet luminosity, typically measured between 1400\AA \, to 1600\AA. The evolution of the UV LF as a function of redshift depends on a number of galaxy properties that can subsequently be used to trace how the galaxy population changes with time. These include star formation rates \citep[e.g.][]{Bower2012,Paardekooper2013}, active galactic nuclei (AGN) emission \citep[e.g.][]{Aird2015,Matsuoka2018c,Giallongo2019,Niida2020,Harikane2021,Adams2021} and dust obscuration \citep[e.g.][]{Clay2015,Bowler2015,Tamura2019,Bowler2020}. Ultraviolet emission from galaxies is also responsible for a key phase of the Universe's evolution, a process known as reionisation \citep[see e.g.][and references therein]{Robertson2015,Finkelstein2019}. There is still debate with regards to whether AGN, massive galaxies or the more numerous, fainter galaxies provide greater contributions towards the budget of ionising photons \citep{Madau2015,Yoshiura2017,Bosch2018,Hassan2018,Parsa2018,Dayal2020,Naidu2020}.

In the first year, several luminosity functions of distant galaxies have been measured using new JWST data.  This includes studies such as \citet{Donnan2022,Finkelstein2022c,Bouwens2022c,Gonzalez2023,Castellano2023,McLeod2023,Leung2023} and \citet{Finkelstein2023}.  In general, these studies probe the luminosity function in the rest-frame UV up to $z \sim 12$, but there are some cases of conflicting results.  There are several reasons for this, including (1) The limited survey volumes used to measure the UV LF, due to the few early observations available from JWST; (2) Potential systematics in previous measurements, such as contaminant lower-z galaxies. What is needed is a combination of different datasets that samples a large area and in which the data has been reduced and analyzed in a consistent fashion.  

 In this paper we conduct such a consistent reduction and analysis of the deepest NIRCam pointings that have been undertaken in the first few months of operations of JWST. This dataset includes observations from the Prime Extragalactic Areas for Reionization Science \citep[PEARLS, PI: R. Windhorst \& H.Hammel, PID: 1176 \& 2738][]{Windhorst2023}, The Cosmic Evolution Early Release Science Survey \citep[CEERS, PID: 1345, PI: S. Finkestein, see also][]{Bagley2022}, The Grism Lens Amplified Survey from Space \citep[GLASS, PID: 1324, PI: T. Treu;][]{Treu2022}, The Next Generation Deep Extragalactic Exploratory Public Survey \citep[NGDEEP, PID: 2079, PI: Finkelstein, Papovich and Pirzkal;][]{Bagley2023}, the JWST Advanced Deep Extragalactic Survey\citep[JADES, PID: 1180, PI: D. Eisenstein][]{Eisenstein2023} and the Early Release Observations of SMACS-0723 \citep{Pontoppidan2022}. We describe each of these surveys in more detail in \S 2.1. This collection is called the EPOCHS dataset version 1 (Conselice et al. In Prep), which will be used to study the properties of the first galaxies to form in the Universe.  We use these images and photometry to conduct a photometric search for candidate galaxies with redshifts $z>6.5$, using the deep, near-infrared coverage now available to provide strong constraints on the Lyman Break $\lambda_{\rm rest} < 1216$\AA\,\citep{Guhathakurta1990,Steidel1992,Steidel1996}. This work presents measurements of the ultraviolet luminosity function and star formation rate density using this new sample. Future works will target other physical properties of these galaxies, such as their star forming properties (Austin et al. In Prep) and stellar masses (Harvey et al. In Prep)
 
 The outline of this paper is as follows.  In \S 2, we describe the imaging data used along with the reduction processes followed to obtain robust photometry. In \S 3 we present our SED modelling and selection procedures, assess their reliability, compare to other samples and show some highlights of unique objects. In \S 4 we use our sample to measure the UV LF in four redshift bins in the range $7.5<z<13.5$. In \S 5 we discuss these results in the context of the UV LF's evolution and the implications on the star formation rate density in the early Universe. We summarise our findings and conclusions in \S 6. We assume a standard cosmology with $H_0=70$\,km\,s$^{-1}$\,Mpc$^{-1}$, $\Omega_{\rm M}=0.3$ and $\Omega_{\Lambda} = 0.7$. All magnitudes listed follow the AB magnitude system \citep{Oke1974,Oke1983}.

\section{Data Reduction and Products} \label{sec:data}

\subsection{The Surveys and Fields Used}

The data we use for this analysis originates from several sources, including the Early Release Science programmes of CEERS, GLASS, NGDEEP alongside the Early Release Observations of SMACS~0723. We also use the current PEARLS Survey GTO fields of El Gordo, MACS-0416 and the North Ecliptic Pole Time Domain Field \citep[][]{Windhorst2023}.  These data primarily include observations taken with the \textit{Near Infrared Camera} \citep[NIRCam;][]{rieke05, rieke08, rieke15,Rieke2022}. In order to generate a dataset which has been consistently handled, we reprocess all of the NIRCam imaging from their lowest-level, raw form obtained from the MAST database. 

A more general discussion of these fields and their suitability for finding high redshift galaxies is presented in Conselice et al. In Prep. 
 Whilst the public fields of CEERS, NGDEEP, GLASS and SMACS have been extensively used in the past, this is not the case for the PEARLS datasets. For the three lensing fields of SMACS~0723, MACS-0416 and El Gordo, one of the two NIRCam modules is positioned to cover the lensing cluster. The second module is located 3 arcminutes to the side, with its precise location determined by the position angle at which the observations were taken at. While we reduce both modules in these fields, we do not include the cluster regions in this study so as to simplify the later analysis. The inclusion of these clusters would require the use of strong gravitational lensing models and consideration of contamination from intra-cluster light (ICL). A more detailed look into lensed objects located behind the clusters is being coordinated as a parallel effort to this work \citep[][Bhatawdekar et al In Prep]{Diego2023,Frye2023,Kamienski2023,Carleton2023}. 

\subsubsection{PEARLS}

For this work, we include recent GTO observations from the Prime Extragalactic Areas for Reionization Science (PEARLS, PI: R. Windhorst \& H.Hammel, PID: 1176 \& 2738) survey \citep{Windhorst2023}. These currently consist of two fields targeting gravitationally lensing galaxy clusters and one blank field. The two clusters we use are MACS-0416 and El Gordo. The blank (non-cluster) field is located near the North Ecliptic Pole and is called the Time Domain Field \citep[NEP-TDF][]{Jansen2018}. This field consists of four sets of partially overlapping pairs of NIRCam pointings, with each pair of NIRCam pointings referred to as a `spoke'. As a GTO survey, PEARLS has proprietary time on its data, but the first set of NIRCam imaging of the NEP-TDF were made publicly available immediately. El Gordo and the first epoch of MACS-0416 are also now public. For each of the two clusters, the cluster proper is centred in one NIRCam module, with the second NIRCam module placed in a random, three arcminute distance, parallel location depending on date and angle of the observation undertaken. MACS-0416 was observed three times at different epochs and position angles, generating a deep, single module image of the cluster and three small parallels. This work utilises all three epochs of the MACS-0416 observations, meaning there are three NIRCam modules worth of data offset from the primary lensing cluster. All PEARLS observations used 8 NIRCam photometric bands, F090W, F115W, F150W, F200W, F277W, F356W, F410M and F444W. Within the NEP-TDF field, deep HST/ACS imaging in the F606W filter is also present through the GO-15278 (PI: R.~Jansen) and GO-16252/16793 (PIs: R.~Jansen \&
N.~Grogin) between 2017 October 1 and 2022 October 31. Mosaics of these data, astrometrically
aligned to Gaia/DR3 and resampled on 0\farcs03 pixels, were made
available pre-publication by R.~Jansen \& R.~O'Brien (private comm.).

\subsubsection{SMACS-0723}

Observations of the SMACS-0723 galaxy cluster were undertaken as part of the Early Release Observations 10 (ERO) programme \citep[PID: 2736, PI: K. Pontoppidan,][]{Pontoppidan2022} For this study, we utilise the 6-band NIRCam photometry taken in the F090W, F150W, F200W, F277W, F356W, and F444W broad-band filters. While we include high-z candidates from these observations in our final catalogues, we {\em do not} use this field when measuring the UV luminosity function within this paper. While a number of high redshift galaxies have been previously identified in this field \citep{Adams2023,Atek2022,Yan2022}, the lack of the F115W photometric band leads to significant photo-z scatter in the redshift range of $7<z<10$. NIRSpec observations of four, high-redshift sources have shown that photo-z's were systematically overestimated, largely due to the lack of constraining power on the Lyman break positioning \citep{Carnall2022,Trussler2022}. The use of this field in measuring the UV LF and other population based statistics is subsequently at risk of scatter between redshift bins due to these uncertainties.

\subsubsection{GLASS}

The Grism Lens Amplified Survey from Space (GLASS) observation programme focuses primarily on the Abell 2744 galaxy cluster with a selection of JWST instrumentation \citep[PID: 1324, PI: T. Treu,][]{Treu2022}. But in parallel to these observations, the GLASS programme has generated one of the deepest NIRCam imaging sets publicly available. GLASS contains two overlapping parallel NIRCam observations in seven filters F090W, F115W, F150W, F200W, F277W, F356W and F444W. This field has already provided several strongly detected high-redshift candidates up to $z=12.5$ \citep{Castellano2022,Naidu2022,Castellano2023}. This study makes use of both epochs of GLASS NIRCam observations. A wider reduction of the Abell-2744 region (including the UNCOVER programme) will be included in future analysis \citep[see][for details about UNCOVER]{Bezanson2022}.

\subsubsection{CEERS}
    
This study also makes use of both primary observing runs (July 2022 \& December 2022) of the Cosmic Evolution Early Release Science Survey \citep[CEERS, PID: 1345, PI: S. Finkelstein, see also][]{Bagley2022}. This consists of 10 NIRCam pointings mosaiced over the Extended Groth Strip \citep[EGS:][]{Groth2994} with 7 photometric bands (F115W, F150W, F200W, F277W, F356W, F410M and F444W). This field provides the single largest area used in our study at 64.15 square arcminutes after masking. The field does lack the F090W band and so we include HST CANDELS imaging of the F606W and F814W filters that was reduced by the CEERS team (their HDR1) following \citet{Koekemoer2011} in order to cover this bluer wavelength range.

\subsubsection{NGDEEP}

This study makes use of the The Next Generation Deep Extragalactic Exploratory Public Survey \citep[NGDEEP, PID: 2079, PIs: S.\@ Finkelstein, Papovich and Pirzkal,][]{Bagley2023,Leung2023} and our reduction is updated relative to initial work conducted in \citet{Austin2023}. In particular, flat field improvements in the NIRCam pipeline had led to great improvements in the red images, though wisp subtraction remains the main limiting factor in the blue images. NGDEEP consists of NIRCam imaging that was run in parallel to NIRISS spectroscopy of the Hubble Ultra Deep Field (HUDF). The NIRCam imaging covers part of the HUDF-Par2 parallel field and consists of 6 broadband filters: F115W, F150W, F200W, F277W, F356W and F444W, all with average depths of $m_{\rm AB}>29.5$. Due to the lack of data below 1 micron, we add in existing HST data from F606W and F814W through the use of the most recent GOODS-S mosaic (v2.5) from the Hubble Legacy Fields team \citep{Illingworth2016,Whitaker2019}.

\subsubsection{JADES}

The first data release (DR1) of the The JWST Advanced Deep Extragalactic Survey \citep[JADES, PID:1180, PI: D. Eisenstein][]{Eisenstein2023,Bunker2023b,Hainline2023a} was conducted in June 2023, including full mosaics using the pmap1084 calibrations \citep{Reike2023}. The release consists of 6 overlapping NIRCam pointings from their DEEP tier which uses the following filters: F090W, F115W, F150W, F200W, F277W, F335M, F356W, F410M and F444W. The field location is around the Hubble Ultra-Deep Field (HUDF) in GOODS-S. In this work, we rereduce this data using our own pipeline for consistency with our other fields. We find that our reductions have average depths that are around 0.1 magnitudes shallower than the official reductions, with one small region of the field affected by residual wisping in the F150W and F200W bands. We employ the same HST/ACS F606W mosaic of the wider GOODS-S region as used in our NGDEEP analysis.

\subsection{The Image Reduction Pipeline}

All of the uncalibrated lower-level JWST data products (`uncal.fits') in this study have been processed following our modified version of the JWST official pipeline. This is a similar process to that used in \citet{Leo2022} and \citet{Adams2023}, but with minor changes based on improvements developed over the first year of handling JWST data. The full pipeline can be summarised as follows: (1) We use version 1.8.2 and CRDS v1084, which contained the most up-to-date NIRCam calibrations at the time of writing and includes the third round of in-flight calibrations. We note that v1084 leads to substantial improvements in the flat fielding of the red NIRCam bands over the previous major version v0995. (2) We subtract templates of wisps, artefacts present in the F150W and F200W imaging, between stage 1 and stage 2 of the pipeline. (3) After stage 2 of the pipeline, we apply the 1/f noise correction derived by Chris Willott.\footnote{\url{https://github.com/chriswillott/jwst}} (4) We skip the sky subtraction step from stage 3 of the pipeline. Instead, we perform background subtraction on each NIRCam frame between stage 2 and stage 3 (`cal.fits' files), allowing for quicker assessment of the background subtraction performance and fine-tuning. This consists of an initial constant/flat background subtraction followed by a 2-dimensional background subtraction using {\tt photutils} \citep{larry_bradley_2022_6825092}. (5) After Stage 3 of the pipeline, we align the final F444W image onto a GAIA-derived WCS \citep{GAIADR2,GAIADR3} using \texttt{tweakreg}, part of the DrizzlePac python package\footnote{\url{https://github.com/spacetelescope/drizzlepac}}, and then match all remaining filters to this derived WCS, ensuring the individual images are aligned to one another. We then pixel-match the images to the F444W image with the use of \texttt{reproject} \citep{Hoffmann2021}.\footnote{\url{https://reproject.readthedocs.io/en/stable/}} The final resolution of the drizzled images is 0.03 arcseconds/pixel. We note that this reduction differs from the official PEARLS and CEERS reductions in \citet{Windhorst2023} and \citet{Bagley2022}. We compare the photometry extracted from the different reduction pipelines in \S 2.4.

\subsection{Source Photometry and Cataloguing}

\begin{table*}[]
\caption{Average $5\sigma$ depths for point sources in 0.32 arcsecond diameter apertures. Depths are calculated by placing random apertures in regions of the image that are empty in the final segmentation maps used to make our catalogues. We perform this process on each of the fields used in this study. The North Ecliptic Pole Time Domain Field (NEP-TDF), El Gordo and MACS-0416 fields are components of the wider PEARLS survey, whilst SMACS-0723, GLASS, CEERS, JADES-DR1 and NGDEEP are public observations. Source catalogues use local depths calculated with the Normalised Mean Absolute Deviation (NMAD) of the nearest 200 empty apertures, deriving final photometric errors on a source-by-source basis. The four NEP-TDF spokes and ten CEERS pointings have mean depths consistent with one another to within 0.1 mags, with the exception of CEERS-P9 which has an additional exposure in F115W and F444W increasing the depths in these two bands by more than 0.2 mags. Where applicable, HST depths are also shown. For the two fields within GOODS-S (NGDEEP and JADES), HST data has two tiers of depth, we show the average depth for each tier. The bottom row shows the total area used in square arcminutes, after the masking was applied and cluster modules omitted.}
\centering
\begin{tabular}{l|ccc|cccccc}
 & & \textbf{PEARLS} & & &  & \textbf{Public} & \\
Band & NEP-TDF & El Gordo & MACS-0416 & SMACS-0723 & GLASS &  CEERS & NGDEEP & JADES DR1 \\ \hline
F606W & 28.70      & --    & --    & -- & --    &  28.50   &  29.2-30.3   &  29.05-30.50 \\
F814W & --      & --    & --     & -- & --      & 28.30  & 28.8-30.5  & -- \\ \hline
F090W & 28.50      & 28.25    & 28.65    & 28.75 & 29.15    &  --   &  --   &  29.60 \\
F115W & 28.50      & 28.25    & 28.60     & -- & 29.10      & 28.85  & 29.80  & 29.80 \\
F150W & 28.50      & 28.20    & 28.50     & 28.80 & 28.85   & 28.60  & 29.50  & 29.70   \\
F200W & 28.65      & 28.45    & 28.65     & 28.95 & 29.00  & 28.80   & 29.50  & 29.70  \\
F277W & 29.20      & 28.95    & 29.15    & 29.45 & 29.55     & 29.20 & 30.30  & 30.20      \\
F335M & --      & --    & --    & -- & --    & --  &  -- &  29.60   \\
F356W & 29.30      & 29.00    & 29.35    & 29.55 & 29.60    & 29.30  &  30.20 & 30.15     \\
F410M & 28.55      & 28.45    & 28.75    & -- & --       & 28.50   &  --  & 29.65 \\
F444W & 28.90      & 28.85    & 29.05    & 29.30 & 29.85   & 28.85 & 30.20 & 30.00 \\ \hline
Area (arcmin$^2$) & 57.32 & 3.90 & 12.30 & 4.31 & 9.76 & 64.15 & 6.31 & 22.98	\\
\end{tabular} \label{tab:5sig}
\end{table*}

With the final mosaics of the field completed, we carry out source identification and extraction. For this process, we use the code {\tt SExtractor} \citep{Bertin1996}. We run in dual-image mode with a weighted stack of the red wide band images (F277W, F356W, F444W) used for object selection, from which we carry out forced aperture photometry for multi-band measurements. This photometry is calculated within 0.32 arcsecond diameter circular apertures and we include an aperture correction derived from simulated \texttt{WebbPSF} point spread functions for each band used \citep{Perrin2012,Perrin2014}. This diameter was chosen to enclose the central/brightest $70-80$ per cent of the flux of a point source, but small enough to avoid contamination.  This enables us to balance the use of the highest signal pixels when calculating fluxes, while avoiding the dependence on a correction derived from the PSF model that is as high (or higher) than the measurement made.

We calculate the depth of our final images by placing circular apertures in `empty' regions of the image. An `empty' aperture is one where no pre-existing source in the image, as identified in our source extraction, is located within 1 arcsecond of the central coordinate of our aperture. These apertures are used to derive an average depth for each field as well as calculate local depths across each field. The final photometric errors for each source are calculated by using the normalised median absolute deviation \citep[NMAD:][]{hoaglin2000understanding} of the nearest 200 empty apertures to calculate the local depth. This process is required in order to generate realistic photometric errors, which {\tt SExtractor} is known to underestimate. The average field depths in each photometric band are presented in Table \ref{tab:5sig}.

Each of our fields is masked by eye. These masks cover diffraction spikes, the few remaining snowballs, regions of high intensity intra-cluster medium (in the NIRCam modules containing any foreground cluster) and a buffer around the edges of the images. The edges of the images are typically shallower due to the dithering patterns used by JWST observations. The total amount of unmasked area used in this study is listed alongside the average depths of each field in Table \ref{tab:5sig}.

\subsection{Comparing Image Processing Pipelines}

At the time of writing, the PEARLS team has developed an official internal pipeline, dubbed ProCess, which has been used to generate official NIRCam reductions for each of its fields using the ProFound software package \citep[][Robotham et al. In Prep]{Robotham2018}. The key difference is ProFound conducts its own implementation of background subtraction and wisp removal \citep[See Appendix A of][ and \citet{Robotham2023} for more details and comparisons]{Windhorst2023}. Similarly, the CEERS collaboration has also generated a slightly different pipeline and publicly released reductions of four NIRCam pointings in their data release v0.5 and v0.6 \citep{Bagley2022}. To assess if these reduction procedures result in systematic differences when measuring the final photometry, we comparison our photometry with that extracted using SExtractor with the same settings on each of the images made by the PEARLS and CEERS teams. It should be noted that the CEERS images are produced using pmap0995, and so we compare pmap0995 version of our CEERS images to theirs in order to properly assess differences in the pipelines, no other changes other than CRDS version are used.

To conduct our comparisons, we first generate a catalogue of galaxies selected in each band and cross-match to the one derived from the F444W image. We then measure the separation of the matches to check if each of our images are correctly aligned with each other. It is important that the alignment between these images is very good for extracting correct colours with forced photometry. We find that the typical source centroid offset is less than 0.02 arcseconds across the fields analysed, or less than two thirds of a pixel.

After this, we compare both aperture and total (AUTO) photometry extracted from our final F444W selected catalogues to those extracted from the PEARLS team's reductions. We calculate the mean magnitude offset of sources between apparent magnitudes of $22<m<26$ in each photometric band, ensuring we are using bright, but non-saturated, sources. We find that our photometry is accurate relative to other reductions, with mean offsets of less than 0.03 magnitudes in blue photometric bands and 0.01 magnitudes in red photometric bands. Similarly, astrometry is consistent to less than 0.07 arcseconds. Our reductions currently only use GAIA stars in the field to calibrate the WCS, while the PEARLS reductions are matched to GAIA-aligned catalogues generated from wider-field, ground-based imaging. The PEARLS reductions thus use a larger set of secondary targets calibrated to the GAIA system when constructing the JWST astrometry and are expected to have higher performance compared to our own. However, this finding shows that this offset is at the $<2$ pixel level. The comparison to the official CEERS collaborations reductions reaches the same conclusion. Future astrometric improvements, e.g. using complementary ground-based data, will be implemented in the second generation of the EPOCHS sample in order to aid spectroscopic follow up.

\section{A Robust Sample of Ultra-High Redshift sources} \label{sec:method}

With the imaging and cataloguing completed, we fit our NIRCam-derived spectral energy distributions for all sources identified in order to derive photometric redshifts for each source. We conduct this process with the use of EAZY-py, the python implementation of the EAZY photometric redshift code\citep{Brammer2008}. We include the newly developed templates presented in \citet{Larson2022}. These templates expand upon one of the default template sets, which use 12 templates generated using the Flexible Stellar Population Synthesis (FSPS) code \citep{Conroy2010}, to include galaxies which are bluer and have stronger emission lines, which we expect are more appropriate for $z>8$ galaxy SEDs. We also test the SED templates used in the JADES works \citep{Hainline2023a}, but find that the \citet{Larson2022} templates are slightly better performing when compared to current spectroscopic results. We also tested the use of a second rapid SED fitting code LePhare \citep{Ilbert2006,Arnouts1999} with the use of \citep{Bruzual2003} templates, but find that emission line strengths often struggled to fit medium band excesses which could lead to the rejection of legitimate sources when quality cuts are applied.

Some photo-z codes have mechanisms for fine-tuning the zero points of the photometric bands in order to optimise the results against a provided spectroscopic sample. We do not employ these techniques because each of the individual chips that make up the NIRCam modules (8 in the blue and 2 in the red) have their own independent calibrations and photometric zero point offsets. We would thus require each of these zero-point modifications to be made on a chip-by-chip basis as opposed to the final mosaic. The number of objects with spectroscopic redshifts within the area of a single chip is inherently small due to the small field of view that each covers.  As a result, the zero point modifications we could apply may be biased towards certain galaxy colours, depending upon the type of spectroscopically confirmed galaxies within each module. Following discussion with members of the community, residual zero point errors were anticipated to be approximately 5 per cent. We subsequently implemented a minimum 10 per cent error on the measured photometry in order to account for potential zero point issues within the NIRCam reduction pipeline in addition to other systematics such minor reduction imperfections and errors in the template SED shapes themselves.

To select a robust sample of high redshift galaxies, we employ the following criteria:

\begin{enumerate}
    \item $\leq 3 \sigma$ detection in all bands blueward of the expected Lyman break. 
    \item $\geq 5\sigma$ detection in the first 2 bands directly redward of the Lyman break, and $\geq 2\sigma$ detection in all other redward bands, excluding f410M. If the galaxy appears only in the long wavelength NIRCam photometry (i.e. a f200W or higher dropout), we increase the requirement to 7 a sigma detection to minimise the potential for selecting red camera artefacts which can appear to look like noisy $16<z<18$ sources.
    \item $\int^{1.10\times z_{phot}}_{0.90\times z_{phot}} \ P(z) \ dz \ \geq \ 0.6 $ to ensure the majority of the redshift PDF is located within the primary peak.
    \item $\chi^2_{red} < 3 (6)$ for best-fitting SED to be classed as robust (good).
    \item $\Delta \chi^2 \geq$ 4 between high-z and low-z EAZY \ runs (where maximum redshift is set to 6). This ensures that the high-z solution is much more statisically probable. 
    \item If the 50\% encircled flux radius ({\tt FLUX\_RADIUS} parameter in SExtractor) is smaller than the FWHM of the PSF in the F444W band, then we require that $\Delta \chi^2 \geq$ 4 between the best-fitting high-z galaxy solution and the best-fitting brown dwarf template (see below for further discussion).
    \item 50\% encircled flux radius is $\geq$1.5 pixels in the long-wavelength wideband NIRCam photometry (f277W, f356W, f444W). This avoids detecting hot pixels in the LW detectors as f200W dropouts. 
\end{enumerate}

Targets must pass all of the above criteria to fall within our parent sample of high-z galaxies. We limit our parent high-z sample to having a redshift above $z>6.5$.

\begin{figure}
\centering
\includegraphics[width=1.02\columnwidth]{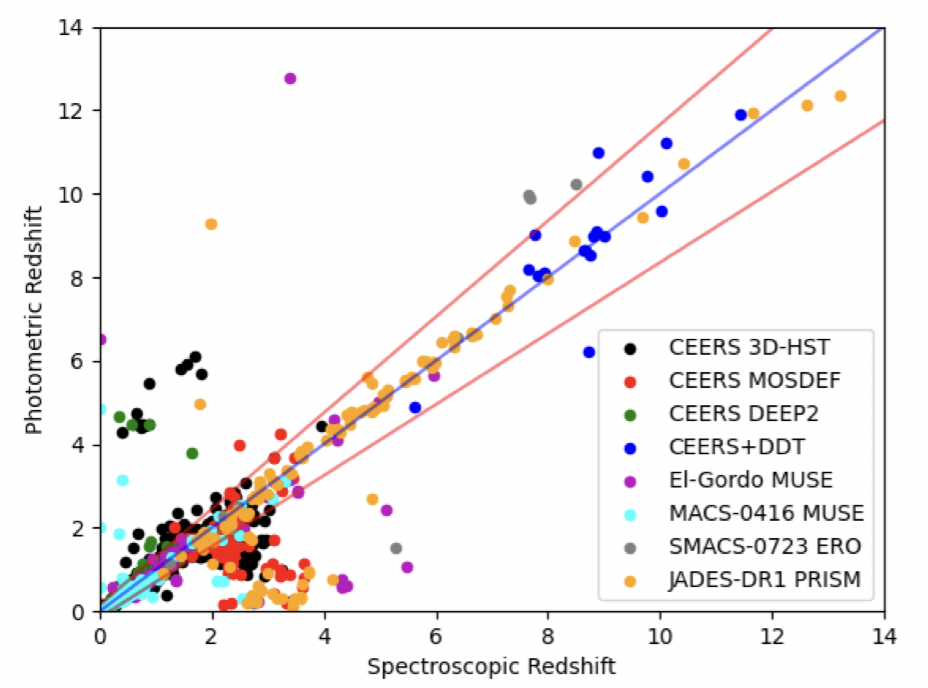}\caption{Diagnostic plots showing the performance of our EAZY photometric redshift procedure when applied to our NIRCam photometry. The CEERS DDT refers to the Directors Discretionary Time proposal in the CEERS field \citep{arrabal2023}, the other CEERS sources are those from \citet{Tang2023} and \citet{Haro2023b}. CEERS data includes HST F606W and F814W imaging due to its lack coverage below 1 micron. Despite being largely limited to NIR wavelength range, photo-z's generally perform well with the greatest spread around $2<z<3$. The blue line shows the ideal case of photometric redshifts matching spectroscopic redshifts, the red lines show 15 percent offsets in $1+z$, often used to describe the performance of photometric redshifts.}
\label{fig:photoz}
\end{figure}

\subsection{Identifying and Removing Brown Dwarfs}

Despite the wavelength coverage provided by JWST, local brown dwarf stars can be misidentified as high-redshift galaxies and enter high-z samples if not properly accounted for. In order to address this, we conduct SED fitting of brown dwarf templates from the Sonora Bobcat model set \citep{Marley2022,Hainline2023b}. We then remove galaxies identified as being both better fit by a brown dwarf template and have a size consistent with the FWHM of the PSF in the F444W band. This process removed 56 objects from our sample. The majority of these (39 objects) had galaxy photo-z's of $6.5<z<7.5$. This procedure is only run on objects initially identified as being robust high-z candidates and not on every single source, so it does not amount to a complete brown dwarf sample though this may be explored in future work. As a sanity check, we fit the objects identified as brown dwarfs in the CEERS field by \citet{Hainline2023b}, which all did not pass all of our quality checks to enter our high-z sample in the first place, and find that we would have still identified these objects as potential brown dwarfs.

\subsection{Testing the selection procedure}

To validate the photometric redshifts, we compare our photo-z measurements to catalogues of spectroscopic redshifts available in these fields. Within SMACS-0723, we compare to the 10 objects covered by NIRSpec in \citet{Carnall2022}. In El-Gordo and MACS-0416 we compare to MUSE observations reported by \citet{Caminha2017,Caminha2022}. Within CEERS, we combine catalogues obtained from the DEEP2 \citep{Newman2013}, 3D-HST \citep{Momcheva2016} and MOSDEF/MOSFIRE \citep{Kriek2015} programmes to construct a catalogue of 1500+ spectroscopic redshifts. A summary of our findings is presented in Figure \ref{fig:photoz}.

For the SMACS-0723 cluster, we find that 6/10 sources from \citet{Carnall2022} have photometric redshift solutions in our catalogues within 15 per cent of the spectroscopic solution. Three of these outliers are high-redshift sources, with photometric redshifts $z_{phot}>9$ while the spectroscopic redshift is around $z\sim7.6$. While correctly identified as being high redshift, the lack of F115W leads to large uncertainties on the Lyman Break positioning. Without this constraint, photo-z codes favour fitting the observed F444W excess of these sources with a Balmer break at $z\sim9.5$, as opposed to [OIII]/H$\beta$ line emission at $z\sim7.6$ \citep[see][for a detailed discussion]{Trussler2022}. While the data available in this field has proven its capabilities in correctly identifying high-z from low-z targets, these broad uncertainties in the redshift range of $7<z<10$ make its use in the measurement of the UV LF inherently risky due to the low number statistics combined with the potential for sources to scatter between redshift bins.

Within El Gordo, we test our photo-z procedure by matching our photometric sample to 402 spectroscopic redshifts obtained by the MUSE instrument \citep{Caminha2022}. The sample contains a large number of cluster members at $z\sim0.87$, as well as a wide range of galaxy redshifts from $0<z<6$. We obtain cross matches for 238 sources with confidence flags indicating a secure redshift. Our EAZY photometric procedure obtains 218/238 (92 per cent) redshifts. Similar to El Gordo, we match our catalogues of MACS-0416 to MUSE spectroscopic catalogues \citep{Caminha2017}. We find 261 cross matches, of which 235/261 (90\%) have photometric redshifts within 15 per cent of the spectroscopic redshift.
 
Within CEERS, we compare our photometric redshifts to spectroscopic redshifts obtained from the DEEP2 survey \citep{Newman2013}, 3D-HST \citep{Momcheva2016} and MOSDEF/MOSFIRE surveys \citep{Kriek2015} of the Extended Goth Strip (EGS). Using DEEP2, we obtain 280 cross-matches to our CEERS catalogues, of these we find 265/280 (95\%) have photometric redshifts that agree with the spectroscopic redshifts to within 15 per cent. All of these sources are at $z_{\rm spec}<1.6$. With MOSDEF, we find 117/166 (70\%) cross matches have redshifts that agree within 15 per cent. These objects have higher spectroscopic redshifts of $1.4<z<4$. Finally, we obtain 1135/1249 (91\%) photometric redshifts within 15 per cent of the spectroscopic solution using 3D-HST. This sample spans a redshift range of $0<z<4$. 

From the CEERS survey itself, we compare to 19 high-z spec-z galaxies \citep{Haro2023b,Tang2023} and find a 79\% spec-z recovery rate. Of the failures, two are fainter than our $5\sigma$ limits, one is a partially blended source and one is still classed as high-z but our redshift was too high ($z=10.9$ instead of the $z=8.88$ spec-z). We successfully recover a number of significant high-z sources, such as the the AGN candidate at $z=8.68$ identified in \citep{Larson2023} with $z_{phot}=8.65$ and `Maisie's galaxy' at $z=11.44$ with $z_{phot}=11.90$ \citep{Finkelstein2022c} In total, 11/15 identified high-z galaxies enter our final sample, with two galaxies in our masking near image edges and two more below our detection limits. The CEERS team implements smaller extraction apertures (0.2 arcsec diameter) than in this study (0.32 arcsec), enabling fainter sources to be identified for spectroscopic follow up, though with increased risk of influence of bad pixels and reliance on correct modelling of the PSF. Consequently, it is a common occurrence for CEERS selections to identify sources which are approximately $4\sigma$ in our work.

Finally, we test the set-up of our photometric redshift codes on the early photometry and spectroscopy released from the JADES collaboration \citep{Robertson2022,CurtisLake2022}. Using the DR1 PRISM catalogue, there are 144 spectroscopic cross matches with our catalogues. Of these, 97/124 (78\%) unmasked objects have photometric redshifts within 15\% and all 15 objects at $z>6.5$ have the correct redshifts. 12/15 meet our robust criteria, with those rejected failing our detection thresholds. Further investigation identifies that two of these three objects lie in a small region of poor wisp subtraction, leading to shallow local depths.

\subsection{Completeness  Analysis}
\begin{figure*}
\centering
\includegraphics[width=2\columnwidth]{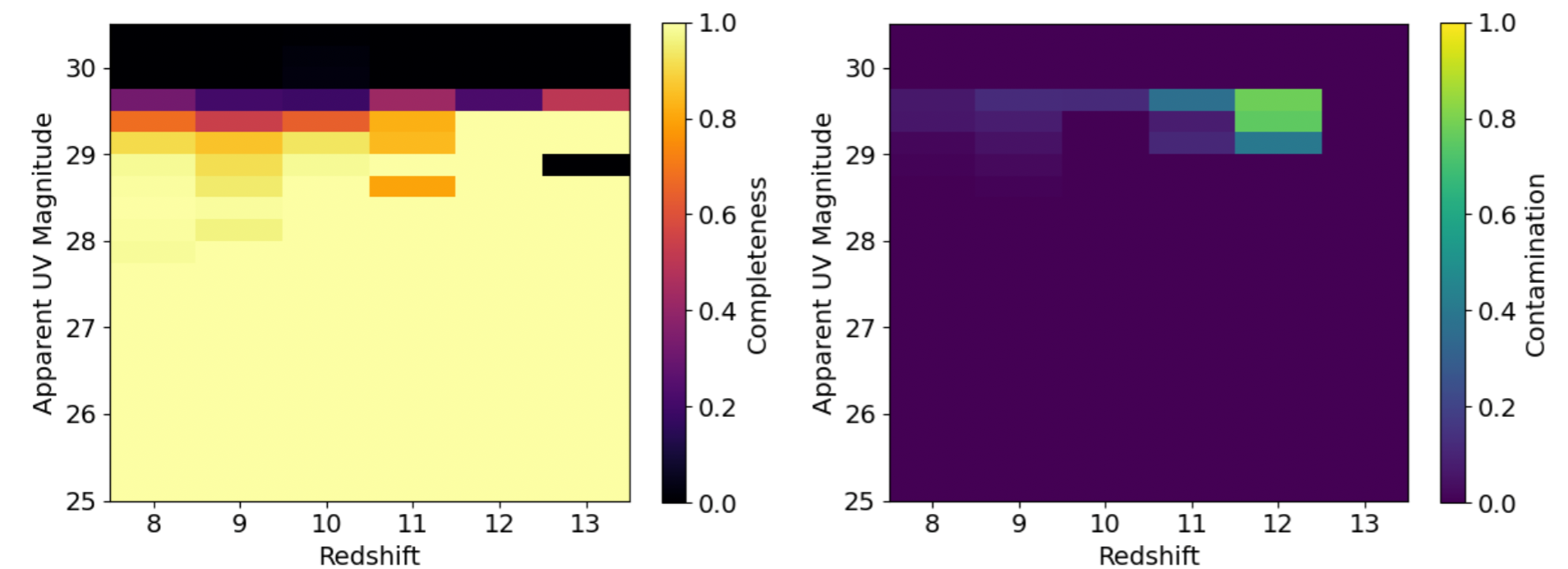}
\caption{An example of the selection completeness simulation run using the JAGUAR SAM with the survey set-up of JADES. The colour bar indicates the fractional completeness recovered and the contamination fraction of $z<5$ objects entering the sample. This forms the completeness factor $C_i$ which is used in Equation 1. Bins with contamination above 10\% or completeness less than 50\% are not used in the measurement of the UV LF as they are considered unreliable. Other fields show similar results but shifted to brighter luminosities on account of their shallower depths.}
\label{fig:complete}
\end{figure*}

\subsubsection{Source Detection}

We carry out completeness tests on our data using a simple simulation to extract artificial sources that are inserted into the images. We insert galaxies with a Sérsic index of $n=1$, and absolute luminosities ranging from $-16$ to $-24$ into the F444W image used for our selection. Half-light radii of the artificial sources are varied and follow the distribution of theoretical predicted sizes of distant galaxies from the BlueTides simulation \citep{Bluetides-I, Bluetides-II,Marshall2022}. The maximum half-light radius of our simulated galaxies was placed at 1000 pc. With the exquisite resolution provided by JWST, galaxies at ultra-high redshifts are often slightly larger than the PSF, which can be a key factor that must be considered when trying to understand completeness.

The simulations we  carry are are performed by sweeping the magnitude range in 0.2 mag intervals and with redshift intervals of 0.5, placing 1000 artificial galaxies in the field of interest for each interval, and then running SExtractor with the same settings as for our unmodified images. A size-luminosity relation of $r \propto L^{0.5}$ was assumed throughout \citep{Grazian2012}, with a reference radius of 800 pc at $M_{UV} = -21$. This allows us to efficiency within our set up for detecting galaxies at various redshifts, and what fraction enter our catalogues.  This is necessary for using our galaxy counts to produce a luminosity function.  This procedure is carried out for each individual field. 

\subsubsection{Selection Procedure}

To test our photometric redshifts and selection procedure further, we employ the use of simulated photometric catalogues generated by the JAGUAR semi-analytic model \citep{Williams18}. This model is selected due to its performance in replicating colours of known high-z galaxies \citep{Wilkins2023}. In order to have reasonable numbers of luminous high-z objects, we use five realisations of the simulation, producing around 550 square arcminutes of simulated area (10 times that of any one field used in this work). From these simulations, we add photometric errors and scatter the true photometry in accordance to the average depths for each of our survey fields. We then run our photometric redshift and selection pipeline on the resultant catalogues to assess the completeness and contamination (which we define as sources at $z<5$) that fall into luminosity bins ($\Delta m = 0.25$) in the photometric band containing the rest-frame UV per redshift interval of $\Delta z = 1$. 

We find that our selection criteria is the dominant limitation in completeness as opposed to source detection. This is largely because we utilise a stack of the deepest bands in the reddest wavelengths for detection but employ $5\sigma$ cuts on the blue bands in our selection process. As is anticipated, the completeness is high (85-95\%) for the most luminous sources and rapidly plummets as the $5\sigma$ criteria is reached. Of interest is the contamination of low-z sources and how this behaves with apparent magnitude. For bright sources, this is unsurprisingly very low (0-5\%). However, we find that contamination begins to rise just before completeness drops off. With upwards of $\sim10\%$ levels of contamination in the 0.25 mag width magnitude bin just before the $5\sigma$ limits are reached. We show an example of this simulations applied to the JADES survey design in Figure \ref{fig:complete}.

For the purpose of measuring the ultraviolet luminosity function in this work, we only use the parameter space identified as being greater than 50 per cent complete in our simulations. This is to avoid depending on using corrections that are as large, or larger, than any measurements being undertaken. {We additionally do not consider the use of a field in the UV LF when the contamination level is greater than 10\%. This conservative limit is selected (and a cut implemented as opposed to correcting for it) because the contaminating sources themselves are poorly understood and their relative distribution within the JAGUAR simulation may not replicate reality. We therefore cut at the point we start to see an uptick in contamination. Ultimately, obtaining spectroscopy of contaminating sources, in addition to real high-z sources, will be key to understanding what sources really are falling within selections and how numerous they may be.

\subsection{The Final Sample}

Applying this procedure to our catalogues for each field, we obtain a final sample of 1046 high-redshift candidates ($z>6.5$) over an area of 181 arcmin$^2$. Of these, 413 have best fitting redshifts between $7.5<z<13.5$. This is one of the largest samples (in area and raw numbers) of high-z, JWST selected candidates to date. The full list of candidates will be presented in the companion paper Conselice et al. (in prep). To provide some context for this work, we present a selection of SED fits for some `highlight' sources, as well as some random sources, in Appendix B.

\subsubsection{The Most UV Luminous Sources}

The parent galaxy sample contains over 1000 sources located within the first billion years of the history of the Universe. Here, we highlight a few examples of extreme sources used in the measurement of the UV LF at $z>7.5$.

The known AGN candidate at $z=8.68$ identified in \citet{Larson2023} remains the most intrinsically luminous source in our compilation of JWST fields with $M_{\rm UV} = -22.28$. The next brightest source at $z>7.5$ is identified as NEP-3:15371 in our catalogues with $M_{\rm UV} = -21.34$ and $z=9.00\pm0.1$. The North Ecliptic Pole Time Domain Field contains a relatively large number of UV luminous sources at $z\sim8.1-8.3$ including 5 sources with $M_{\rm UV} < -21.0$ which may be indicative of an overdensity. At $z>10$, the brightest source is NEP-3:18221 with $z=11.5\pm0.4$ and $M_{\rm UV} = -21.03$ which is followed by the two sources located within the GLASS field at $z=10.7$ and $z=12.4$ identified in \citet{Naidu2022} and \citet{Castellano2022}. This indicates how fortuitous the GLASS galaxy candidates were, given this study probes nearly 20 times the area and identifies only a single intrinsically brighter source at $z>10$. SED examples for these luminous galaxies are presented in Appendix B.

\subsection{Comparisons to Samples from Other Studies}

\begin{table}
\caption{The overlap of sources in our final catalogues compared to those of other studies that have analysed the same fields. The primary value indicates objects which would meet our strict selection criteria. Those that are in brackets indicate the inclusion of potential candidates which have matching primary photo-z solutions (within 15\%) but fail some of our criteria. For example, they have a significant secondary solution, are just below $5\sigma$ detected or have a $3\sigma$ detection bluewards of the anticipated Lyman break in our analysis. More details are provided in Appendix A.}
\hspace{-1cm}\begin{tabular}{l|lll}
Study          & Sources & Recovered & Overlap \%  \\ 
 & & & Selected ($z_{\rm phot}$) \\ \hline
Naidu+22       & 2            & 2 (2) & 100 (100)         \\
Atek+22        & 10           & 2 (5) & 20 (50)           \\
Finkelstein+22 & 26           & 12 (18) & 46 (69)          \\
Donnan+22      & 43           & 16 (28) & 37 (65)            \\
Harikane+22    & 21           & 4 (6) & 19 (29)             \\
Yan+22    & 88           & 0 (11) & 0 (13)             \\
Bouwens+22b     & 33           & 6 (20)  & 18 (61)            \\
McLeod+23     & 32           & 24 (28)  & 75 (88)            \\
Finkelstein+23     & 88           & 29 (62)  & 33 (70)            \\\hline
\end{tabular} \label{tab:compare}
\end{table}

We compare our sample to those obtained from the work of \citet{Atek2022,Naidu2022,Castellano2022,Finkelstein2022c,Donnan2022,Yan2022,Harikane2023,Bouwens2022c,McLeod2023} and \citet{Finkelstein2023}. We focus on examining which sources appear in both our catalogues and those generated by these works, the matching radius used is 0.3 arcseconds. A summary of our findings is presented in Table \ref{tab:compare}. We find that on average, only about 20-40 per cent of sources appear across multiple early studies, a conclusion also reached by \citet{Bouwens2022c} for their own sample. For the sources that are selected by other works and not in our own, we find that about half of them do have a primary high-redshift solution and half of them are fainter than our detection limits. Since the shape of the UV LF means that there are more galaxies at fainter luminosities, the scattering of sources above and below these detection limits can be significant. This is especially so when considering different studies that use different apertures for measuring photometry \citep[resulting in different limiting magnitudes, see e.g.][which uses 0.1 arcsec radius apertures]{Finkelstein2022c}. We note one main exception for our comparisons with early studies is \citet{Bouwens2022c}, which selects 10 luminous objects at $z\sim6.5$ which have a Lyman break located within the F090W JWST band and hence fail our criteria of having a fully clean band beyond the break. The numbers in Table \ref{tab:compare} are consequently not as first appears for this reason. More detailed comparisons are presented in Appendix A. 

Perhaps reassuringly, more recent studies using newer calibrations and implementing improved knowledge of high-z selection and contamination management have greater degrees of overlap. Our strongest sample overlap is with the work by \citet{McLeod2023}, with 75\% of their sources entering our final sample and with 88\% of sources agreeing on primary photo-z solutions. Also recently, the CEERS team published their official list of high-z candidates \citep{Finkelstein2023}. While only 33\% enter our final sample, the primary cause of this is the aperture size choice. This means there are a large fraction of objects which lie just below our $5\sigma$ limits and this is the primary reason they do not make our sample, we tend to agree (70\%) that these sources have strong high-z solutions.

\section{The Ultraviolet Luminosity Function}

\subsection{Background}

Using our final list of candidate galaxies, fit SEDs, and an understanding of our completeness, we construct an initial ultraviolet luminosity function (UV LF) at $z>8$ from these JWST data. To conduct this measurement, we use the 1/$V_{\rm max}$ method \citep{Schmidt1968,rowanrobinson1968}. The maximum observable redshift for each source $z_{\text{det}}$ is calculated by iteratively shifting the best-fit SED of each source in small steps of $\delta z = 0.01$ and convolving with the filter sets until it no longer meets our rest-frame UV detection criteria. The object can thus be detected within a co-moving volume ($V_{\rm max}$) which is the volume contained within a section of a spherical shell between the radius $z_{\text{min}} < z < z_{\text{max}}$. Here, $z_{\text{min}}$ is the lower boundary of the redshift bin being considered and $z_{\text{max}}$ is either the maximum bound of the redshift bin or $z_{\text{det}}$ if it is found to be lower. We also experiment with the Lynden-Bell $C^-$ method \citep{Lynden-Bell1971,Woodroofe1985,Wang1986,Efron1992} and find results are generally consistent within 0.5$\sigma$.

To calculate the $M_{\rm UV}$ value for each galaxy, we convolve a 100\AA\, wide top-hat filter centred on the rest-frame 1500\AA\, for the best-fitting SED \citep{Finkelstein2015,Adams2021,Finkelstein2022c}. These SEDs are generated using aperture-corrected photometry, which can underestimate the total flux for resolved sources. We therefore apply a simple correction factor using the ratio between the aperture and total photometry ({\tt MAG\_AUTO}) measured by SExtractor in the photometric band containing the rest-frame 1500\AA.

For the measurement of the UV LF, we discard the SMACS-0723 field. With the absence of the F115W photometric band, there is a much greater degree of uncertainty on candidate redshifts between $8<z<10$ \citep[see e.g.][]{Trussler2022}. Since we are not using the NIRCam modules containing lensing clusters in this study, the SMACS-0723 field provides minimal cosmic volume, but comes with elevated risk of objects scattering between redshift bins.

\subsection{The UV LF Construction}

Implementing the above, the rest frame UV LF ($\Phi(M)$) is calculated using:

\begin{equation}\label{eqn:lf}
\Phi(M) d \log(M) = \frac{1}{\Delta M } \sum_i^N \frac{1}{C_{i,f} V_{\rm max,i}} ,
\end{equation}
where $\Delta M$ is the width of the magnitude bins and $C_{i,f}$ is the completeness correction for a galaxy $i$ depending on its location within the fields, $f$.  We only use fields and redshift ranges in this calculation if their completeness is greater than 50 per cent and contamination less than 10 per cent. This ensures that our measurements are not dominated by any derived corrections (similar to the aperture choice when doing PSF corrections). The magnitude bin widths used are 0.5 magnitudes, increasing to 1.0 magnitude for the faintest bins.

The measured uncertainty of the LF is given by:

\begin{equation}
\delta \Phi(M) = \frac{1}{\Delta M} \sqrt{\sum_i^N \left(\frac{1}{C_{i,f} V_{\rm max,i}}\right)^2} .
\end{equation}

For small numbers of objects ($N<5$), the above formula overestimates the Poisson error (leading to instances of e.g. $1\pm1$). In this regime, we directly calculate the Poisson confidence interval using an improved, but still conservative, estimator based on the $\chi^2$ distribution $I = [0.5\chi^2_{2N,a/2}, 0.5\chi^2_{2(N+1),1-a/2}]$ \citep{Ulm1990}.

We measure the UV LF in four primary redshift bins: $7.5<z<8.5$, $8.5<z<9.5$, $9.5<z<11.5$ and $11.5<z<13.5$. Taking into account the completeness criteria of each field, these four redshift bins have a total of 166, 59, 27 and 9 galaxies respectively. We note that our full sample of galaxies selected is larger than that used to measure the UV LF. This is due to the nature of the shape of the UV LF, where there is a greater number density of fainter sources occupying the $<50$ per cent completeness regime. The final results are presented in Table \ref{Tab:Points}.

\begin{table}
\caption{The measured rest-frame UV LF and its error margin in the redshift bins $7.5<z<8.5$, $8.5<z<9.5$, $9.5<z<11.5$ and $11.5<z<13.5$. Column 1 shows the bin in absolute UV magnitude ($\lambda_{\rm rest}=1500$\AA). Column 2 shows the number density of objects and column 3 shows the uncertainties in the number density, which are calculated with equation 2 and summed in quadrature with the derived cosmic variance.}
\centering
\begin{tabular}{lcc}
\hline
$M_{\rm UV}$     & $\Phi (10^{-5})$                        & $\delta\Phi (10^{-5})$                  \\
$[\textrm{mag}]$ & $[\textrm{mag}^{-1} \textrm{Mpc}^{-3}]$ & $[\textrm{mag}^{-1} \textrm{Mpc}^{-3}]$ \\ \hline
$z=8$          &          &            \\
-21.35          &  2.277        &  1.226          \\
-20.85          &  9.974        &  3.137          \\
-20.35          &  13.12        &  3.840          \\
-19.85          &  28.64        &  7.247          \\
-19.35          &  54.04        &  22.19          \\
-18.60          &  69.35        &  28.41          \\\hline
$z=8$-NoNEP          &          &            \\
-21.35          &  0.837        &  0.711          \\
-20.85          &  7.935        &  3.086          \\
-20.35          &  9.090        &  3.409          \\
-19.85          &  20.39        &  5.967          \\
-19.35          &  45.46        &  19.97          \\
-18.60          &  67.41        &  27.73          \\\hline
$z=9$          &          &            \\
-22.05          &  0.628        &  0.536          \\
-21.55          &  0.628        &  0.536          \\
-21.05          &  1.257        &  0.851          \\
-20.55          &  6.427        &  2.534          \\
-20.05          &  10.76        &  5.825          \\
-19.55          &  18.22        &  9.442          \\
-18.80          &  42.45        &  20.32          \\ \hline
$z=10.5$          &          &            \\
-20.95          &  0.721        &  0.480          \\
-20.45          &  1.855        &  0.888          \\
-19.75          &  3.331        &  1.314          \\
-19.45          &  6.674        &  3.960          \\
-18.70          &  9.996        &  6.155          \\ \hline
$z=12.5$          &          &            \\
-20.75          &  0.852        &  0.570          \\
-20.25          &  2.148        &  1.375          \\
-19.50          &  5.923        &  3.884          \\ \hline
\end{tabular} \label{Tab:Points}
\end{table}

\subsection{Cosmic Variance}

Pushing the redshift frontier, the implications of cosmic variance on the measurement of the UV LF are unclear \citep{Dawoodbhoy2023}. To obtain a ballpark figure for the systematic errors introduced by cosmic variance, we utilise the cosmic variance calculations introduced in \citet{Driver2010}. We calculate the variance from each individual field using its approximate dimensions (e.g. we treat all 10 pointings of CEERS as a single large field) and combine them together using a volume weighted sum in quadrature \citep{Moster2011}. Across our UV LFs, the impact of cosmic variance is estimated to be approximately $17\text{---}21$ per cent at the bright end, and up to $42\%$ per cent at the faint end (where there is only the NGDEEP and JADES fields). Cosmic variance is thus a significant contributor to the overall uncertainty on the measured UV LFs, even when combining several, well-separated JWST fields. We add this error in quadrature with the Poisson error and these are included in the errors reported in Table \ref{Tab:Points}.

\subsection{Fitting the measured UV LF}

To fit our UV LFs, we use both the Schechter \citep{Schechter1976} (Sch) and Double Power Law (DPL) functional forms. We use the Levenberg-Marquardt (LM) approach to minimise the $\chi^2$ and obtain the best fitting parameters. We then compare this result to one obtained using a full Markov-Chain-Monte-Carlo (MCMC) using {\tt emcee} \citep{emcee} with wide and uniform priors. The results of these fits are presented in Table \ref{tab:fits}. Our observations do not probe enough area to provide constraints on the UV LF much brighter than the 'knee` at $M_{\rm UV} < -21.5$. To assist in our fitting, we include the use of the ground-based, UltraVISTA derived data points presented in \citet{Donnan2022}. The MCMC methodology is found to produce larger, but likely more realistic, errors than the simpler LM approach.  Overall, the addition of this extra data points does not change the fitted results, but does give us a better accuracy in their measurements.

Due to the few data points present at $z=10.5$ and $z=12.5$, we find that leaving all variables free leads to unconstrained models (see e.g. the DPL fit for $z=10.5$ in Table \ref{tab:fits}). We thus fix the faint-end slope for $z=10.5$ and all but the normalisation for $z=12.5$ for both the Schechter and DPL functional forms. For the faint-end slope we fix to $\alpha=-2.1$ and the other parameters are fixed to the best solution from the previous redshift bin.While we provide the various fit values in Table \ref{tab:fits}, greater depths and volumes are required in order to properly constrain the UV LF at $z\geq12$ and generate realistic model fits.

\begin{table*}
\caption{The best-fit parameters for the Schechter (Sch) and Double Power Law (DPL) functional forms for both the Levenberg–Marquard (LM) approach and the MCMC approach. For the MCMC, we show the median and standard error obtained from the posterior. The individual MCMC sample with highest probability matches the best fitting parameters of the LM methodology. For the $z=8$ redshift bin, our results do not probe as faint as some previous HST studies, we thus include the faintest 3 data points from \citet{McLure2013} to improves the constraints on the faint end. Values presented with an asterisk indicate the parameter is fixed to this value in the fitting procedure.}
\centering
\begin{tabular}{ll|llll}
Redshift & Method   & $\log10(\Phi)$ & $M^*$ & $\alpha$ & $\beta$ \\ \hline
8        & Sch/LM   & $-3.19 \pm 0.21$    & $-20.00\pm 0.22$    & $-1.74 \pm 0.23$      & --      \\
8       & Sch/MCMC & $-3.24^{+0.25}_{-0.37}$    & $-20.02^{+0.27}_{-0.35}$   & $-1.78^{+0.30}_{-0.30}$      &  --    \\ 
8        & DPL/LM   & $-3.95 \pm 0.31$    & $-20.76\pm 0.36$    & $-2.13 \pm 0.18$      & $-5.87 \pm 1.43$      \\
8        & DPL/MCMC &  $-3.76^{+0.31}_{-0.27}$   & $-20.50^{+0.39}_{-0.26}$   & $-2.04^{+0.24}_{-0.20}$      &  $-5.05^{+0.64}_{-0.62}$    \\ \hline
8-NoNEP        & Sch/LM   & $-3.70 \pm 0.33$    & $-20.43\pm 0.31$    & $-2.13 \pm 0.22$      & --      \\
8-NoNEP       & Sch/MCMC & $-3.75^{+0.39}_{-0.53}$    & $-20.45^{+0.39}_{-0.49}$   & $-2.16^{+0.29}_{-0.27}$      &  --    \\ 
8-NoNEP        & DPL/LM   & $-3.94 \pm 0.53$    & $-20.54\pm 0.62$    & $-2.23 \pm 0.26$      & $-4.69 \pm 0.90$      \\
8-NoNEP        & DPL/MCMC &  $-3.88^{+0.54}_{-0.42}$   & $-20.45^{+0.70}_{-0.43}$   & $-2.20^{+0.32}_{-0.23}$      &  $-4.56^{+0.66}_{-0.82}$    \\ \hline
9        & Sch/LM   & $-3.34 \pm 0.43$    & $-19.80\pm 0.55$    & $-1.50 \pm 0.85$      & --      \\
9        & Sch/MCMC & $-3.92^{+0.61}_{-1.20}$    & $-20.33^{+0.63}_{-1.07}$    & $-2.08^{+0.73}_{-0.72}$ & --      \\
9        & DPL/LM  & $-4.08 \pm 0.61$    & $-20.66\pm 0.59$    & $-1.91 \pm 0.73$      & $-6.0 \pm 2.24$      \\
9        & DPL/MCMC & $-4.14^{+0.44}_{-0.87}$ & $-20.59^{+0.54}_{-0.60}$ & $-2.06^{+0.65}_{-0.76}$     & $-5.19^{+1.01}_{-1.15}$     \\
9        & Sch/LM   & $-3.81 \pm 0.23$    & $-20.31\pm 0.22$    & $-2.1^*$      & --      \\
9        & Sch/MCMC & $-3.90^{+0.26}_{-0.34}$    & $-20.35^{+0.27}_{-0.31}$    & $-2.1^*$ & --      \\
9        & DPL/LM  & $-4.37 \pm 0.16$    & $-21.00\pm 0.20$    & $-2.1^*$      & $-6.0 \pm 5.5$      \\
9        & DPL/MCMC & $-4.16^{+0.28}_{-0.25}$ & $-20.60^{+0.43}_{-0.24}$ & $-2.1^*$     & $-5.35^{+1.00}_{-1.08}$     \\ \hline
10.5        & Sch/LM   & $-4.83 \pm 0.57$    & $-21.29 \pm 0.61$    & $-1.93 \pm 0.48$ & --      \\
10.5        & Sch/MCMC & $-4.71^{+0.58}_{-0.91}$    & $-20.98^{+0.89}_{-0.92}$ & $-1.89^{+0.54}_{-0.52}$       & --     \\
10.5        & DPL/LM   & $-4.41\pm1.36$    & $-20.39\pm2.1$    & $-1.65\pm1.31$      & $-3.67\pm1.16$      \\
10.5        & DPL/MCMC & $-4.93^{+0.62}_{-0.82}$    & $-20.95^{+0.92}_{-0.99}$   &  $-2.07^{+0.62}_{-0.57}$     & $-4.46^{+1.25}_{-1.64}$     \\
10.5        & Sch/LM   & $-5.04 \pm 0.25$    & $-21.49\pm 0.37$    & $-2.1^*$ & --      \\
10.5        & Sch/MCMC & $-5.00^{+0.38}_{-0.34}$    & $-21.35^{+0.65}_{-0.45}$ & $-2.1^*$       &  --    \\
10.5        & DPL/LM   & $-5.04 \pm 0.55$    & $-21.26 \pm 0.92$    & $-2.1^*$      & $-4.21 \pm 1.36$      \\
10.5        & DPL/MCMC & $-5.02^{+0.47}_{-0.39}$    & $-21.10^{+0.78}_{-0.64}$   &  $-2.1^*$     & $-4.45^{+0.97}_{-1.02}$     \\\hline
12.5        & Sch/LM   & $-4.99 \pm 0.17$    & $-21.35^*$    & $-2.1^*$ & --      \\
12.5        & Sch/MCMC & $-5.12 ^{+0.21}_{-0.63}$    & $-21.35^*$ & $-2.1^*$       &      \\
12.5        & DPL/LM   & $-4.99 \pm 0.04$    & $-21.1^*$    & $-2.1^*$      & $-4.4^*$      \\
12.5        & DPL/MCMC & $-5.13^{+0.21}_{-0.65}$    & $-21.1^*$   &  $-2.1^*$     & $-4.4^*$     \\\hline
\end{tabular} \label{tab:fits}
\end{table*}

\subsection{The Measured $z=8$ and $z=9$ UV LF}

\begin{figure*}
\centering
\includegraphics[width=1.03\columnwidth]{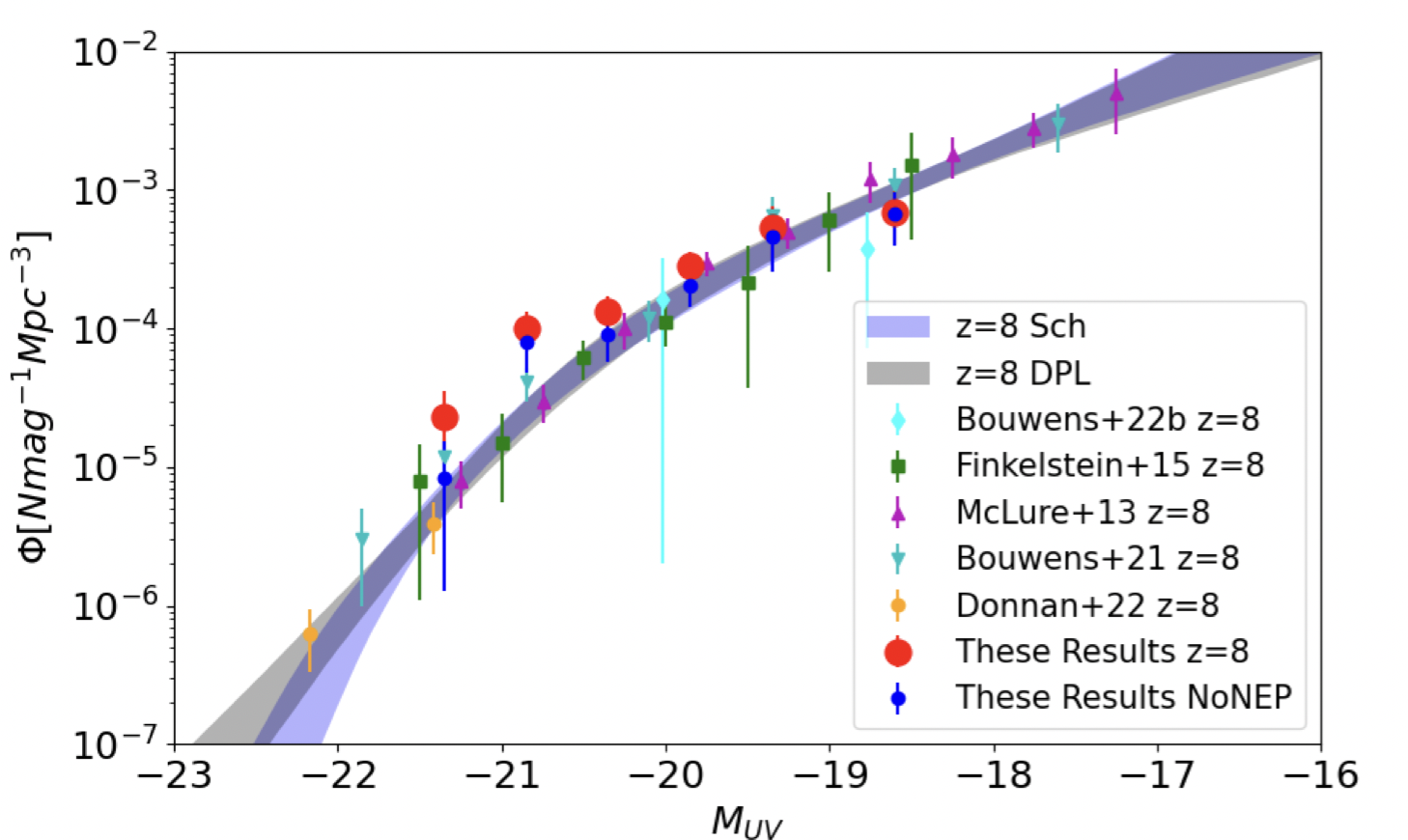}
\includegraphics[width=0.96\columnwidth]{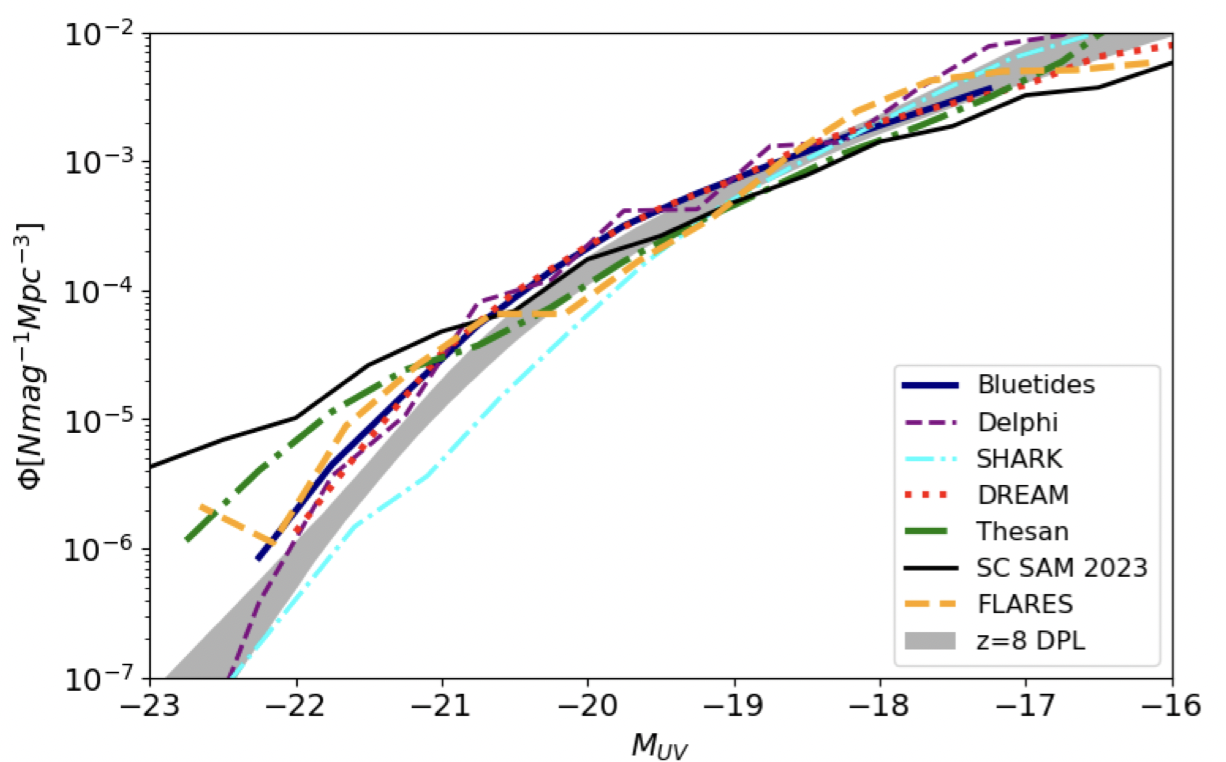}\caption{The UV LF at $7.5<z<8.5$ as measured in this study. We compare our measurements (large red points) to a selection of recent observations (left) and predictions from simulations (right). The shaded blue and black regions indicate the best fitting Schechter and DPL functions to our data plus the UltraVISTA data from \citet{Donnan2022} and faint-end data from \citet{McLure2013} to further constrain the UV LF shape. We compare with observations from \citet{McLure2013,Finkelstein2015,Bouwens2021,Donnan2022} and \citet{bouwens2022}. The simulation results presented on the right side include Bluetides \citep{Wilkins2017}, Delphi \citep{dayal2014,dayal2022}, DREAM \citep{Drakos2022}, Thesan \citep{Kannan2022}, Santa Cruz semi-analytic Model \citep{Yung2023}, SHARK 2.0 \citep{Lagos2023} and FLARES \citep{Vijayan2021,Wilkins2023}.}
\label{fig:UVLF_Z8}
\end{figure*}

Since we are using only blank-field NIRCam data in this study, the $7.5<z<8.5$ redshift bin does not include any benefits which can be provided by gravitational lensing. As a result, the JWST observations in this redshift range are more comparable to the observations previously undertaken by the HST and presently do not extend our knowledge of the faint end \citep[see e.g.][for lensing assisted studies using the HST]{Livermore2017,Ishigaki2018,Bhatawdekar2019}. This measurement thus serves as a sanity check against more established luminosity function measurements that lie comfortably within the capabilities of both JWST and HST.

We present our JWST UV LF measurements at $z=8$, which uses 166 sources, in Figure \ref{fig:UVLF_Z8} and find it is systematically high relative to past HST results, particularly at the bright end. Investigations reveal that the North Ecliptic Pole Time Domain Field (NEP-TDF) contributes a disproportionate amount to these bins. Of interest, is that the brightest $z\sim8$ candidates are largely located within the third spoke of NEP-TDF observations and may be indicative of a potential overdensity. A more detailed clustering analysis of the field is left to future work. Given the NEP-TDF may have an overdensity, we have produced two sets of measurements of the $z=8$ UV LF, one with and one without this field. When the field is omitted, the excess is largely removed and the UV LF well replicates the previous HST observations from \citet{McLure2013}, \citet{Finkelstein2015} and \citet{Bouwens2021}. This validates both the JWST observations and methodologies employed within this study and indicates that there is no significant tension between JWST selected samples of $z=8$ galaxies and those derived using HST data. The dynamical range of luminosity presently probed by this dataset is insufficient to strongly constrain the shape of the UV LF. We therefore utilise the faint-end data points ($M_{\rm UV} > -19$) from \citet{McLure2013} to enable higher quality fits of the UV LF at this redshift.  This extra data does not significantly change the fits we obtain, but reduces their uncertainty.  The inclusion of gravitational lenses is likely required for JWST-based UV LFs to push fainter at $z\sim8$.

The $z=9$ UV LF is presented in Figure \ref{fig:UVLF_Z9} and is measured using 59 sources. Here, the combined depth and area covered by these JWST programmes enable our observations to overlap with the results from the ground-based VISTA programmes which cover square degrees of the sky \citep{Donnan2022}. We find that our measurements are in close agreement with other JWST-based studies and are slightly higher than recent HST observations by \citet{Bouwens2021}. The bright end of the UV LF is elevated due to the inclusion of the AGN candidate from \citet{Larson2023} and this brightest data point is discounted when it comes to model fitting. Compared to \citet{Harikane2023}, the total area in this study is a factor of two higher and the majority of our area is deeper by upwards of half a magnitude in the blue NIRCam bands which contain the rest-frame UV emission at these redshifts. Our effective volumes at lower luminosities is thus significantly greater, providing a larger number of sources (59 at $z=9$ verses 13) and minimising cosmic variance. This provides measurements that not only agree with the \citet{Harikane2023} results, but build upon them by reducing the uncertainties. Towards the faint end, we find a slightly lower number density compared to observations by \citet{Donnan2022}. This is possibly due to SMACS-0723 photo-z uncertainties and the tendency for SED fitting codes to use Balmer breaks to explain any F444W excesses. Our fits match the ultra-deep MIRI GTO (PID: 1283, PI's H. Nørgaard-Nielsen, G. {\"O}stlin) parallel observations conducted in \citet{Gonzalez2023}. Most recently, \citet{Finkelstein2023} conducted a full search of high-z galaxies in the CEERS field. We find close agreement with their results, only slightly lower number densities in our observations across the overlapping luminosity range.

\begin{figure*}
\centering
\includegraphics[width=1.02\columnwidth]{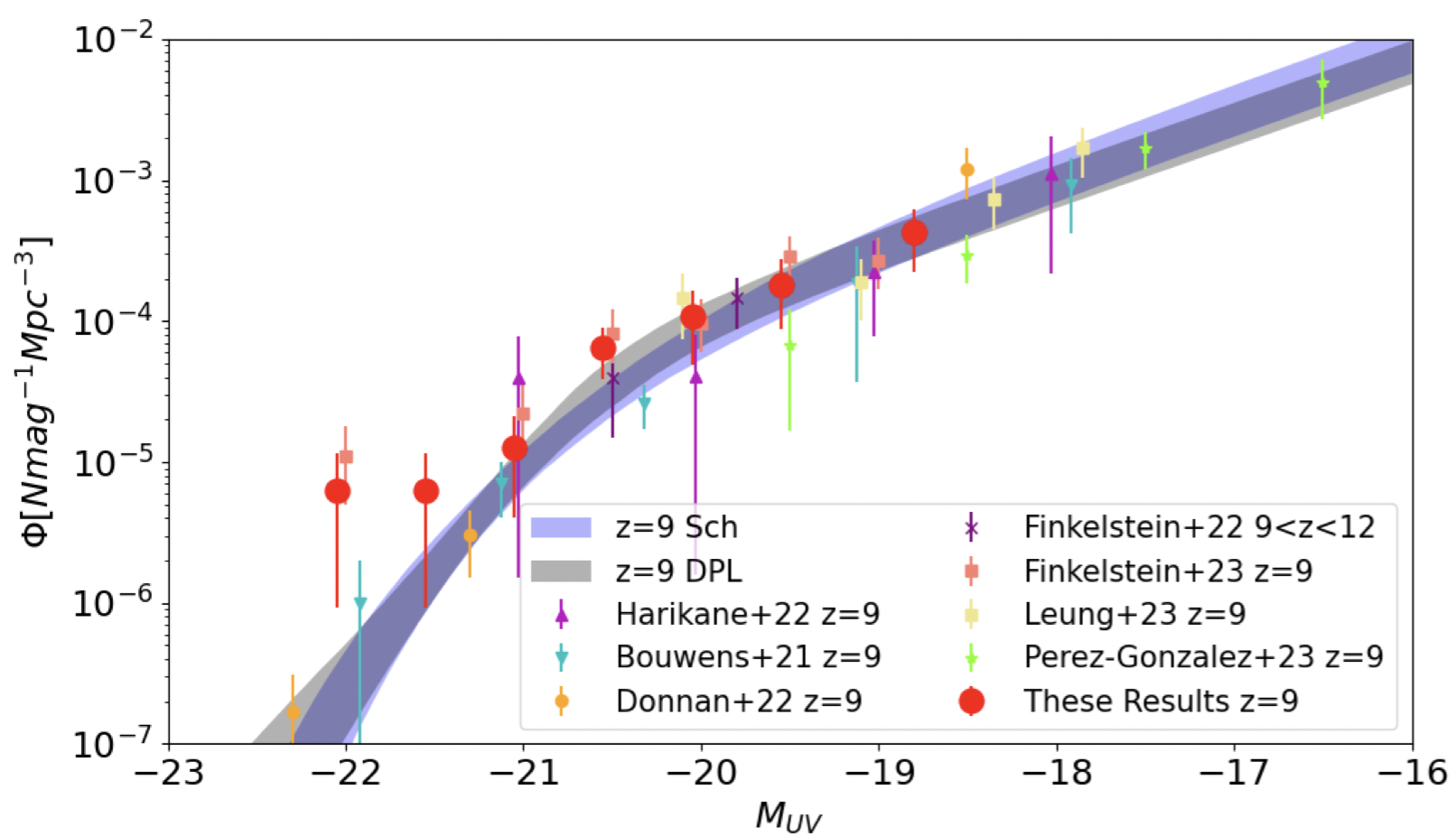}
\includegraphics[width=0.95\columnwidth]{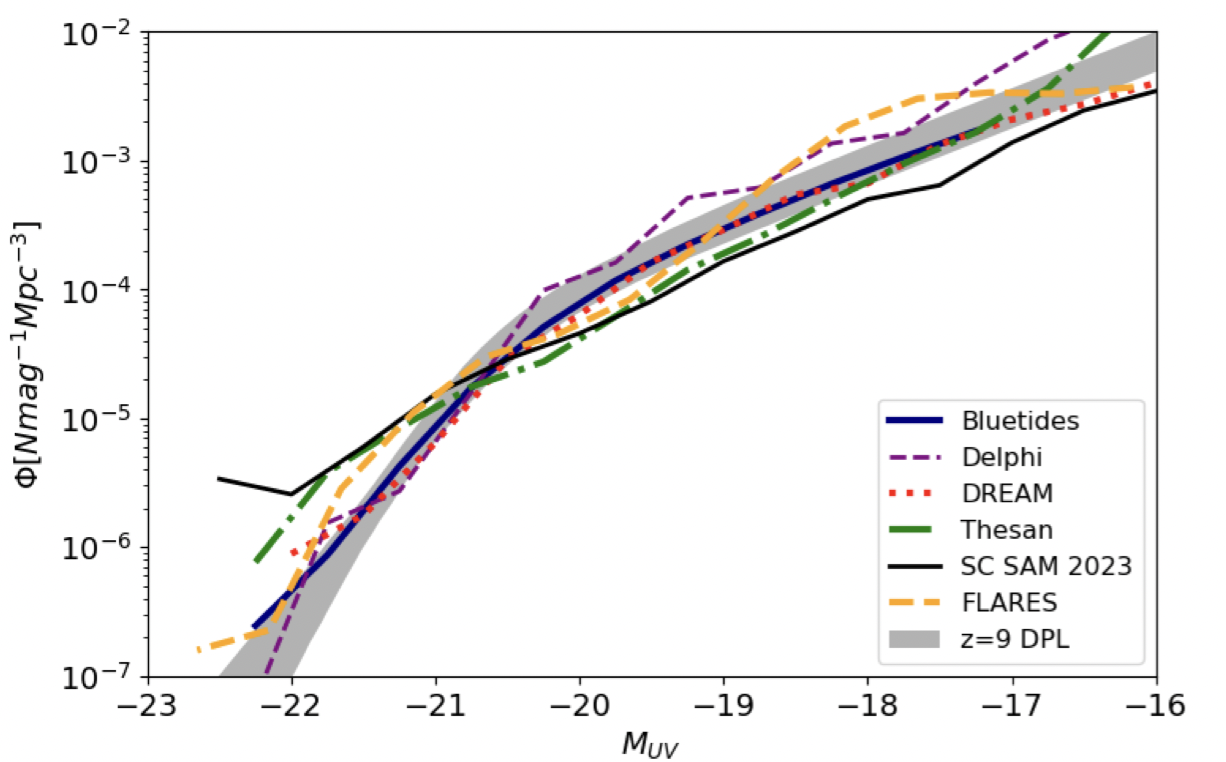} \caption{UV LF at $8.5<z<9.5$ as measured in this study (large red points). We compare our measurements to a selection of recent observations (left) and predictions from simulations (right). The shaded blue and black regions indicate the best fitting Schechter and DPL functions to our data plus the UltraVISTA data from \citet{Donnan2022}. We compare with observations from \citet{Bouwens2021,Donnan2022,Harikane2023,Finkelstein2022c,Finkelstein2023,Leung2023} and \citet{Gonzalez2023}.}
\label{fig:UVLF_Z9}
\end{figure*}

\subsection{The Measured $z=10$ and $z=12$ UV LF}

\begin{figure*}
\centering
\includegraphics[width=1.02\columnwidth]{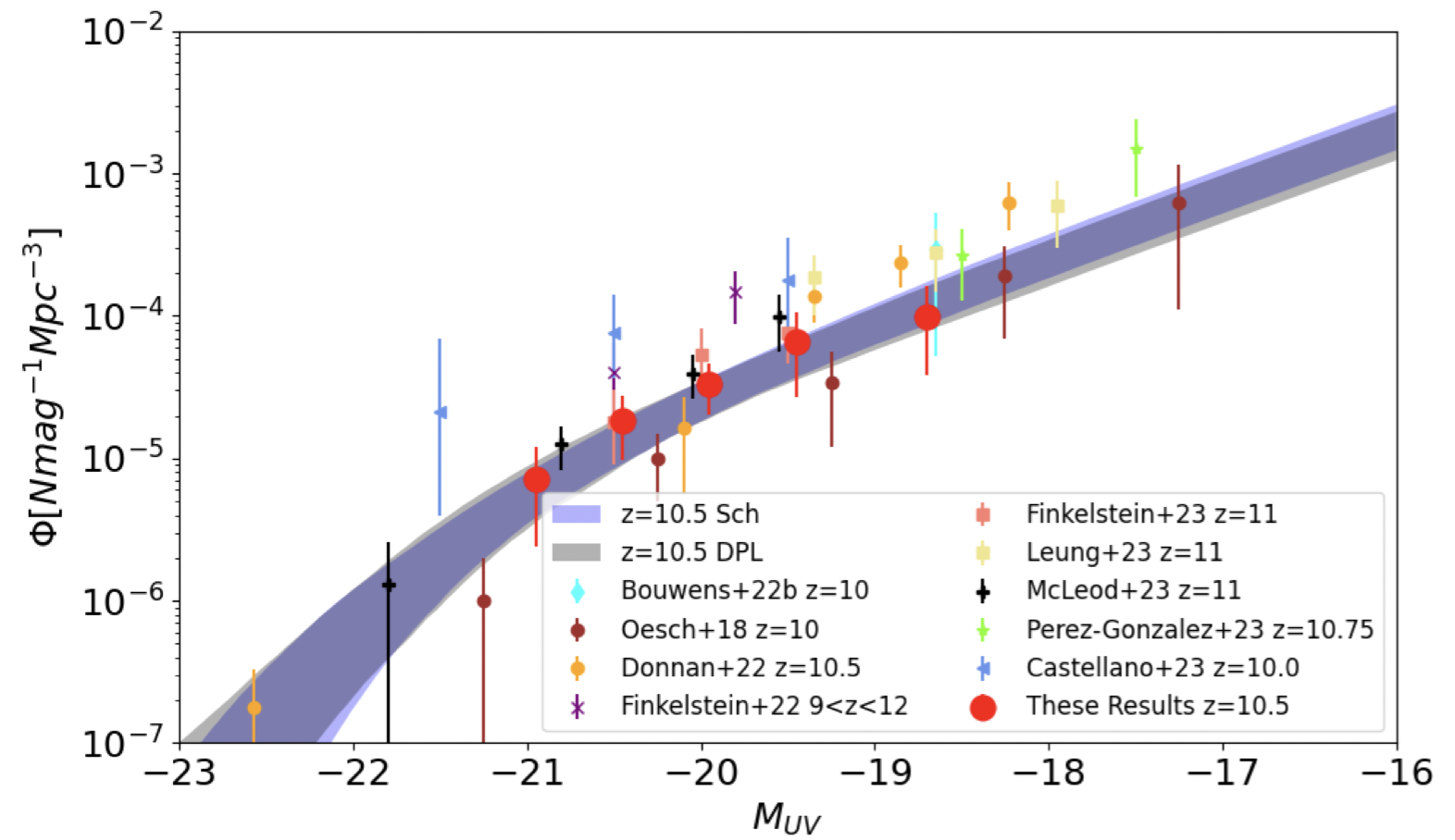}
\includegraphics[width=0.98\columnwidth]{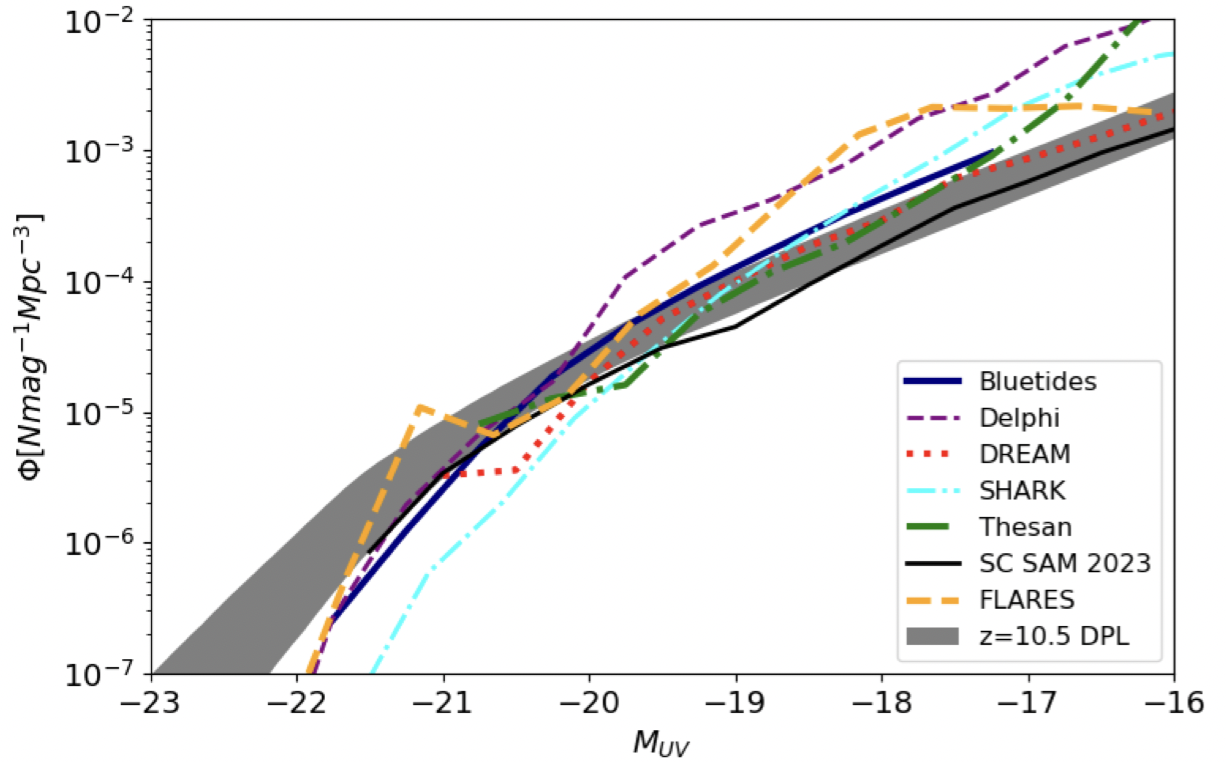}\caption{The UV LF at $9.5<z<11.5$ as measured in this study (large red points). We compare our measurements to a selection of recent observations (left) and predictions from simulations (right). The shaded blue and black regions indicate the best fitting Schechter and DPL functions to our data plus the UltraVISTA data from \citet{Donnan2022}. We compare with observations from \citet{Oesch2018,bouwens2022,Donnan2022,Finkelstein2022c,Castellano2023,Finkelstein2023,Leung2023} and \citet{Gonzalez2023}.}
\label{fig:UVLF_Z10}
\end{figure*}

Our $z=10$ UV LF, presented in Figure \ref{fig:UVLF_Z10}, uses a total of 27 sources in its measurement. This redshift range has sparked some initial discussion within the observing community regarding the measured number density of sources and the implications for the ultraviolet luminosity density. The work of \citet{Oesch2018}, using HST derived results, observed a significant evolution of the UV LF compared to that measured at $z=8$ and $z=9$. However, \citet{Donnan2022} and \citet{McLeod2023}, using JWST, found a much greater number density of galaxies over the same luminosity range. It is presently unclear if the difference between JWST-based results and HST-based results are the consequence of selection procedures or cosmic variance. Many early JWST UV LF measurements have the CEERS survey contribute a large proportion of their volume \citep[see discussion in][]{Willott2023} and if this was overdense, it could explain these observed differences. In this work, the CEERS field provides about 36\% of our cosmic volume and our number densities are found to be slightly smaller than in other JWST-based works. The inclusion of even more area to these depths (e.g. the full JADES survey and the PRIMER survey) and the inclusion of gravitational lenses will ultimately be required in order complete our picture of the $z=10.5$ galaxy population.

While our sample uses only 9 galaxies to measure the $11.5<z<13.5$ UV LF (see Figure \ref{fig:UVLF_Z12}), its constraints can already provide us with an insight into the evolution of the very first galaxies. At $z\sim12.5$, we find that our measured number density agrees well with the observations of both \citet{Donnan2022,Harikane2023,McLeod2023} and \citet{Finkelstein2023}, which sample similar redshift bins. Of note is the high number density of luminous UV sources, which implies a mild evolution in the number density of such sources. This extends the findings of \citet{Bowler2020}, who find the bright end of the UV LF evolves mildly to higher redshifts ($8<z<10$) compared to galaxies in the faint end. \citet{bouwens2022} finds a number density of sources that is approximately an order of magnitude greater than the three other studies used for comparison. As noted within that work, the volume used is small, but the use of medium-band filters enables tighter photometric-redshift constraints. It is possible the disagreement is the consequence of large-scale structure, since the medium-band survey covered only 4.6 arcmin$^2$. These number densities are high relative to the expectation from hydrodynamical simulations, yet a selection of these sources have already been spectroscopically confirmed (e.g. Maisie's galaxy and those from JADES). More spectroscopy and more area will be required in order to prove the UV Luminosity Function evolves slower than anticipated at $z>12$.

\begin{figure*}
\centering
\includegraphics[width=1.00\columnwidth]{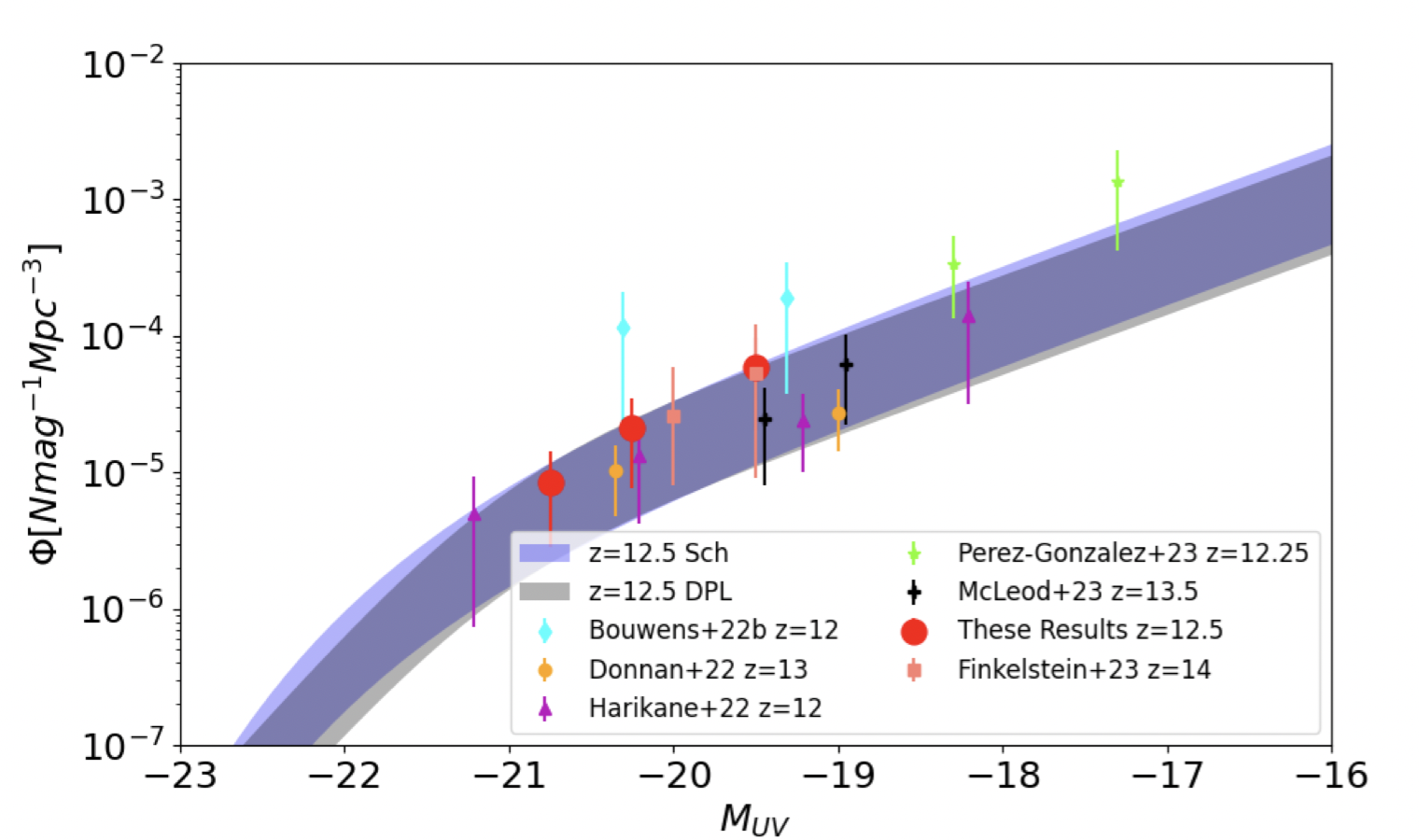}
\includegraphics[width=1.00\columnwidth]{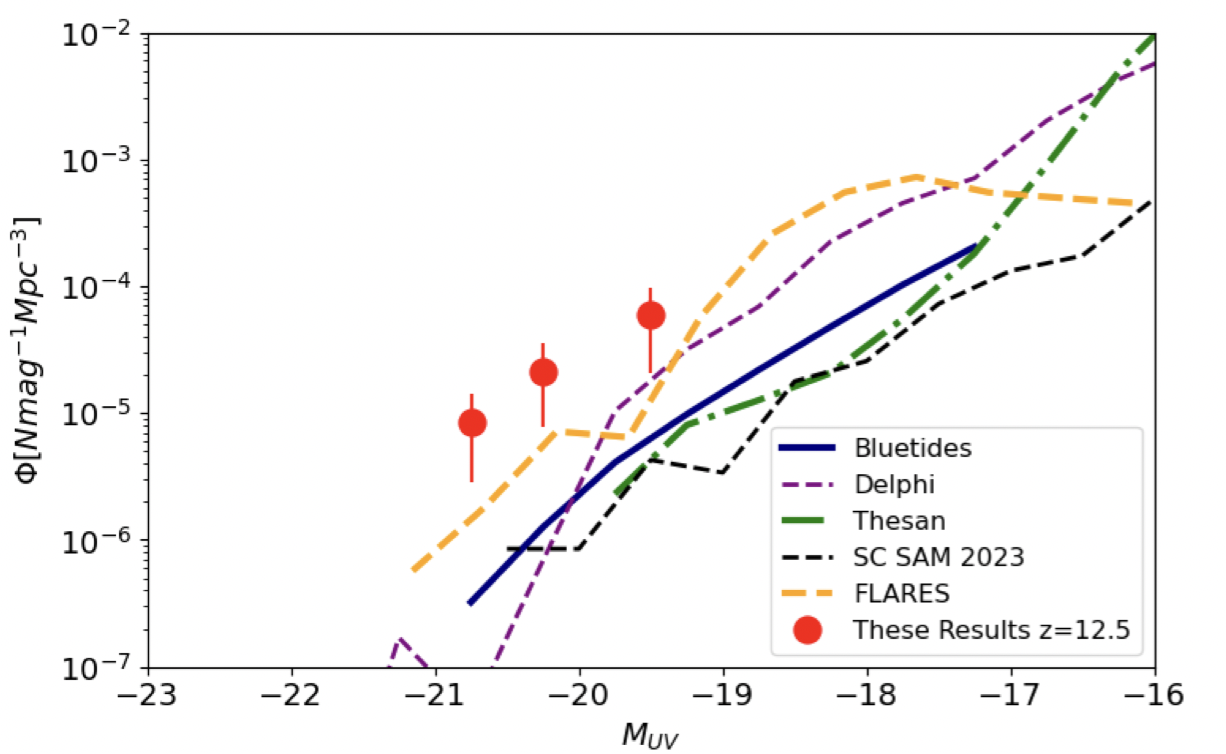}\caption{The UV LF at $11.5<z<13.5$ as measured in this study (large red points). We compare our measurements to a selection of recent observations (left) and predictions from simulations (right). We display here the best fitting Schechter and DPL functional forms but emphasise that uncertainties on these fits are large. Observational data are from \citet{Bouwens2022c,Donnan2022,Harikane2023,Finkelstein2022c,Finkelstein2023} and \citet{Gonzalez2023}.}
\label{fig:UVLF_Z12}
\end{figure*}

\subsection{Pushing Towards the First Galaxies}

To date, there have been a small selection of $z\sim16$ JWST candidates presented \citep{Harikane2023,Donnan2022,Yan2022,Austin2023}. An initial front runner for an extreme redshift galaxy was a $z=16.4$ candidate in the CEERS survey. The first spectroscopy of this $z\sim16$ candidate was recently attempted as a part of a directors discretionary time proposal (PID:2750, PI: P. Arrabal Haro). Initial analysis shows that the $z=16.4$ candidate is in fact a dusty, extreme line emitter at $z\sim4.9$ \citep{arrabal2023}, as predicted independently by \citet{Zavala2022} and \citet{Naidu2022}. In this work, this $z=16.4$ candidate is initially identified as being very high-z, but later rejected as a $z=4.7$ solution, close to the now known spec-z, has a chi square comparable to the initial high-z solution.

Our work identifies a total of 22 candidate $z>13.5$ galaxies meeting our criteria using the template set from \citet{Larson2022}. Of these 22 galaxies, only 6 also meet the criteria when employing the JADES templates \citep{Hainline2023a}. These are a $z=13.9$ and $z=15.8$ galaxy in NGDEEP \citep[see also][]{Austin2023,Leung2023}, a $z=14.1$ candidate in NEP (though by-eye examination reveals a suspiciously low F410M measurement), a $z=14.6$ and $z=18.1$ candidate in CEERS pointing 9 and finally a $z=16.1$ galaxy in JADES DR1. We do not produce a UV LF measurement using these extreme candidates as their reliability is presently unknown. Either spectroscopy or additional medium band coverage will be required in order to increase the confidence that these systems are real. Spectroscopic studies between $10<z<13.5$ have shown photometric selections of galaxies are effective overall and so we limit our UV LF measurements to this regime \citep{CurtisLake2022,Bunker2023,arrabal2023,Wang2023}.

\section{Discussion}

\subsection{Comparing the measured UV LF and its evolution to theory}

\begin{figure}
\centering
\includegraphics[width=0.95\columnwidth]{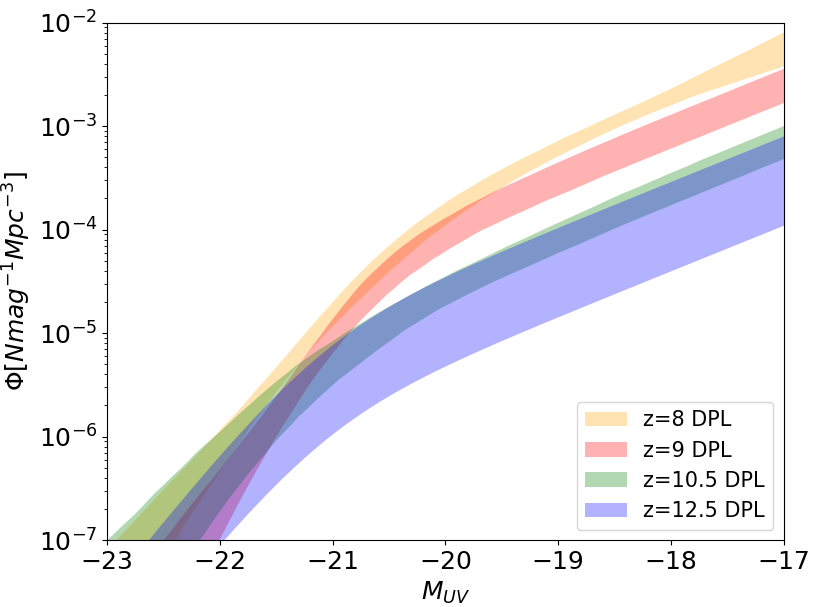}\caption{The measured time evolution of the UV LF. Here, we present the best-fitting DPL functions for our four redshift bins alongside each other. We find that the bright-end of the UV LF has relatively minor to negligible evolution within our errors \citep[agreeing with][]{Bowler2020}, while the faint-end decreases in normalisation towards higher redshifts.}
\label{fig:UVEvo}
\end{figure}

We present the time evolution of our DPL fits to the UV LF in Figure \ref{fig:UVEvo}. To provide better context for the UV LFs presented in this study, and the time evolution observed, we compare our observations and fits to predictions from a selection of recent simulations. These include Bluetides \citep{Feng2016,Wilkins2017}, Delphi \citep{dayal2014,dayal2022}, DREAM \citep{Drakos2022}, Thesan \citep{Kannan2022}, Santa Cruz semi-analytic Model \citep{Yung2023}, SHARK V2.0 \citep{Lagos2023} and FLARES \citep{Lovell2020,Vijayan2021,Wilkins2023}. The distribution of UV LFs generated by these simulations agree more with each other and with the observations at lower redshifts ($z=8$), but show an increasing spread towards higher redshifts. Primary trends are that the Delphi and FLARES simulations tend to predict greater numbers of UV faint galaxies ($-18<M_{\rm UV}<-17$) with Thesan also jumping up in number density at $M_{\rm UV}\sim -16$, while the Santa Cruz semi-analytic model predicts fewer of these galaxies relative to the other models. Simulations begin to diverge in their predictions at $z\geq10$. However, the observational errors also increase in this regime, matching the spread of simulation predictions. This means that we are presently not yet able to confidently favour one physical model over another.

At the most extreme redshifts probed ($z\sim12.5$), we find that the predicted number density of UV luminous sources ($M_{\rm UV}\sim-20.5$) varies by up to 2 dex between simulations. It is immediately apparent that much greater volumes at such redshifts are required in order to fully constrain model predictions in this early epoch. However, current measurements indicate that the number density of UV luminous sources are at the upper end of the prediction ranges, which agrees most strongly with the FLARES simulation.

\subsection{The Star Formation Rate Density in the Early Universe}

\begin{figure*}
\centering
\includegraphics[width=1.5\columnwidth]{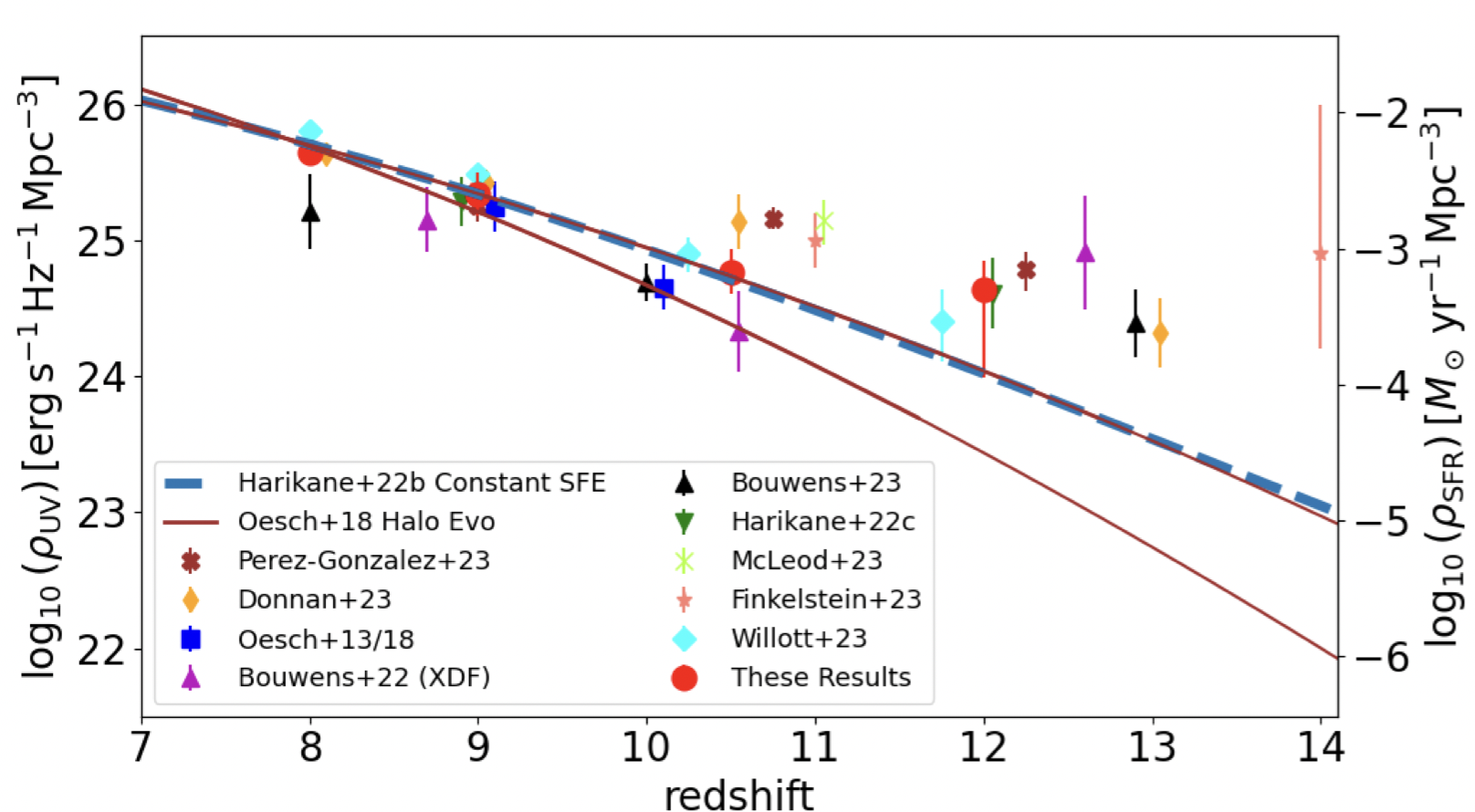}
\caption{The measured UV luminosity density and associated star formation rate density (SFRD), calculated by integrating UV LFs generated by randomly sampling the MCMC distribution for the DPL fits. The integration limits go down to $M_{\rm UV} < -17$. We also show a selection of results from other recent studies covering similar redshift ranges. Small shifts of 0.1 in redshift have been applied to aid readability. The two thin solid lines show the upper and lower bounds of the the dark matter halo evolutionary models used in \citet{Oesch2018}. The blue dashed line shows the constant star formation efficiency model of \citet{Harikane2023}. While some early studies have made a measurement at $z>14$, these estimations were largely based on the $z=16.4$ candidate in the CEERS field which later found to be at $z=4.9$. Our $11.5<z<13.5$ bin has been placed at its average sample redshift of $z=12$.}
\label{fig:SFRD}
\end{figure*}

Using the best-fit DPL models to our UV LF measurements, we integrate the luminosity functions as faint as $M_{\rm UV} \leq -17$ to obtain the luminosity density of ultraviolet light. This measurable has previously been used as a proxy for the star formation rate density (SFRD). To do so, we replicate the procedure of \citet{Harikane2023}, which uses the relations derived in \citet{Madau2014}. Here we use the simple conversion factor of $k=1.15\times10^{-28} [M_\odot {\rm yr}^{-1}/({\rm erg}\, {\rm s}^{-1} {\rm Hz}^{-1})]$. The results of this integral and conversion are presented in Figure \ref{fig:SFRD} alongside recent estimations from the literature. We find that the SFRD estimation at lower redshifts ($z=8-9$) agree well between past studies. Of significant interest is our results at higher redshift. Our data point at $z=10.5$ is lower than the results many other JWST-based studies and is well matched to the extrapolation of theoretical models, such as the constant star forming efficiency model from \citet{Harikane2022}. This is in agreement with the recent results from the CANUCS survey \citep{Willott2023}. At the highest redshifts ($z=12$), our results show a high UV luminosity density. However, this excess relative to simple models is only significant to $1\sigma$ and is consistent with the findings from \citet{Willott2023}. Consequently, significant deviations from simple models (e.g. constant star forming efficiency) may not necessarily be required in order to explain these excesses. Ultimately, larger samples of galaxies at $z>12$ and more spectroscopy will be required in order to limit the impact of cosmic variance and increase confidence in selection procedures.

If the evolution of the UV luminosity density can be described with a simple linear evolution \citep[as used in][]{Donnan2022} in $\log(\rho_{\rm UV})$ with redshift (resulting in an exponential evolution in SFRD), then we obtain a fit of $\log(\rho_{\rm UV}) = -0.36\pm0.01 z + 28.55\pm0.05$. This is in strong agreement with the results from \citet{Willott2023} and much steeper than the derived slope from \citet{McLeod2023} ($-0.23\pm0.04$). If we compare our observations to the predicted SFRD of \citet{Harikane2023}, which is based on a constant star formation efficiency, we find our results up to $z=12$ are consistent with such an evolution. Tension with such a model presently arises if the $z\sim15-16$ samples of galaxies are introduced.

The ultraviolet luminosity density also provides insight into the process of reionisation. We use the rest-frame ultraviolet slope ($\beta_{\rm UV}$), which is $\beta_{\rm UV} = -2.4\pm0.3$ for our sample at $z=8\text{---}10$ \citep[Austin et al. In Prep, see also][]{Cullen2023}, to estimate the sample average escape fraction of photons following the prescription derived in \citet{Chisholm2022}. This provides an average escape fraction of $\sim 11$ per cent, with bluer galaxies as high as $\sim$ 25 per cent and redder galaxies as low as 5 per cent. We then apply this to the critical ultraviolet luminosity density required for the galaxy population to drive runaway reionisation (where ionisation happens at a faster rate than recombination). We follow prescription derived in \citet{Shull2012} to determine at what redshift this critical luminosity density is achieved. We assume an 11 per cent escape fraction, gas temperature of $2\times10^4$K, and a redshift dependent clumping factor ($C_H$) following a power law of $C_H = (2.9)[(1+z)/6]^{-1.1}$ \citep{Shull2012}. Applying this to our luminosity functions, integrated down to $M_{\rm UV}\leq-10$, we find that the critical luminosity density is achieved around $z=8$. The faint end of the UV LF may well not extend to such faint luminosiites and there has been debate regarding if it turns over at $M_{\rm UV}\sim-14$ \citep{Bouwens2017,Livermore2017,Atek2018,Zhang2022}. In this situation, the critical luminosity density will be reached at lower redshifts $z\sim6-7$. Repeating this calculation following the prescription described in \citet{Madau1999,Bolton2007} and \citet{Duncan2015} and a constant clumping factor of $C_H=3$, we find the critical density is often 40-50\% lower than with \citet{Shull2012}, leading to a slightly earlier reionisation. This is a population average result, individual systems will likely begin reionising their surroundings at earlier times.  We will consider reionisation in more detail using our sample in future work. 

\section{Summary and Conclusions}

In this paper we present a detailed analysis of the first results of searching for and characterising distant galaxies with the James Webb Space Telescope using both PEARLS GTO and ERO/Public data. Our total sample, which we call the EPOCHS sample, spans an area of 180 arcmin$^2$ to near-infrared depths greater than $m\sim28.5$, forming one of the largest JWST-based search for high redshift galaxies to date. Applying a set of strict selection criteria to the data, we identify over 1000 candidate high-redshift galaxies ($z>6.5$).  

We use this sample to measure the UV LF of galaxies at $z>7.5$.  Other papers in this series will examine the individual properties of these galaxies, including their star formation rates and their stellar masses as well as morphologies and structures.  Our main conclusions can be summarised as follows:

\begin{enumerate}
    \item We find that the UV LF measured using JWST at $z\sim8$ and $z\sim9$ agrees well with past observations with HST, though there is a notable excess of $z=8$ galaxies in the NEP-TDF field of the PEARLS survey. This shows that there is no significant tension between the results obtained from the photometric selection of high-z galaxies in the wavelength/redshift regime where both telescopes can independently and comfortably find high redshift systems. 
    
    \item We observe a moderate number density of $z=10$ sources that lies between early JWST results and past HST results. This may indicate that early JWST results which were reliant on small volumes or were dominated by the CEERS survey area may have been subject to cosmic variance effects \citep[See also:][]{Willott2023}.
    
    \item We find that the measured number density of sources at $z\sim12$ matches recently reported values using JWST, but lies on the upper end of the range predicted by a selection of different simulations. This indicates that there may be more UV luminous sources at this epoch than what models predict.
    
    \item We find that the star formation rate density (SFRD) matches well with measurements from past studies covering the same redshift range as ours, similar to the luminosity function. The $z=10.5$ SFRD is a perfect match to expectations from a constant star forming efficiency. Our UV LFs find comparable SFRD compared to past studies at $z>12$ and it remains slightly higher than expectations from a constant star formation efficiency. However, it must be noted that sample sizes at such high redshifts remain very small, despite the collated nature of our EPOCHS sample.  More deep observations are needed in larger areas to address this point in more detail.
    
    \item We find that simulations of the high redshift Universe tend to agree more with each other at lower redshifts ($z=8-9$), but however diverge beyond HST's redshift frontier ($z=10-12$). Presently, observational constraints are insufficient to distinguish which model is more favourable. However, the initially high number densities of UV luminous sources with $M_{\rm UV}\sim-20.75$ at $z\sim12.5$ are up to 2 dex higher than expected from some models of the models.
\end{enumerate}

Future improvements to these UV LF estimations are already on the horizon. Once the COSMOS-Web and PRIMER surveys are completed, the area/depth combinations will bridge the deep, pencil-beam surveys currently conducted by JWST with the wide area VISTA surveys that have been previously undertaken. This will pin down the shape around the `knee' of the UV LF at these high redshifts. On the faint end, the completed NGDEEP and JADES programmes will provide deeper and wider blank fields that reach fainter and increase sample sizes, allowing us to fully constrain the faint-end slope ($\alpha$). Additionally, the large number of gravitational lenses that are being targeted by PEARLS, UNCOVER, CANUCS and others will provide an alternative methodology to these ultra-deep fields in order to find and study the more `typical' galaxies that exist at these early times. To avoid limitations from cosmic variance, it is likely that many of these programmes will need to be combined in order to provide the most complete picture of the early Universe. Beyond JWST, the LSST, Euclid, and Roman telescopes will provide shallower, but much wider area datasets (in the 10+ square degrees regime), which will enable us to properly constrain the number density of the most luminous systems and identify the first unobscured AGN to flare up in the early Universe.

The JADES CEERS and UNCOVER teams have now provided some of the first spectroscopic confirmation of galaxies at $z>11$ \citep{CurtisLake2022,Bunker2023,arrabal2023,Haro2023b,Wang2023}. However, knowledge regarding the success rates of these measurements is still incomplete (e.g. number of sources where spectra provide ambiguous results due to insufficient lines/breaks, the number of sources found to really be lower-z etc.). Increasing the number of ultra-high-z sources with spectroscopic confirmation (or rejection) will be a key step in validating early findings using photometric approaches, and will ultimately be how a robust final UV LF, and other physical properties, will be measured at these early times. 

All of the raw JWST data used in this work can be accessed via this MAST DOI: \href{https://archive.stsci.edu/doi/resolve/resolve.html?doi=10.17909/5h64-g193}{DOI 10.17909/5h64-g193}. The authors note that some data from PEARLS is within its proprietary period at the time of writing, but these will all be made accessible over 2024. Full catalogues of sources will be made public alongside the release of Conselice et al. In prep in early 2024.



\vspace{20pt}

We acknowledge support from the ERC Advanced Investigator Grant EPOCHS (788113), as well as two studentships from STFC. LF acknowledges financial support from Coordenação de Aperfeiçoamento de Pessoal de Nível Superior - Brazil (CAPES) in the form of a PhD studentship. R.A.W., S.H.C., and R.A.J. acknowledge support from NASA JWST Interdisciplinary Scientist grants NAG5 12460, NNX14AN10G and 80NSSC18K0200 from GSFC. AZ acknowledges support by Grant No. 2020750 from the United States-Israel Binational Science Foundation (BSF) and Grant No. 2109066 from the United States National Science Foundation (NSF), and by the Ministry of Science \& Technology, Israel. CC is supported by the National Natural Science Foundation of China, No. 11803044, 11933003, 12173045. This work is sponsored (in part) by the Chinese Academy of Sciences (CAS), through a grant to the CAS South America Center for Astronomy (CASSACA). We acknowledge the science research grants from the China Manned Space Project with NO. CMS-CSST-2021-A05. WPM acknowledges that support for this work was provided by the National Aeronautics and Space Administration through Chandra Award Numbers GO8-19119X, GO9-20123X, GO0-21126X and GO1-22134X issued by the Chandra X-ray Center, which is operated by the Smithsonian Astrophysical Observatory for and on behalf of the National Aeronautics Space Administration under contract NAS8-03060. MAM acknowledges the support of a National Research Council of Canada Plaskett Fellowship, and the Australian Research Council Centre of Excellence for All Sky Astrophysics in 3 Dimensions (ASTRO 3D), through project number CE17010001. CNAW acknowledges funding from the JWST/NIRCam contract NASS-0215 to the University of Arizona. M.N. acknowledges INAF-Mainstreams 1.05.01.86.20. CNAW acknowledges support from the NIRCam Development Contract NAS5-02105
from NASA Goddard Space Flight Center to the University of Arizona.


This work is based on observations made with the NASA/ESA \textit{Hubble Space Telescope} (HST) and NASA/ESA/CSA \textit{James Webb Space Telescope} (JWST) obtained from the \texttt{Mikulski Archive for Space Telescopes} (\texttt{MAST}) at the \textit{Space Telescope Science Institute} (STScI), which is operated by the Association of Universities for Research in Astronomy, Inc., under NASA contract NAS 5-03127 for JWST, and NAS 5–26555 for HST. The observations used in this work are associated with JWST programs 1176 and 2738. In addition, public datasets from JWST programs 2736 (SMACS-0723 ERO), 1324 (GLASS), 1345 (CEERS) and 2079 (NGDEEP) are also used within the work presented. The authors thank all involved in the construction and operations of the telescope as well as those who designed and executed these observations, their number are too large to list here and without each of their continued efforts, such work would not be possible. The authors also thank Steven Willner for his constructive comments and the anonymous referee for the improvements that they suggested. This work is dedicated to the memory of our dedicated colleague and co-author Mario Nonino, who sadly passed during the completion of this work.

The authors thank Anthony Holloway and Sotirios Sanidas for their providing their expertise in high performance computing and other IT support throughout this work. Data used in this publication (including catalogues and imaging) will be made publicly available once initial works are completed with students involved in its reduction and analysis. The anticipated timescale is early Summer 2024. Products using GTO data will be made available as and when exclusive access periods lapse. This work makes use of {\tt astropy} \citep{Astropy2013,Astropy2018,Astropy2022}, {\tt matplotlib} \citep{Hunter2007}, {\tt reproject}, {\tt DrizzlePac} \citep{Hoffmann2021}, {\tt SciPy} \citep{2020SciPy-NMeth} and {\tt photutils} \citep{larry_bradley_2022_6825092}.



\bibliography{mnras_template,main}

\begin{thebibliography}{}
\expandafter\ifx\csname natexlab\endcsname\relax\def\natexlab#1{#1}\fi
\providecommand{\url}[1]{\href{#1}{#1}}
\providecommand{\dodoi}[1]{doi:~\href{http://doi.org/#1}{\nolinkurl{#1}}}
\providecommand{\doeprint}[1]{\href{http://ascl.net/#1}{\nolinkurl{http://ascl.net/#1}}}
\providecommand{\doarXiv}[1]{\href{https://arxiv.org/abs/#1}{\nolinkurl{https://arxiv.org/abs/#1}}}

\bibitem[{{Adams} {et~al.}(2021){Adams}, {Bowler}, {Jarvis}, {H{\"a}u{\ss}ler},
  \& {Lagos}}]{Adams2021}
{Adams}, N.~J., {Bowler}, R.~A.~A., {Jarvis}, M.~J., {H{\"a}u{\ss}ler}, B., \&
  {Lagos}, C.~D.~P. 2021, \mnras, 506, 4933, \dodoi{10.1093/mnras/stab1956}

\bibitem[{{Adams} {et~al.}(2023){Adams}, {Conselice}, {Ferreira}, {Austin},
  {Trussler}, {Juod{\v{z}}balis}, {Wilkins}, {Caruana}, {Dayal}, {Verma}, \&
  {Vijayan}}]{Adams2023}
{Adams}, N.~J., {Conselice}, C.~J., {Ferreira}, L., {et~al.} 2023, \mnras, 518,
  4755, \dodoi{10.1093/mnras/stac3347}

\bibitem[{{Aird} {et~al.}(2015){Aird}, {Coil}, {Georgakakis}, {Nandra},
  {Barro}, \& {P{\'e}rez-Gonz{\'a}lez}}]{Aird2015}
{Aird}, J., {Coil}, A.~L., {Georgakakis}, A., {et~al.} 2015, \mnras, 451, 1892,
  \dodoi{10.1093/mnras/stv1062}

\bibitem[{{Arnouts} {et~al.}(1999){Arnouts}, {Cristiani}, {Moscardini},
  {Matarrese}, {Lucchin}, {Fontana}, \& {Giallongo}}]{Arnouts1999}
{Arnouts}, S., {Cristiani}, S., {Moscardini}, L., {et~al.} 1999, \mnras, 310,
  540, \dodoi{10.1046/j.1365-8711.1999.02978.x}

\bibitem[{{Arrabal Haro} {et~al.}(2023{\natexlab{a}}){Arrabal Haro},
  {Dickinson}, {Finkelstein}, {Kartaltepe}, {Donnan}, {Burgarella}, {Carnall},
  {Cullen}, {Dunlop}, {Fern{\'a}ndez}, {Fujimoto}, {Jung}, {Krips}, {Larson},
  {Papovich}, {P{\'e}rez-Gonz{\'a}lez}, {Amor{\'\i}n}, {Bagley}, {Buat},
  {Casey}, {Chworowsky}, {Cohen}, {Ferguson}, {Giavalisco}, {Huertas-Company},
  {Hutchison}, {Kocevski}, {Koekemoer}, {Lucas}, {McLeod}, {McLure}, {Pirzkal},
  {Seill{\'e}}, {Trump}, {Weiner}, {Wilkins}, \& {Zavala}}]{arrabal2023}
{Arrabal Haro}, P., {Dickinson}, M., {Finkelstein}, S.~L., {et~al.}
  2023{\natexlab{a}}, \nat, 622, 707, \dodoi{10.1038/s41586-023-06521-7}

\bibitem[{{Arrabal Haro} {et~al.}(2023{\natexlab{b}}){Arrabal Haro},
  {Dickinson}, {Finkelstein}, {Fujimoto}, {Fern{\'a}ndez}, {Kartaltepe},
  {Jung}, {Cole}, {Burgarella}, {Chworowsky}, {Hutchison}, {Morales},
  {Papovich}, {Simons}, {Amor{\'\i}n}, {Backhaus}, {Bagley}, {Bisigello},
  {Calabr{\`o}}, {Castellano}, {Cleri}, {Dav{\'e}}, {Dekel}, {Ferguson},
  {Fontana}, {Gawiser}, {Giavalisco}, {Harish}, {Hathi}, {Hirschmann},
  {Holwerda}, {Huertas-Company}, {Koekemoer}, {Larson}, {Lucas}, {Mobasher},
  {P{\'e}rez-Gonz{\'a}lez}, {Pirzkal}, {Rose}, {Santini}, {Trump}, {de la
  Vega}, {Wang}, {Weiner}, {Wilkins}, {Yang}, {Yung}, \& {Zavala}}]{Haro2023b}
---. 2023{\natexlab{b}}, \apjl, 951, L22, \dodoi{10.3847/2041-8213/acdd54}

\bibitem[{{Astropy Collaboration} {et~al.}(2013){Astropy Collaboration},
  {Robitaille}, {Tollerud}, {Greenfield}, {Droettboom}, {Bray}, {Aldcroft},
  {Davis}, {Ginsburg}, {Price-Whelan}, {Kerzendorf}, {Conley}, {Crighton},
  {Barbary}, {Muna}, {Ferguson}, {Grollier}, {Parikh}, {Nair}, {Unther},
  {Deil}, {Woillez}, {Conseil}, {Kramer}, {Turner}, {Singer}, {Fox}, {Weaver},
  {Zabalza}, {Edwards}, {Azalee Bostroem}, {Burke}, {Casey}, {Crawford},
  {Dencheva}, {Ely}, {Jenness}, {Labrie}, {Lim}, {Pierfederici}, {Pontzen},
  {Ptak}, {Refsdal}, {Servillat}, \& {Streicher}}]{Astropy2013}
{Astropy Collaboration}, {Robitaille}, T.~P., {Tollerud}, E.~J., {et~al.} 2013,
  \aap, 558, A33, \dodoi{10.1051/0004-6361/201322068}

\bibitem[{{Astropy Collaboration} {et~al.}(2018){Astropy Collaboration},
  {Price-Whelan}, {Sip{\H{o}}cz}, {G{\"u}nther}, {Lim}, {Crawford}, {Conseil},
  {Shupe}, {Craig}, {Dencheva}, {Ginsburg}, {VanderPlas}, {Bradley},
  {P{\'e}rez-Su{\'a}rez}, {de Val-Borro}, {Aldcroft}, {Cruz}, {Robitaille},
  {Tollerud}, {Ardelean}, {Babej}, {Bach}, {Bachetti}, {Bakanov}, {Bamford},
  {Barentsen}, {Barmby}, {Baumbach}, {Berry}, {Biscani}, {Boquien}, {Bostroem},
  {Bouma}, {Brammer}, {Bray}, {Breytenbach}, {Buddelmeijer}, {Burke},
  {Calderone}, {Cano Rodr{\'\i}guez}, {Cara}, {Cardoso}, {Cheedella}, {Copin},
  {Corrales}, {Crichton}, {D'Avella}, {Deil}, {Depagne}, {Dietrich}, {Donath},
  {Droettboom}, {Earl}, {Erben}, {Fabbro}, {Ferreira}, {Finethy}, {Fox},
  {Garrison}, {Gibbons}, {Goldstein}, {Gommers}, {Greco}, {Greenfield},
  {Groener}, {Grollier}, {Hagen}, {Hirst}, {Homeier}, {Horton}, {Hosseinzadeh},
  {Hu}, {Hunkeler}, {Ivezi{\'c}}, {Jain}, {Jenness}, {Kanarek}, {Kendrew},
  {Kern}, {Kerzendorf}, {Khvalko}, {King}, {Kirkby}, {Kulkarni}, {Kumar},
  {Lee}, {Lenz}, {Littlefair}, {Ma}, {Macleod}, {Mastropietro}, {McCully},
  {Montagnac}, {Morris}, {Mueller}, {Mumford}, {Muna}, {Murphy}, {Nelson},
  {Nguyen}, {Ninan}, {N{\"o}the}, {Ogaz}, {Oh}, {Parejko}, {Parley}, {Pascual},
  {Patil}, {Patil}, {Plunkett}, {Prochaska}, {Rastogi}, {Reddy Janga},
  {Sabater}, {Sakurikar}, {Seifert}, {Sherbert}, {Sherwood-Taylor}, {Shih},
  {Sick}, {Silbiger}, {Singanamalla}, {Singer}, {Sladen}, {Sooley},
  {Sornarajah}, {Streicher}, {Teuben}, {Thomas}, {Tremblay}, {Turner},
  {Terr{\'o}n}, {van Kerkwijk}, {de la Vega}, {Watkins}, {Weaver}, {Whitmore},
  {Woillez}, {Zabalza}, \& {Astropy Contributors}}]{Astropy2018}
{Astropy Collaboration}, {Price-Whelan}, A.~M., {Sip{\H{o}}cz}, B.~M., {et~al.}
  2018, \aj, 156, 123, \dodoi{10.3847/1538-3881/aabc4f}

\bibitem[{{Astropy Collaboration} {et~al.}(2022){Astropy Collaboration},
  {Price-Whelan}, {Lim}, {Earl}, {Starkman}, {Bradley}, {Shupe}, {Patil},
  {Corrales}, {Brasseur}, {N{\"o}the}, {Donath}, {Tollerud}, {Morris},
  {Ginsburg}, {Vaher}, {Weaver}, {Tocknell}, {Jamieson}, {van Kerkwijk},
  {Robitaille}, {Merry}, {Bachetti}, {G{\"u}nther}, {Aldcroft},
  {Alvarado-Montes}, {Archibald}, {B{\'o}di}, {Bapat}, {Barentsen},
  {Baz{\'a}n}, {Biswas}, {Boquien}, {Burke}, {Cara}, {Cara}, {Conroy},
  {Conseil}, {Craig}, {Cross}, {Cruz}, {D'Eugenio}, {Dencheva}, {Devillepoix},
  {Dietrich}, {Eigenbrot}, {Erben}, {Ferreira}, {Foreman-Mackey}, {Fox},
  {Freij}, {Garg}, {Geda}, {Glattly}, {Gondhalekar}, {Gordon}, {Grant},
  {Greenfield}, {Groener}, {Guest}, {Gurovich}, {Handberg}, {Hart},
  {Hatfield-Dodds}, {Homeier}, {Hosseinzadeh}, {Jenness}, {Jones}, {Joseph},
  {Kalmbach}, {Karamehmetoglu}, {Ka{\l}uszy{\'n}ski}, {Kelley}, {Kern},
  {Kerzendorf}, {Koch}, {Kulumani}, {Lee}, {Ly}, {Ma}, {MacBride}, {Maljaars},
  {Muna}, {Murphy}, {Norman}, {O'Steen}, {Oman}, {Pacifici}, {Pascual},
  {Pascual-Granado}, {Patil}, {Perren}, {Pickering}, {Rastogi}, {Roulston},
  {Ryan}, {Rykoff}, {Sabater}, {Sakurikar}, {Salgado}, {Sanghi}, {Saunders},
  {Savchenko}, {Schwardt}, {Seifert-Eckert}, {Shih}, {Jain}, {Shukla}, {Sick},
  {Simpson}, {Singanamalla}, {Singer}, {Singhal}, {Sinha}, {Sip{\H{o}}cz},
  {Spitler}, {Stansby}, {Streicher}, {{\v{S}}umak}, {Swinbank}, {Taranu},
  {Tewary}, {Tremblay}, {de Val-Borro}, {Van Kooten}, {Vasovi{\'c}}, {Verma},
  {de Miranda Cardoso}, {Williams}, {Wilson}, {Winkel}, {Wood-Vasey}, {Xue},
  {Yoachim}, {Zhang}, {Zonca}, \& {Astropy Project Contributors}}]{Astropy2022}
{Astropy Collaboration}, {Price-Whelan}, A.~M., {Lim}, P.~L., {et~al.} 2022,
  \apj, 935, 167, \dodoi{10.3847/1538-4357/ac7c74}

\bibitem[{{Atek} {et~al.}(2018){Atek}, {Richard}, {Kneib}, \&
  {Schaerer}}]{Atek2018}
{Atek}, H., {Richard}, J., {Kneib}, J.-P., \& {Schaerer}, D. 2018, \mnras, 479,
  5184, \dodoi{10.1093/mnras/sty1820}

\bibitem[{{Atek} {et~al.}(2023){Atek}, {Shuntov}, {Furtak}, {Richard}, {Kneib},
  {Mahler}, {Zitrin}, {McCracken}, {Charlot}, {Chevallard}, \&
  {Chemerynska}}]{Atek2022}
{Atek}, H., {Shuntov}, M., {Furtak}, L.~J., {et~al.} 2023, \mnras, 519, 1201,
  \dodoi{10.1093/mnras/stac3144}

\bibitem[{{Austin} {et~al.}(2023){Austin}, {Adams}, {Conselice}, {Harvey},
  {Ormerod}, {Trussler}, {Li}, {Ferreira}, {Dayal}, \&
  {Juod{\v{z}}balis}}]{Austin2023}
{Austin}, D., {Adams}, N., {Conselice}, C.~J., {et~al.} 2023, \apjl, 952, L7,
  \dodoi{10.3847/2041-8213/ace18d}

\bibitem[{{Bagley} {et~al.}(2023{\natexlab{a}}){Bagley}, {Finkelstein},
  {Koekemoer}, {Ferguson}, {Arrabal Haro}, {Dickinson}, {Kartaltepe},
  {Papovich}, {P{\'e}rez-Gonz{\'a}lez}, {Pirzkal}, {Somerville}, {Willmer},
  {Yang}, {Yung}, {Fontana}, {Grazian}, {Grogin}, {Hirschmann}, {Kewley},
  {Kirkpatrick}, {Kocevski}, {Lotz}, {Medrano}, {Morales}, {Pentericci},
  {Ravindranath}, {Trump}, {Wilkins}, {Calabr{\`o}}, {Cooper}, {Costantin}, {de
  la Vega}, {Hilbert}, {Hutchison}, {Larson}, {Lucas}, {McGrath}, {Ryan},
  {Wang}, \& {Wuyts}}]{Bagley2022}
{Bagley}, M.~B., {Finkelstein}, S.~L., {Koekemoer}, A.~M., {et~al.}
  2023{\natexlab{a}}, \apjl, 946, L12, \dodoi{10.3847/2041-8213/acbb08}

\bibitem[{{Bagley} {et~al.}(2023{\natexlab{b}}){Bagley}, {Pirzkal},
  {Finkelstein}, {Papovich}, {Berg}, {Lotz}, {Leung}, {Ferguson}, {Koekemoer},
  {Dickinson}, {Kartaltepe}, {Kocevski}, {Somerville}, {Yung}, {Backhaus},
  {Casey}, {Castellano}, {Ch{\'a}vez Ortiz}, {Chworowsky}, {Cox}, {Dav{\'e}},
  {Davis}, {Estrada-Carpenter}, {Fontana}, {Fujimoto}, {Gardner}, {Giavalisco},
  {Grazian}, {Grogin}, {Hathi}, {Hutchison}, {Jaskot}, {Jung}, {Kewley},
  {Kirkpatrick}, {Larson}, {Matharu}, {Natarajan}, {Pentericci},
  {P{\'e}rez-Gonz{\'a}lez}, {Ravindranath}, {Rothberg}, {Ryan}, {Shen},
  {Simons}, {Snyder}, {Trump}, \& {Wilkins}}]{Bagley2023}
{Bagley}, M.~B., {Pirzkal}, N., {Finkelstein}, S.~L., {et~al.}
  2023{\natexlab{b}}, arXiv e-prints, arXiv:2302.05466,
  \dodoi{10.48550/arXiv.2302.05466}

\bibitem[{Bertin \& Arnouts(1996)}]{Bertin1996}
Bertin, E., \& Arnouts, S. 1996, \aaps, 117, 393

\bibitem[{{Bezanson} {et~al.}(2022){Bezanson}, {Labbe}, {Whitaker}, {Leja},
  {Price}, {Franx}, {Brammer}, {Marchesini}, {Zitrin}, {Wang}, {Weaver},
  {Furtak}, {Atek}, {Coe}, {Cutler}, {Dayal}, {van Dokkum}, {Feldmann},
  {Forster Schreiber}, {Fujimoto}, {Geha}, {Glazebrook}, {de Graaff}, {Greene},
  {Juneau}, {Kassin}, {Kriek}, {Khullar}, {Maseda}, {Mowla}, {Muzzin},
  {Nanayakkara}, {Nelson}, {Oesch}, {Pacifici}, {Pan}, {Papovich}, {Setton},
  {Shapley}, {Smit}, {Stefanon}, {Taylor}, \& {Williams}}]{Bezanson2022}
{Bezanson}, R., {Labbe}, I., {Whitaker}, K.~E., {et~al.} 2022, arXiv e-prints,
  arXiv:2212.04026, \dodoi{10.48550/arXiv.2212.04026}

\bibitem[{{Bhatawdekar} {et~al.}(2019){Bhatawdekar}, {Conselice},
  {Margalef-Bentabol}, \& {Duncan}}]{Bhatawdekar2019}
{Bhatawdekar}, R., {Conselice}, C.~J., {Margalef-Bentabol}, B., \& {Duncan}, K.
  2019, \mnras, 486, 3805, \dodoi{10.1093/mnras/stz866}

\bibitem[{{Bolton} \& {Haehnelt}(2007)}]{Bolton2007}
{Bolton}, J.~S., \& {Haehnelt}, M.~G. 2007, \mnras, 382, 325,
  \dodoi{10.1111/j.1365-2966.2007.12372.x}

\bibitem[{{Bosch-Ramon}(2018)}]{Bosch2018}
{Bosch-Ramon}, V. 2018, \aap, 617, L3, \dodoi{10.1051/0004-6361/201833952}

\bibitem[{{Bouwens} {et~al.}(2023){Bouwens}, {Illingworth}, {Oesch},
  {Stefanon}, {Naidu}, {van Leeuwen}, \& {Magee}}]{Bouwens2022c}
{Bouwens}, R., {Illingworth}, G., {Oesch}, P., {et~al.} 2023, \mnras, 523,
  1009, \dodoi{10.1093/mnras/stad1014}

\bibitem[{{Bouwens} {et~al.}(2017){Bouwens}, {Oesch}, {Illingworth}, {Ellis},
  \& {Stefanon}}]{Bouwens2017}
{Bouwens}, R.~J., {Oesch}, P.~A., {Illingworth}, G.~D., {Ellis}, R.~S., \&
  {Stefanon}, M. 2017, \apj, 843, 129, \dodoi{10.3847/1538-4357/aa70a4}

\bibitem[{{Bouwens} {et~al.}(2011){Bouwens}, {Illingworth}, {Labbe}, {Oesch},
  {Trenti}, {Carollo}, {van Dokkum}, {Franx}, {Stiavelli}, {Gonz{\'a}lez},
  {Magee}, \& {Bradley}}]{Bouwens2011}
{Bouwens}, R.~J., {Illingworth}, G.~D., {Labbe}, I., {et~al.} 2011, \nat, 469,
  504, \dodoi{10.1038/nature09717}

\bibitem[{Bouwens {et~al.}(2015)Bouwens, Illingworth, Oesch, Trenti, Labbé,
  Bradley, Carollo, Dokkum, Gonzalez, Holwerda, Franx, Spitler, Smit, \&
  Magee}]{Bouwens2015}
Bouwens, R.~J., Illingworth, G.~D., Oesch, P.~A., {et~al.} 2015, \apj, 803, 1,
  \dodoi{10.1088/0004-637X/803/1/34}

\bibitem[{{Bouwens} {et~al.}(2021){Bouwens}, {Oesch}, {Stefanon},
  {Illingworth}, {Labb{\'e}}, {Reddy}, {Atek}, {Montes}, {Naidu},
  {Nanayakkara}, {Nelson}, \& {Wilkins}}]{Bouwens2021}
{Bouwens}, R.~J., {Oesch}, P.~A., {Stefanon}, M., {et~al.} 2021, \aj, 162, 47,
  \dodoi{10.3847/1538-3881/abf83e}

\bibitem[{{Bouwens} {et~al.}(2022){Bouwens}, {Smit}, {Schouws}, {Stefanon},
  {Bowler}, {Endsley}, {Gonzalez}, {Inami}, {Stark}, {Oesch}, {Hodge},
  {Aravena}, {da Cunha}, {Dayal}, {Looze}, {Ferrara}, {Fudamoto}, {Graziani},
  {Li}, {Nanayakkara}, {Pallottini}, {Schneider}, {Sommovigo}, {Topping}, {van
  der Werf}, {Algera}, {Barrufet}, {Hygate}, {Labb{\'e}}, {Riechers}, \&
  {Witstok}}]{bouwens2022}
{Bouwens}, R.~J., {Smit}, R., {Schouws}, S., {et~al.} 2022, \apj, 931, 160,
  \dodoi{10.3847/1538-4357/ac5a4a}

\bibitem[{{Bower} {et~al.}(2012){Bower}, {Benson}, \& {Crain}}]{Bower2012}
{Bower}, R.~G., {Benson}, A.~J., \& {Crain}, R.~A. 2012, \mnras, 422, 2816,
  \dodoi{10.1111/j.1365-2966.2012.20516.x}

\bibitem[{Bowler {et~al.}(2015)Bowler, Dunlop, McLure, McCracken,
  Milvang-Jensen, Furusawa, Taniguchi, Fèvre, Fynbo, Jarvis, \&
  Häußler}]{Bowler2015}
Bowler, R.~A., Dunlop, J.~S., McLure, R.~J., {et~al.} 2015, \mnras, 452, 1817,
  \dodoi{10.1093/mnras/stv1403}

\bibitem[{{Bowler} {et~al.}(2020){Bowler}, {Jarvis}, {Dunlop}, {McLure},
  {McLeod}, {Adams}, {Milvang-Jensen}, \& {McCracken}}]{Bowler2020}
{Bowler}, R.~A.~A., {Jarvis}, M.~J., {Dunlop}, J.~S., {et~al.} 2020, \mnras,
  493, 2059, \dodoi{10.1093/mnras/staa313}

\bibitem[{Bradley {et~al.}(2022)Bradley, Sipőcz, Robitaille, Tollerud,
  Vinícius, Deil, Barbary, Wilson, Busko, Donath, Günther, Cara, Lim,
  Meßlinger, Conseil, Bostroem, Droettboom, Bray, Bratholm, Barentsen, Craig,
  Rathi, Pascual, Perren, Georgiev, de~Val-Borro, Kerzendorf, Bach, Quint, \&
  Souchereau}]{larry_bradley_2022_6825092}
Bradley, L., Sipőcz, B., Robitaille, T., {et~al.} 2022, astropy/photutils:
  1.5.0, 1.5.0,  Zenodo, \dodoi{10.5281/zenodo.6825092}

\bibitem[{{Brammer} {et~al.}(2008){Brammer}, {van Dokkum}, \&
  {Coppi}}]{Brammer2008}
{Brammer}, G.~B., {van Dokkum}, P.~G., \& {Coppi}, P. 2008, \apj, 686, 1503,
  \dodoi{10.1086/591786}

\bibitem[{Bruzual \& Charlot(2003)}]{Bruzual2003}
Bruzual, G., \& Charlot, S. 2003, \mnras, 344, 1000,
  \dodoi{10.1046/j.1365-8711.2003.06897.x}

\bibitem[{{Bunker} {et~al.}(2023{\natexlab{a}}){Bunker}, {Saxena}, {Cameron},
  {Willott}, {Curtis-Lake}, {Jakobsen}, {Carniani}, {Smit}, {Maiolino},
  {Witstok}, {Curti}, {D'Eugenio}, {Jones}, {Ferruit}, {Arribas}, {Charlot},
  {Chevallard}, {Giardino}, {de Graaff}, {Looser}, {L{\"u}tzgendorf}, {Maseda},
  {Rawle}, {Rix}, {Del Pino}, {Alberts}, {Egami}, {Eisenstein}, {Endsley},
  {Hainline}, {Hausen}, {Johnson}, {Rieke}, {Rieke}, {Robertson}, {Shivaei},
  {Stark}, {Sun}, {Tacchella}, {Tang}, {Williams}, {Willmer}, {Baker}, {Baum},
  {Bhatawdekar}, {Bowler}, {Boyett}, {Chen}, {Circosta}, {Helton}, {Ji},
  {Kumari}, {Lyu}, {Nelson}, {Parlanti}, {Perna}, {Sandles}, {Scholtz},
  {Suess}, {Topping}, {{\"U}bler}, {Wallace}, \& {Whitler}}]{Bunker2023}
{Bunker}, A.~J., {Saxena}, A., {Cameron}, A.~J., {et~al.} 2023{\natexlab{a}},
  \aap, 677, A88, \dodoi{10.1051/0004-6361/202346159}

\bibitem[{{Bunker} {et~al.}(2023{\natexlab{b}}){Bunker}, {Cameron},
  {Curtis-Lake}, {Jakobsen}, {Carniani}, {Curti}, {Witstok}, {Maiolino},
  {D'Eugenio}, {Looser}, {Willott}, {Bonaventura}, {Hainline}, {Uebler},
  {Willmer}, {Saxena}, {Smit}, {Alberts}, {Arribas}, {Baker}, {Baum},
  {Bhatawdekar}, {Bowler}, {Boyett}, {Charlot}, {Chen}, {Chevallard},
  {Circosta}, {DeCoursey}, {de Graaff}, {Egami}, {Eisenstein}, {Endsley},
  {Ferruit}, {Giardino}, {Hausen}, {Helton}, {Hviding}, {Ji}, {Johnson},
  {Jones}, {Kumari}, {Laseter}, {Luetzgendorf}, {Maseda}, {Nelson}, {Parlanti},
  {Perna}, {Rawle}, {Rix}, {Rieke}, {Robertson}, {Rodriguez Del Pino},
  {Sandles}, {Scholtz}, {Sharpe}, {Skarbinski}, {Stark}, {Sun}, {Tacchella},
  {Topping}, {Villanueva}, {Wallace}, {Williams}, \& {Woodrum}}]{Bunker2023b}
{Bunker}, A.~J., {Cameron}, A.~J., {Curtis-Lake}, E., {et~al.}
  2023{\natexlab{b}}, arXiv e-prints, arXiv:2306.02467,
  \dodoi{10.48550/arXiv.2306.02467}

\bibitem[{{Caminha} {et~al.}(2017){Caminha}, {Grillo}, {Rosati}, {Balestra},
  {Mercurio}, {Vanzella}, {Biviano}, {Caputi}, {Delgado-Correal}, {Karman},
  {Lombardi}, {Meneghetti}, {Sartoris}, \& {Tozzi}}]{Caminha2017}
{Caminha}, G.~B., {Grillo}, C., {Rosati}, P., {et~al.} 2017, \aap, 600, A90,
  \dodoi{10.1051/0004-6361/201629297}

\bibitem[{{Caminha} {et~al.}(2023){Caminha}, {Grillo}, {Rosati}, {Liu},
  {Acebron}, {Bergamini}, {Caputi}, {Mercurio}, {Tozzi}, {Vanzella}, {Demarco},
  {Frye}, {Rosani}, \& {Sharon}}]{Caminha2022}
---. 2023, \aap, 678, A3, \dodoi{10.1051/0004-6361/202244897}

\bibitem[{{Carleton} {et~al.}(2023){Carleton}, {Cohen}, {Frye}, {Pigarelli},
  {Zhang}, {Windhorst}, {Diego}, {Conselice}, {Cheng}, {Driver}, {Foo},
  {Bhatawdekar}, {Kamieneski}, {Jansen}, {Yan}, {Summers}, {Robotham},
  {Willmer}, {Koekemoer}, {Tompkins}, {Coe}, {Grogin}, {Marshall}, {Nonino},
  {Pirzkal}, \& {Ryan}}]{Carleton2023}
{Carleton}, T., {Cohen}, S.~H., {Frye}, B.~L., {et~al.} 2023, \apj, 953, 83,
  \dodoi{10.3847/1538-4357/ace343}

\bibitem[{{Carnall} {et~al.}(2023){Carnall}, {Begley}, {McLeod}, {Hamadouche},
  {Donnan}, {McLure}, {Dunlop}, {Milvang-Jensen}, {Bondestam}, {Cullen},
  {Jewell}, \& {Pollock}}]{Carnall2022}
{Carnall}, A.~C., {Begley}, R., {McLeod}, D.~J., {et~al.} 2023, \mnras, 518,
  L45, \dodoi{10.1093/mnrasl/slac136}

\bibitem[{{Castellano} {et~al.}(2022){Castellano}, {Fontana}, {Treu},
  {Santini}, {Merlin}, {Leethochawalit}, {Trenti}, {Vanzella}, {Mestric},
  {Bonchi}, {Belfiori}, {Nonino}, {Paris}, {Polenta}, {Roberts-Borsani},
  {Boyett}, {Brada{\v{c}}}, {Calabr{\`o}}, {Glazebrook}, {Grillo}, {Mascia},
  {Mason}, {Mercurio}, {Morishita}, {Nanayakkara}, {Pentericci}, {Rosati},
  {Vulcani}, {Wang}, \& {Yang}}]{Castellano2022}
{Castellano}, M., {Fontana}, A., {Treu}, T., {et~al.} 2022, \apjl, 938, L15,
  \dodoi{10.3847/2041-8213/ac94d0}

\bibitem[{{Castellano} {et~al.}(2023){Castellano}, {Fontana}, {Treu}, {Merlin},
  {Santini}, {Bergamini}, {Grillo}, {Rosati}, {Acebron}, {Leethochawalit},
  {Paris}, {Bonchi}, {Belfiori}, {Calabr{\`o}}, {Correnti}, {Nonino},
  {Polenta}, {Trenti}, {Boyett}, {Brammer}, {Broadhurst}, {Caminha}, {Chen},
  {Filippenko}, {Fortuni}, {Glazebrook}, {Mascia}, {Mason}, {Menci},
  {Meneghetti}, {Mercurio}, {Metha}, {Morishita}, {Nanayakkara}, {Pentericci},
  {Roberts-Borsani}, {Roy}, {Vanzella}, {Vulcani}, {Yang}, \&
  {Wang}}]{Castellano2023}
---. 2023, \apjl, 948, L14, \dodoi{10.3847/2041-8213/accea5}

\bibitem[{{Chisholm} {et~al.}(2022){Chisholm}, {Saldana-Lopez}, {Flury},
  {Schaerer}, {Jaskot}, {Amor{\'\i}n}, {Atek}, {Finkelstein}, {Fleming},
  {Ferguson}, {Fern{\'a}ndez}, {Giavalisco}, {Hayes}, {Heckman}, {Henry}, {Ji},
  {Marques-Chaves}, {Mauerhofer}, {McCandliss}, {Oey}, {{\"O}stlin},
  {Rutkowski}, {Scarlata}, {Thuan}, {Trebitsch}, {Wang}, {Worseck}, \&
  {Xu}}]{Chisholm2022}
{Chisholm}, J., {Saldana-Lopez}, A., {Flury}, S., {et~al.} 2022, \mnras, 517,
  5104, \dodoi{10.1093/mnras/stac2874}

\bibitem[{{Clay} {et~al.}(2015){Clay}, {Thomas}, {Wilkins}, \&
  {Henriques}}]{Clay2015}
{Clay}, S.~J., {Thomas}, P.~A., {Wilkins}, S.~M., \& {Henriques}, B. M.~B.
  2015, \mnras, 451, 2692, \dodoi{10.1093/mnras/stv818}

\bibitem[{{Conroy} \& {Gunn}(2010)}]{Conroy2010}
{Conroy}, C., \& {Gunn}, J.~E. 2010, \apj, 712, 833,
  \dodoi{10.1088/0004-637X/712/2/833}

\bibitem[{{Cullen} {et~al.}(2023){Cullen}, {McLeod}, {McLure}, {Dunlop},
  {Donnan}, {Carnall}, {Keating}, {Magee}, {Arellano-Cordova}, {Bowler},
  {Begley}, {Flury}, {Hamadouche}, \& {Stanton}}]{Cullen2023}
{Cullen}, F., {McLeod}, D.~J., {McLure}, R.~J., {et~al.} 2023, arXiv e-prints,
  arXiv:2311.06209, \dodoi{10.48550/arXiv.2311.06209}

\bibitem[{{Curtis-Lake} {et~al.}(2023){Curtis-Lake}, {Carniani}, {Cameron},
  {Charlot}, {Jakobsen}, {Maiolino}, {Bunker}, {Witstok}, {Smit}, {Chevallard},
  {Willott}, {Ferruit}, {Arribas}, {Bonaventura}, {Curti}, {D'Eugenio},
  {Franx}, {Giardino}, {Looser}, {L{\"u}tzgendorf}, {Maseda}, {Rawle}, {Rix},
  {Rodr{\'\i}guez del Pino}, {{\"U}bler}, {Sirianni}, {Dressler}, {Egami},
  {Eisenstein}, {Endsley}, {Hainline}, {Hausen}, {Johnson}, {Rieke},
  {Robertson}, {Shivaei}, {Stark}, {Tacchella}, {Williams}, {Willmer},
  {Bhatawdekar}, {Bowler}, {Boyett}, {Chen}, {de Graaff}, {Helton}, {Hviding},
  {Jones}, {Kumari}, {Lyu}, {Nelson}, {Perna}, {Sandles}, {Saxena}, {Suess},
  {Sun}, {Topping}, {Wallace}, \& {Whitler}}]{CurtisLake2022}
{Curtis-Lake}, E., {Carniani}, S., {Cameron}, A., {et~al.} 2023, Nature
  Astronomy, 7, 622, \dodoi{10.1038/s41550-023-01918-w}

\bibitem[{{Dawoodbhoy} {et~al.}(2023){Dawoodbhoy}, {Shapiro}, {Ocvirk},
  {Lewis}, {Aubert}, {Sorce}, {Ahn}, {Iliev}, {Park}, {Teyssier}, \&
  {Yepes}}]{Dawoodbhoy2023}
{Dawoodbhoy}, T., {Shapiro}, P.~R., {Ocvirk}, P., {et~al.} 2023, \mnras, 524,
  6231, \dodoi{10.1093/mnras/stad2331}

\bibitem[{{Dayal} {et~al.}(2014){Dayal}, {Ferrara}, {Dunlop}, \&
  {Pacucci}}]{dayal2014}
{Dayal}, P., {Ferrara}, A., {Dunlop}, J.~S., \& {Pacucci}, F. 2014, \mnras,
  445, 2545, \dodoi{10.1093/mnras/stu1848}

\bibitem[{{Dayal} {et~al.}(2020){Dayal}, {Volonteri}, {Choudhury}, {Schneider},
  {Trebitsch}, {Gnedin}, {Atek}, {Hirschmann}, \& {Reines}}]{Dayal2020}
{Dayal}, P., {Volonteri}, M., {Choudhury}, T.~R., {et~al.} 2020, \mnras, 495,
  3065, \dodoi{10.1093/mnras/staa1138}

\bibitem[{{Dayal} {et~al.}(2022){Dayal}, {Ferrara}, {Sommovigo}, {Bouwens},
  {Oesch}, {Smit}, {Gonzalez}, {Schouws}, {Stefanon}, {Kobayashi}, {Bremer},
  {Algera}, {Aravena}, {Bowler}, {da Cunha}, {Fudamoto}, {Graziani}, {Hodge},
  {Inami}, {De Looze}, {Pallottini}, {Riechers}, {Schneider}, {Stark}, \&
  {Endsley}}]{dayal2022}
{Dayal}, P., {Ferrara}, A., {Sommovigo}, L., {et~al.} 2022, \mnras, 512, 989,
  \dodoi{10.1093/mnras/stac537}

\bibitem[{{Diego} {et~al.}(2023){Diego}, {Meena}, {Adams}, {Broadhurst}, {Dai},
  {Coe}, {Frye}, {Kelly}, {Koekemoer}, {Pascale}, {Willner}, {Zackrisson},
  {Zitrin}, {Windhorst}, {Cohen}, {Jansen}, {Summers}, {Tompkins}, {Conselice},
  {Driver}, {Yan}, {Grogin}, {Marshall}, {Pirzkal}, {Robotham}, {Ryan},
  {Willmer}, {Bradley}, {Caminha}, {Caputi}, {Carleton}, \&
  {Kamieneski}}]{Diego2023}
{Diego}, J.~M., {Meena}, A.~K., {Adams}, N.~J., {et~al.} 2023, \aap, 672, A3,
  \dodoi{10.1051/0004-6361/202245238}

\bibitem[{{Donnan} {et~al.}(2023){Donnan}, {McLeod}, {Dunlop}, {McLure},
  {Carnall}, {Begley}, {Cullen}, {Hamadouche}, {Bowler}, {Magee}, {McCracken},
  {Milvang-Jensen}, {Moneti}, \& {Targett}}]{Donnan2022}
{Donnan}, C.~T., {McLeod}, D.~J., {Dunlop}, J.~S., {et~al.} 2023, \mnras, 518,
  6011, \dodoi{10.1093/mnras/stac3472}

\bibitem[{{Drakos} {et~al.}(2022){Drakos}, {Villasenor}, {Robertson}, {Hausen},
  {Dickinson}, {Ferguson}, {Furlanetto}, {Greene}, {Madau}, {Shapley}, {Stark},
  \& {Wechsler}}]{Drakos2022}
{Drakos}, N.~E., {Villasenor}, B., {Robertson}, B.~E., {et~al.} 2022, \apj,
  926, 194, \dodoi{10.3847/1538-4357/ac46fb}

\bibitem[{{Driver} \& {Robotham}(2010)}]{Driver2010}
{Driver}, S.~P., \& {Robotham}, A. S.~G. 2010, \mnras, 407, 2131,
  \dodoi{10.1111/j.1365-2966.2010.17028.x}

\bibitem[{{Duncan} \& {Conselice}(2015)}]{Duncan2015}
{Duncan}, K., \& {Conselice}, C.~J. 2015, \mnras, 451, 2030,
  \dodoi{10.1093/mnras/stv1049}

\bibitem[{{Duncan} {et~al.}(2014){Duncan}, {Conselice}, {Mortlock}, {Hartley},
  {Guo}, {Ferguson}, {Dav{\'e}}, {Lu}, {Ownsworth}, {Ashby}, {Dekel},
  {Dickinson}, {Faber}, {Giavalisco}, {Grogin}, {Kocevski}, {Koekemoer},
  {Somerville}, \& {White}}]{Duncan2014}
{Duncan}, K., {Conselice}, C.~J., {Mortlock}, A., {et~al.} 2014, \mnras, 444,
  2960, \dodoi{10.1093/mnras/stu1622}

\bibitem[{{Efron} \& {Petrosian}(1992)}]{Efron1992}
{Efron}, B., \& {Petrosian}, V. 1992, \apj, 399, 345, \dodoi{10.1086/171931}

\bibitem[{{Eisenstein} {et~al.}(2023){Eisenstein}, {Willott}, {Alberts},
  {Arribas}, {Bonaventura}, {Bunker}, {Cameron}, {Carniani}, {Charlot},
  {Curtis-Lake}, {D'Eugenio}, {Endsley}, {Ferruit}, {Giardino}, {Hainline},
  {Hausen}, {Jakobsen}, {Johnson}, {Maiolino}, {Rieke}, {Rieke}, {Rix},
  {Robertson}, {Stark}, {Tacchella}, {Williams}, {Willmer}, {Baker}, {Baum},
  {Bhatawdekar}, {Boyett}, {Chen}, {Chevallard}, {Circosta}, {Curti},
  {Danhaive}, {DeCoursey}, {de Graaff}, {Dressler}, {Egami}, {Helton},
  {Hviding}, {Ji}, {Jones}, {Kumari}, {L{\"u}tzgendorf}, {Laseter}, {Looser},
  {Lyu}, {Maseda}, {Nelson}, {Parlanti}, {Perna}, {Pusk{\'a}s}, {Rawle},
  {Rodr{\'\i}guez Del Pino}, {Sandles}, {Saxena}, {Scholtz}, {Sharpe},
  {Shivaei}, {Silcock}, {Simmonds}, {Skarbinski}, {Smit}, {Stone}, {Suess},
  {Sun}, {Tang}, {Topping}, {{\"U}bler}, {Villanueva}, {Wallace}, {Whitler},
  {Witstok}, \& {Woodrum}}]{Eisenstein2023}
{Eisenstein}, D.~J., {Willott}, C., {Alberts}, S., {et~al.} 2023, arXiv
  e-prints, arXiv:2306.02465, \dodoi{10.48550/arXiv.2306.02465}

\bibitem[{{Endsley} {et~al.}(2023){Endsley}, {Stark}, {Whitler}, {Topping},
  {Chen}, {Plat}, {Chisholm}, \& {Charlot}}]{Endsley2022}
{Endsley}, R., {Stark}, D.~P., {Whitler}, L., {et~al.} 2023, \mnras, 524, 2312,
  \dodoi{10.1093/mnras/stad1919}

\bibitem[{{Feng} {et~al.}(2016{\natexlab{a}}){Feng}, {Di-Matteo}, {Croft},
  {Bird}, {Battaglia}, \& {Wilkins}}]{Bluetides-I}
{Feng}, Y., {Di-Matteo}, T., {Croft}, R.~A., {et~al.} 2016{\natexlab{a}},
  \mnras, 455, 2778, \dodoi{10.1093/mnras/stv2484}

\bibitem[{{Feng} {et~al.}(2016{\natexlab{b}}){Feng}, {Di-Matteo}, {Croft},
  {Bird}, {Battaglia}, \& {Wilkins}}]{Feng2016}
---. 2016{\natexlab{b}}, \mnras, 455, 2778, \dodoi{10.1093/mnras/stv2484}

\bibitem[{{Ferreira} {et~al.}(2022){Ferreira}, {Adams}, {Conselice},
  {Sazonova}, {Austin}, {Caruana}, {Ferrari}, {Verma}, {Trussler},
  {Broadhurst}, {Diego}, {Frye}, {Pascale}, {Wilkins}, {Windhorst}, \&
  {Zitrin}}]{Leo2022}
{Ferreira}, L., {Adams}, N., {Conselice}, C.~J., {et~al.} 2022, \apjl, 938, L2,
  \dodoi{10.3847/2041-8213/ac947c}

\bibitem[{Finkelstein {et~al.}(2015)Finkelstein, Ryan, Papovich, Dickinson,
  Song, Somerville, Ferguson, Salmon, Giavalisco, Koekemoer, Ashby, Behroozi,
  Castellano, Dunlop, Faber, Fazio, Fontana, Grogin, Hathi, Jaacks, Kocevski,
  Livermore, McLure, Merlin, Mobasher, Newman, Rafelski, Tilvi, \&
  Willner}]{Finkelstein2015}
Finkelstein, S.~L., Ryan, R.~E., Papovich, C., {et~al.} 2015, \apj, 810, 71,
  \dodoi{10.1088/0004-637X/810/1/71}

\bibitem[{{Finkelstein} {et~al.}(2019){Finkelstein}, {D'Aloisio},
  {Paardekooper}, {Ryan}, {Behroozi}, {Finlator}, {Livermore}, {Upton
  Sanderbeck}, {Dalla Vecchia}, \& {Khochfar}}]{Finkelstein2019}
{Finkelstein}, S.~L., {D'Aloisio}, A., {Paardekooper}, J.-P., {et~al.} 2019,
  \apj, 879, 36, \dodoi{10.3847/1538-4357/ab1ea8}

\bibitem[{{Finkelstein} {et~al.}(2023{\natexlab{a}}){Finkelstein}, {Bagley},
  {Ferguson}, {Wilkins}, {Kartaltepe}, {Papovich}, {Yung}, {Arrabal Haro},
  {Behroozi}, {Dickinson}, {Kocevski}, {Koekemoer}, {Larson}, {Le Bail},
  {Morales}, {P{\'e}rez-Gonz{\'a}lez}, {Burgarella}, {Dav{\'e}}, {Hirschmann},
  {Somerville}, {Wuyts}, {Bromm}, {Casey}, {Fontana}, {Fujimoto}, {Gardner},
  {Giavalisco}, {Grazian}, {Grogin}, {Hathi}, {Hutchison}, {Jha}, {Jogee},
  {Kewley}, {Kirkpatrick}, {Long}, {Lotz}, {Pentericci}, {Pierel}, {Pirzkal},
  {Ravindranath}, {Ryan}, {Trump}, {Yang}, {Bhatawdekar}, {Bisigello}, {Buat},
  {Calabr{\`o}}, {Castellano}, {Cleri}, {Cooper}, {Croton}, {Daddi}, {Dekel},
  {Elbaz}, {Franco}, {Gawiser}, {Holwerda}, {Huertas-Company}, {Jaskot},
  {Leung}, {Lucas}, {Mobasher}, {Pandya}, {Tacchella}, {Weiner}, \&
  {Zavala}}]{Finkelstein2022c}
{Finkelstein}, S.~L., {Bagley}, M.~B., {Ferguson}, H.~C., {et~al.}
  2023{\natexlab{a}}, \apjl, 946, L13, \dodoi{10.3847/2041-8213/acade4}

\bibitem[{{Finkelstein} {et~al.}(2023{\natexlab{b}}){Finkelstein}, {Leung},
  {Bagley}, {Dickinson}, {Ferguson}, {Papovich}, {Akins}, {Arrabal Haro},
  {Dave}, {Dekel}, {Kartaltepe}, {Kocevski}, {Koekemoer}, {Pirzkal},
  {Somerville}, {Yung}, {Amorin}, {Backhaus}, {Behroozi}, {Bisigello}, {Bromm},
  {Casey}, {Chavez Ortiz}, {Cheng}, {Chworowsky}, {Cleri}, {Cooper}, {Davis},
  {de la Vega}, {Elbaz}, {Franco}, {Fontana}, {Fujimoto}, {Giavalisco},
  {Grogin}, {Holwerda}, {Huertas-Company}, {Hirschmann}, {Iyer}, {Jogee},
  {Jung}, {Larson}, {Lucas}, {Mobasher}, {Morales}, {Morley}, {Mukherjee},
  {Perez-Gonzalez}, {Ravindranath}, {Rodighiero}, {Rowland}, {Tacchella},
  {Taylor}, {Trump}, \& {Wilkins}}]{Finkelstein2023}
{Finkelstein}, S.~L., {Leung}, G. C.~K., {Bagley}, M.~B., {et~al.}
  2023{\natexlab{b}}, arXiv e-prints, arXiv:2311.04279,
  \dodoi{10.48550/arXiv.2311.04279}

\bibitem[{{Foreman-Mackey} {et~al.}(2013){Foreman-Mackey}, {Hogg}, {Lang}, \&
  {Goodman}}]{emcee}
{Foreman-Mackey}, D., {Hogg}, D.~W., {Lang}, D., \& {Goodman}, J. 2013, \pasp,
  125, 306, \dodoi{10.1086/670067}

\bibitem[{{Frye} {et~al.}(2023){Frye}, {Pascale}, {Foo}, {Leimbach}, {Garuda},
  {Robles}, {Summers}, {Diaz}, {Kamieneski}, {Furtak}, {Cohen}, {Diego},
  {Beauchesne}, {Windhorst}, {Willner}, {Koekemoer}, {Zitrin}, {Caminha},
  {Caputi}, {Coe}, {Conselice}, {Dai}, {Dole}, {Driver}, {Grogin},
  {Harrington}, {Jansen}, {Kneib}, {Lehnert}, {Lowenthal}, {Marshall},
  {Menanteau}, {Pampliega}, {Pirzkal}, {Polletta}, {Richard}, {Robotham},
  {Ryan}, {Rutkowski}, {Sif{\'o}n}, {Tompkins}, {Wang}, {Yan}, \&
  {Yun}}]{Frye2023}
{Frye}, B.~L., {Pascale}, M., {Foo}, N., {et~al.} 2023, \apj, 952, 81,
  \dodoi{10.3847/1538-4357/acd929}

\bibitem[{{Gaia Collaboration} {et~al.}(2018){Gaia Collaboration}, {Brown},
  {Vallenari}, {Prusti}, {de Bruijne}, {Babusiaux}, {Bailer-Jones}, {Biermann},
  {Evans}, {Eyer}, {Jansen}, {Jordi}, {Klioner}, {Lammers}, {Lindegren},
  {Luri}, {Mignard}, {Panem}, {Pourbaix}, {Randich}, {Sartoretti}, {Siddiqui},
  {Soubiran}, {van Leeuwen}, {Walton}, {Arenou}, {Bastian}, {Cropper},
  {Drimmel}, {Katz}, {Lattanzi}, {Bakker}, {Cacciari}, {Casta{\~n}eda},
  {Chaoul}, {Cheek}, {De Angeli}, {Fabricius}, {Guerra}, {Holl}, {Masana},
  {Messineo}, {Mowlavi}, {Nienartowicz}, {Panuzzo}, {Portell}, {Riello},
  {Seabroke}, {Tanga}, {Th{\'e}venin}, {Gracia-Abril}, {Comoretto},
  {Garcia-Reinaldos}, {Teyssier}, {Altmann}, {Andrae}, {Audard},
  {Bellas-Velidis}, {Benson}, {Berthier}, {Blomme}, {Burgess}, {Busso},
  {Carry}, {Cellino}, {Clementini}, {Clotet}, {Creevey}, {Davidson}, {De
  Ridder}, {Delchambre}, {Dell'Oro}, {Ducourant},
  {Fern{\'a}ndez-Hern{\'a}ndez}, {Fouesneau}, {Fr{\'e}mat}, {Galluccio},
  {Garc{\'\i}a-Torres}, {Gonz{\'a}lez-N{\'u}{\~n}ez}, {Gonz{\'a}lez-Vidal},
  {Gosset}, {Guy}, {Halbwachs}, {Hambly}, {Harrison}, {Hern{\'a}ndez},
  {Hestroffer}, {Hodgkin}, {Hutton}, {Jasniewicz}, {Jean-Antoine-Piccolo},
  {Jordan}, {Korn}, {Krone-Martins}, {Lanzafame}, {Lebzelter}, {L{\"o}ffler},
  {Manteiga}, {Marrese}, {Mart{\'\i}n-Fleitas}, {Moitinho}, {Mora}, {Muinonen},
  {Osinde}, {Pancino}, {Pauwels}, {Petit}, {Recio-Blanco}, {Richards},
  {Rimoldini}, {Robin}, {Sarro}, {Siopis}, {Smith}, {Sozzetti}, {S{\"u}veges},
  {Torra}, {van Reeven}, {Abbas}, {Abreu Aramburu}, {Accart}, {Aerts},
  {Altavilla}, {{\'A}lvarez}, {Alvarez}, {Alves}, {Anderson}, {Andrei},
  {Anglada Varela}, {Antiche}, {Antoja}, {Arcay}, {Astraatmadja}, {Bach},
  {Baker}, {Balaguer-N{\'u}{\~n}ez}, {Balm}, {Barache}, {Barata}, {Barbato},
  {Barblan}, {Barklem}, {Barrado}, {Barros}, {Barstow}, {Bartholom{\'e}
  Mu{\~n}oz}, {Bassilana}, {Becciani}, {Bellazzini}, {Berihuete}, {Bertone},
  {Bianchi}, {Bienaym{\'e}}, {Blanco-Cuaresma}, {Boch}, {Boeche}, {Bombrun},
  {Borrachero}, {Bossini}, {Bouquillon}, {Bourda}, {Bragaglia}, {Bramante},
  {Breddels}, {Bressan}, {Brouillet}, {Br{\"u}semeister}, {Brugaletta},
  {Bucciarelli}, {Burlacu}, {Busonero}, {Butkevich}, {Buzzi}, {Caffau},
  {Cancelliere}, {Cannizzaro}, {Cantat-Gaudin}, {Carballo}, {Carlucci},
  {Carrasco}, {Casamiquela}, {Castellani}, {Castro-Ginard}, {Charlot},
  {Chemin}, {Chiavassa}, {Cocozza}, {Costigan}, {Cowell}, {Crifo}, {Crosta},
  {Crowley}, {Cuypers}, {Dafonte}, {Damerdji}, {Dapergolas}, {David}, {David},
  {de Laverny}, {De Luise}, {De March}, {de Martino}, {de Souza}, {de Torres},
  {Debosscher}, {del Pozo}, {Delbo}, {Delgado}, {Delgado}, {Di Matteo},
  {Diakite}, {Diener}, {Distefano}, {Dolding}, {Drazinos}, {Dur{\'a}n},
  {Edvardsson}, {Enke}, {Eriksson}, {Esquej}, {Eynard Bontemps}, {Fabre},
  {Fabrizio}, {Faigler}, {Falc{\~a}o}, {Farr{\`a}s Casas}, {Federici},
  {Fedorets}, {Fernique}, {Figueras}, {Filippi}, {Findeisen}, {Fonti},
  {Fraile}, {Fraser}, {Fr{\'e}zouls}, {Gai}, {Galleti}, {Garabato},
  {Garc{\'\i}a-Sedano}, {Garofalo}, {Garralda}, {Gavel}, {Gavras}, {Gerssen},
  {Geyer}, {Giacobbe}, {Gilmore}, {Girona}, {Giuffrida}, {Glass}, {Gomes},
  {Granvik}, {Gueguen}, {Guerrier}, {Guiraud}, {Guti{\'e}rrez-S{\'a}nchez},
  {Haigron}, {Hatzidimitriou}, {Hauser}, {Haywood}, {Heiter}, {Helmi}, {Heu},
  {Hilger}, {Hobbs}, {Hofmann}, {Holland}, {Huckle}, {Hypki}, {Icardi},
  {Jan{\ss}en}, {Jevardat de Fombelle}, {Jonker}, {Juh{\'a}sz}, {Julbe},
  {Karampelas}, {Kewley}, {Klar}, {Kochoska}, {Kohley}, {Kolenberg},
  {Kontizas}, {Kontizas}, {Koposov}, {Kordopatis}, {Kostrzewa-Rutkowska},
  {Koubsky}, {Lambert}, {Lanza}, {Lasne}, {Lavigne}, {Le Fustec}, {Le
  Poncin-Lafitte}, {Lebreton}, {Leccia}, {Leclerc}, {Lecoeur-Taibi},
  {Lenhardt}, {Leroux}, {Liao}, {Licata}, {Lindstr{\o}m}, {Lister}, {Livanou},
  {Lobel}, {L{\'o}pez}, {Managau}, {Mann}, {Mantelet}, {Marchal}, {Marchant},
  {Marconi}, {Marinoni}, {Marschalk{\'o}}, {Marshall}, {Martino}, {Marton},
  {Mary}, {Massari}, {Matijevi{\v{c}}}, {Mazeh}, {McMillan}, {Messina},
  {Michalik}, {Millar}, {Molina}, {Molinaro}, {Moln{\'a}r}, {Montegriffo},
  {Mor}, {Morbidelli}, {Morel}, {Morris}, {Mulone}, {Muraveva}, {Musella},
  {Nelemans}, {Nicastro}, {Noval}, {O'Mullane}, {Ord{\'e}novic},
  {Ord{\'o}{\~n}ez-Blanco}, {Osborne}, {Pagani}, {Pagano}, {Pailler},
  {Palacin}, {Palaversa}, {Panahi}, {Pawlak}, {Piersimoni}, {Pineau}, {Plachy},
  {Plum}, {Poggio}, {Poujoulet}, {Pr{\v{s}}a}, {Pulone}, {Racero}, {Ragaini},
  {Rambaux}, {Ramos-Lerate}, {Regibo}, {Reyl{\'e}}, {Riclet}, {Ripepi}, {Riva},
  {Rivard}, {Rixon}, {Roegiers}, {Roelens}, {Romero-G{\'o}mez}, {Rowell},
  {Royer}, {Ruiz-Dern}, {Sadowski}, {Sagrist{\`a} Sell{\'e}s}, {Sahlmann},
  {Salgado}, {Salguero}, {Sanna}, {Santana-Ros}, {Sarasso}, {Savietto},
  {Schultheis}, {Sciacca}, {Segol}, {Segovia}, {S{\'e}gransan}, {Shih},
  {Siltala}, {Silva}, {Smart}, {Smith}, {Solano}, {Solitro}, {Sordo}, {Soria
  Nieto}, {Souchay}, {Spagna}, {Spoto}, {Stampa}, {Steele},
  {Steidelm{\"u}ller}, {Stephenson}, {Stoev}, {Suess}, {Surdej}, {Szabados},
  {Szegedi-Elek}, {Tapiador}, {Taris}, {Tauran}, {Taylor}, {Teixeira},
  {Terrett}, {Teyssandier}, {Thuillot}, {Titarenko}, {Torra Clotet}, {Turon},
  {Ulla}, {Utrilla}, {Uzzi}, {Vaillant}, {Valentini}, {Valette}, {van Elteren},
  {Van Hemelryck}, {van Leeuwen}, {Vaschetto}, {Vecchiato}, {Veljanoski},
  {Viala}, {Vicente}, {Vogt}, {von Essen}, {Voss}, {Votruba}, {Voutsinas},
  {Walmsley}, {Weiler}, {Wertz}, {Wevers}, {Wyrzykowski}, {Yoldas},
  {{\v{Z}}erjal}, {Ziaeepour}, {Zorec}, {Zschocke}, {Zucker}, {Zurbach}, \&
  {Zwitter}}]{GAIADR2}
{Gaia Collaboration}, {Brown}, A.~G.~A., {Vallenari}, A., {et~al.} 2018, \aap,
  616, A1, \dodoi{10.1051/0004-6361/201833051}

\bibitem[{{Gaia Collaboration} {et~al.}(2023){Gaia Collaboration}, {Vallenari},
  {Brown}, {Prusti}, {de Bruijne}, {Arenou}, {Babusiaux}, {Biermann},
  {Creevey}, {Ducourant}, {Evans}, {Eyer}, {Guerra}, {Hutton}, {Jordi},
  {Klioner}, {Lammers}, {Lindegren}, {Luri}, {Mignard}, {Panem}, {Pourbaix},
  {Randich}, {Sartoretti}, {Soubiran}, {Tanga}, {Walton}, {Bailer-Jones},
  {Bastian}, {Drimmel}, {Jansen}, {Katz}, {Lattanzi}, {van Leeuwen}, {Bakker},
  {Cacciari}, {Casta{\~n}eda}, {De Angeli}, {Fabricius}, {Fouesneau},
  {Fr{\'e}mat}, {Galluccio}, {Guerrier}, {Heiter}, {Masana}, {Messineo},
  {Mowlavi}, {Nicolas}, {Nienartowicz}, {Pailler}, {Panuzzo}, {Riclet}, {Roux},
  {Seabroke}, {Sordo}, {Th{\'e}venin}, {Gracia-Abril}, {Portell}, {Teyssier},
  {Altmann}, {Andrae}, {Audard}, {Bellas-Velidis}, {Benson}, {Berthier},
  {Blomme}, {Burgess}, {Busonero}, {Busso}, {C{\'a}novas}, {Carry}, {Cellino},
  {Cheek}, {Clementini}, {Damerdji}, {Davidson}, {de Teodoro}, {Nu{\~n}ez
  Campos}, {Delchambre}, {Dell'Oro}, {Esquej}, {Fern{\'a}ndez-Hern{\'a}ndez},
  {Fraile}, {Garabato}, {Garc{\'\i}a-Lario}, {Gosset}, {Haigron}, {Halbwachs},
  {Hambly}, {Harrison}, {Hern{\'a}ndez}, {Hestroffer}, {Hodgkin}, {Holl},
  {Jan{\ss}en}, {Jevardat de Fombelle}, {Jordan}, {Krone-Martins}, {Lanzafame},
  {L{\"o}ffler}, {Marchal}, {Marrese}, {Moitinho}, {Muinonen}, {Osborne},
  {Pancino}, {Pauwels}, {Recio-Blanco}, {Reyl{\'e}}, {Riello}, {Rimoldini},
  {Roegiers}, {Rybizki}, {Sarro}, {Siopis}, {Smith}, {Sozzetti}, {Utrilla},
  {van Leeuwen}, {Abbas}, {{\'A}brah{\'a}m}, {Abreu Aramburu}, {Aerts},
  {Aguado}, {Ajaj}, {Aldea-Montero}, {Altavilla}, {{\'A}lvarez}, {Alves},
  {Anders}, {Anderson}, {Anglada Varela}, {Antoja}, {Baines}, {Baker},
  {Balaguer-N{\'u}{\~n}ez}, {Balbinot}, {Balog}, {Barache}, {Barbato},
  {Barros}, {Barstow}, {Bartolom{\'e}}, {Bassilana}, {Bauchet}, {Becciani},
  {Bellazzini}, {Berihuete}, {Bernet}, {Bertone}, {Bianchi}, {Binnenfeld},
  {Blanco-Cuaresma}, {Blazere}, {Boch}, {Bombrun}, {Bossini}, {Bouquillon},
  {Bragaglia}, {Bramante}, {Breedt}, {Bressan}, {Brouillet}, {Brugaletta},
  {Bucciarelli}, {Burlacu}, {Butkevich}, {Buzzi}, {Caffau}, {Cancelliere},
  {Cantat-Gaudin}, {Carballo}, {Carlucci}, {Carnerero}, {Carrasco},
  {Casamiquela}, {Castellani}, {Castro-Ginard}, {Chaoul}, {Charlot}, {Chemin},
  {Chiaramida}, {Chiavassa}, {Chornay}, {Comoretto}, {Contursi}, {Cooper},
  {Cornez}, {Cowell}, {Crifo}, {Cropper}, {Crosta}, {Crowley}, {Dafonte},
  {Dapergolas}, {David}, {David}, {de Laverny}, {De Luise}, {De March}, {De
  Ridder}, {de Souza}, {de Torres}, {del Peloso}, {del Pozo}, {Delbo},
  {Delgado}, {Delisle}, {Demouchy}, {Dharmawardena}, {Di Matteo}, {Diakite},
  {Diener}, {Distefano}, {Dolding}, {Edvardsson}, {Enke}, {Fabre}, {Fabrizio},
  {Faigler}, {Fedorets}, {Fernique}, {Fienga}, {Figueras}, {Fournier},
  {Fouron}, {Fragkoudi}, {Gai}, {Garcia-Gutierrez}, {Garcia-Reinaldos},
  {Garc{\'\i}a-Torres}, {Garofalo}, {Gavel}, {Gavras}, {Gerlach}, {Geyer},
  {Giacobbe}, {Gilmore}, {Girona}, {Giuffrida}, {Gomel}, {Gomez},
  {Gonz{\'a}lez-N{\'u}{\~n}ez}, {Gonz{\'a}lez-Santamar{\'\i}a},
  {Gonz{\'a}lez-Vidal}, {Granvik}, {Guillout}, {Guiraud},
  {Guti{\'e}rrez-S{\'a}nchez}, {Guy}, {Hatzidimitriou}, {Hauser}, {Haywood},
  {Helmer}, {Helmi}, {Sarmiento}, {Hidalgo}, {Hilger}, {H{\l}adczuk}, {Hobbs},
  {Holland}, {Huckle}, {Jardine}, {Jasniewicz}, {Jean-Antoine Piccolo},
  {Jim{\'e}nez-Arranz}, {Jorissen}, {Juaristi Campillo}, {Julbe}, {Karbevska},
  {Kervella}, {Khanna}, {Kontizas}, {Kordopatis}, {Korn}, {K{\'o}sp{\'a}l},
  {Kostrzewa-Rutkowska}, {Kruszy{\'n}ska}, {Kun}, {Laizeau}, {Lambert},
  {Lanza}, {Lasne}, {Le Campion}, {Lebreton}, {Lebzelter}, {Leccia}, {Leclerc},
  {Lecoeur-Taibi}, {Liao}, {Licata}, {Lindstr{\o}m}, {Lister}, {Livanou},
  {Lobel}, {Lorca}, {Loup}, {Madrero Pardo}, {Magdaleno Romeo}, {Managau},
  {Mann}, {Manteiga}, {Marchant}, {Marconi}, {Marcos}, {Marcos Santos},
  {Mar{\'\i}n Pina}, {Marinoni}, {Marocco}, {Marshall}, {Martin Polo},
  {Mart{\'\i}n-Fleitas}, {Marton}, {Mary}, {Masip}, {Massari},
  {Mastrobuono-Battisti}, {Mazeh}, {McMillan}, {Messina}, {Michalik}, {Millar},
  {Mints}, {Molina}, {Molinaro}, {Moln{\'a}r}, {Monari}, {Mongui{\'o}},
  {Montegriffo}, {Montero}, {Mor}, {Mora}, {Morbidelli}, {Morel}, {Morris},
  {Muraveva}, {Murphy}, {Musella}, {Nagy}, {Noval}, {Oca{\~n}a}, {Ogden},
  {Ordenovic}, {Osinde}, {Pagani}, {Pagano}, {Palaversa}, {Palicio},
  {Pallas-Quintela}, {Panahi}, {Payne-Wardenaar}, {Pe{\~n}alosa Esteller},
  {Penttil{\"a}}, {Pichon}, {Piersimoni}, {Pineau}, {Plachy}, {Plum}, {Poggio},
  {Pr{\v{s}}a}, {Pulone}, {Racero}, {Ragaini}, {Rainer}, {Raiteri}, {Rambaux},
  {Ramos}, {Ramos-Lerate}, {Re Fiorentin}, {Regibo}, {Richards}, {Rios Diaz},
  {Ripepi}, {Riva}, {Rix}, {Rixon}, {Robichon}, {Robin}, {Robin}, {Roelens},
  {Rogues}, {Rohrbasser}, {Romero-G{\'o}mez}, {Rowell}, {Royer}, {Ruz Mieres},
  {Rybicki}, {Sadowski}, {S{\'a}ez N{\'u}{\~n}ez}, {Sagrist{\`a} Sell{\'e}s},
  {Sahlmann}, {Salguero}, {Samaras}, {Sanchez Gimenez}, {Sanna},
  {Santove{\~n}a}, {Sarasso}, {Schultheis}, {Sciacca}, {Segol}, {Segovia},
  {S{\'e}gransan}, {Semeux}, {Shahaf}, {Siddiqui}, {Siebert}, {Siltala},
  {Silvelo}, {Slezak}, {Slezak}, {Smart}, {Snaith}, {Solano}, {Solitro},
  {Souami}, {Souchay}, {Spagna}, {Spina}, {Spoto}, {Steele},
  {Steidelm{\"u}ller}, {Stephenson}, {S{\"u}veges}, {Surdej}, {Szabados},
  {Szegedi-Elek}, {Taris}, {Taylor}, {Teixeira}, {Tolomei}, {Tonello}, {Torra},
  {Torra}, {Torralba Elipe}, {Trabucchi}, {Tsounis}, {Turon}, {Ulla}, {Unger},
  {Vaillant}, {van Dillen}, {van Reeven}, {Vanel}, {Vecchiato}, {Viala},
  {Vicente}, {Voutsinas}, {Weiler}, {Wevers}, {Wyrzykowski}, {Yoldas}, {Yvard},
  {Zhao}, {Zorec}, {Zucker}, \& {Zwitter}}]{GAIADR3}
{Gaia Collaboration}, {Vallenari}, A., {Brown}, A.~G.~A., {et~al.} 2023, \aap,
  674, A1, \dodoi{10.1051/0004-6361/202243940}

\bibitem[{{Giallongo} {et~al.}(2019){Giallongo}, {Grazian}, {Fiore}, {Kodra},
  {Urrutia}, {Castellano}, {Cristiani}, {Dickinson}, {Fontana}, {Menci},
  {Pentericci}, {Boutsia}, {Newman}, \& {Puccetti}}]{Giallongo2019}
{Giallongo}, E., {Grazian}, A., {Fiore}, F., {et~al.} 2019, \apj, 884, 19,
  \dodoi{10.3847/1538-4357/ab39e1}

\bibitem[{{Grazian} {et~al.}(2012){Grazian}, {Castellano}, {Fontana},
  {Pentericci}, {Dunlop}, {McLure}, {Koekemoer}, {Dickinson}, {Faber},
  {Ferguson}, {Galametz}, {Giavalisco}, {Grogin}, {Hathi}, {Kocevski}, {Lai},
  {Newman}, \& {Vanzella}}]{Grazian2012}
{Grazian}, A., {Castellano}, M., {Fontana}, A., {et~al.} 2012, \aap, 547, A51,
  \dodoi{10.1051/0004-6361/201219669}

\bibitem[{{Groth} \& et~al.(1994)}]{Groth2994}
{Groth}, E.~J., \& et~al. 1994, \baas, 185, 5309

\bibitem[{{Guhathakurta} {et~al.}(1990){Guhathakurta}, {Tyson}, \&
  {Majewski}}]{Guhathakurta1990}
{Guhathakurta}, P., {Tyson}, J.~A., \& {Majewski}, S.~R. 1990, \apjl, 357, L9,
  \dodoi{10.1086/185754}

\bibitem[{{Hainline} {et~al.}(2023{\natexlab{a}}){Hainline}, {Johnson},
  {Robertson}, {Tacchella}, {Helton}, {Sun}, {Eisenstein}, {Simmonds},
  {Topping}, {Whitler}, {Willmer}, {Rieke}, {Suess}, {Hviding}, {Cameron},
  {Alberts}, {Baker}, {Bhatawdekar}, {Boyett}, {Bunker}, {Carniani}, {Charlot},
  {Chen}, {Curti}, {Curtis-Lake}, {D'Eugenio}, {Egami}, {Endsley}, {Hausen},
  {Ji}, {Looser}, {Lyu}, {Maiolino}, {Nelson}, {Puskas}, {Rawle}, {Sandles},
  {Saxena}, {Smit}, {Stark}, {Williams}, {Willott}, \&
  {Witstok}}]{Hainline2023a}
{Hainline}, K.~N., {Johnson}, B.~D., {Robertson}, B., {et~al.}
  2023{\natexlab{a}}, arXiv e-prints, arXiv:2306.02468,
  \dodoi{10.48550/arXiv.2306.02468}

\bibitem[{{Hainline} {et~al.}(2023{\natexlab{b}}){Hainline}, {Helton},
  {Johnson}, {Sun}, {Topping}, {Leisenring}, {Baker}, {Eisenstein}, {Hausen},
  {Hviding}, {Lyu}, {Robertson}, {Tacchella}, {Williams}, {Willmer}, \&
  {Roellig}}]{Hainline2023b}
{Hainline}, K.~N., {Helton}, J.~M., {Johnson}, B.~D., {et~al.}
  2023{\natexlab{b}}, arXiv e-prints, arXiv:2309.03250,
  \dodoi{10.48550/arXiv.2309.03250}

\bibitem[{{Harikane} {et~al.}(2022{\natexlab{a}}){Harikane}, {Ono}, {Ouchi},
  {Liu}, {Sawicki}, {Shibuya}, {Behroozi}, {He}, {Shimasaku}, {Arnouts},
  {Coupon}, {Fujimoto}, {Gwyn}, {Huang}, {Inoue}, {Kashikawa}, {Komiyama},
  {Matsuoka}, \& {Willott}}]{Harikane2021}
{Harikane}, Y., {Ono}, Y., {Ouchi}, M., {et~al.} 2022{\natexlab{a}}, \apjs,
  259, 20, \dodoi{10.3847/1538-4365/ac3dfc}

\bibitem[{{Harikane} {et~al.}(2022{\natexlab{b}}){Harikane}, {Inoue},
  {Mawatari}, {Hashimoto}, {Yamanaka}, {Fudamoto}, {Matsuo}, {Tamura}, {Dayal},
  {Yung}, {Hutter}, {Pacucci}, {Sugahara}, \& {Koekemoer}}]{Harikane2022}
{Harikane}, Y., {Inoue}, A.~K., {Mawatari}, K., {et~al.} 2022{\natexlab{b}},
  \apj, 929, 1, \dodoi{10.3847/1538-4357/ac53a9}

\bibitem[{{Harikane} {et~al.}(2023){Harikane}, {Ouchi}, {Oguri}, {Ono},
  {Nakajima}, {Isobe}, {Umeda}, {Mawatari}, \& {Zhang}}]{Harikane2023}
{Harikane}, Y., {Ouchi}, M., {Oguri}, M., {et~al.} 2023, \apjs, 265, 5,
  \dodoi{10.3847/1538-4365/acaaa9}

\bibitem[{{Hashimoto} {et~al.}(2018){Hashimoto}, {Laporte}, {Mawatari},
  {Ellis}, {Inoue}, {Zackrisson}, {Roberts-Borsani}, {Zheng}, {Tamura},
  {Bauer}, {Fletcher}, {Harikane}, {Hatsukade}, {Hayatsu}, {Matsuda}, {Matsuo},
  {Okamoto}, {Ouchi}, {Pell{\'o}}, {Rydberg}, {Shimizu}, {Taniguchi},
  {Umehata}, \& {Yoshida}}]{Hashimoto2018}
{Hashimoto}, T., {Laporte}, N., {Mawatari}, K., {et~al.} 2018, \nat, 557, 392,
  \dodoi{10.1038/s41586-018-0117-z}

\bibitem[{{Hassan} {et~al.}(2018){Hassan}, {Dav{\'e}}, {Mitra}, {Finlator},
  {Ciardi}, \& {Santos}}]{Hassan2018}
{Hassan}, S., {Dav{\'e}}, R., {Mitra}, S., {et~al.} 2018, \mnras, 473, 227,
  \dodoi{10.1093/mnras/stx2194}

\bibitem[{Hoaglin {et~al.}(1983)Hoaglin, Mosteller, \&
  Tukey}]{hoaglin2000understanding}
Hoaglin, D.~C., Mosteller, F., \& Tukey, J.~W. 1983, Understanding Robust and
  Exploratory Data Analysis (New York: Wiley)

\bibitem[{{Hoffmann} {et~al.}(2021){Hoffmann}, {Mack}, {Avila}, {Martlin},
  {Cohen}, \& {Bajaj}}]{Hoffmann2021}
{Hoffmann}, S.~L., {Mack}, J., {Avila}, R., {et~al.} 2021, in American
  Astronomical Society Meeting Abstracts, Vol.~53, American Astronomical
  Society Meeting Abstracts, 216.02

\bibitem[{{Hubble} \& {Humason}(1931)}]{Hubble1931}
{Hubble}, E., \& {Humason}, M.~L. 1931, \apj, 74, 43, \dodoi{10.1086/143323}

\bibitem[{Hunter(2007)}]{Hunter2007}
Hunter, J.~D. 2007, Computing in Science \& Engineering, 9, 90,
  \dodoi{10.1109/MCSE.2007.55}

\bibitem[{{Ilbert} {et~al.}(2006){Ilbert}, {Arnouts}, {McCracken},
  {Bolzonella}, {Bertin}, {Le F{\`e}vre}, {Mellier}, {Zamorani}, {Pell{\`o}},
  {Iovino}, {Tresse}, {Le Brun}, {Bottini}, {Garilli}, {Maccagni}, {Picat},
  {Scaramella}, {Scodeggio}, {Vettolani}, {Zanichelli}, {Adami}, {Bardelli},
  {Cappi}, {Charlot}, {Ciliegi}, {Contini}, {Cucciati}, {Foucaud}, {Franzetti},
  {Gavignaud}, {Guzzo}, {Marano}, {Marinoni}, {Mazure}, {Meneux}, {Merighi},
  {Paltani}, {Pollo}, {Pozzetti}, {Radovich}, {Zucca}, {Bondi}, {Bongiorno},
  {Busarello}, {de La Torre}, {Gregorini}, {Lamareille}, {Mathez}, {Merluzzi},
  {Ripepi}, {Rizzo}, \& {Vergani}}]{Ilbert2006}
{Ilbert}, O., {Arnouts}, S., {McCracken}, H.~J., {et~al.} 2006, \aap, 457, 841,
  \dodoi{10.1051/0004-6361:20065138}

\bibitem[{{Illingworth} {et~al.}(2016){Illingworth}, {Magee}, {Bouwens},
  {Oesch}, {Labbe}, {van Dokkum}, {Whitaker}, {Holden}, {Franx}, \&
  {Gonzalez}}]{Illingworth2016}
{Illingworth}, G., {Magee}, D., {Bouwens}, R., {et~al.} 2016, arXiv e-prints,
  arXiv:1606.00841, \dodoi{10.48550/arXiv.1606.00841}

\bibitem[{{Ishigaki} {et~al.}(2018){Ishigaki}, {Kawamata}, {Ouchi}, {Oguri},
  {Shimasaku}, \& {Ono}}]{Ishigaki2018}
{Ishigaki}, M., {Kawamata}, R., {Ouchi}, M., {et~al.} 2018, \apj, 854, 73,
  \dodoi{10.3847/1538-4357/aaa544}

\bibitem[{{Jansen} \& {Windhorst}(2018)}]{Jansen2018}
{Jansen}, R.~A., \& {Windhorst}, R.~A. 2018, \pasp, 130, 124001,
  \dodoi{10.1088/1538-3873/aae476}

\bibitem[{{Jiang} {et~al.}(2021){Jiang}, {Kashikawa}, {Wang}, {Walth}, {Ho},
  {Cai}, {Egami}, {Fan}, {Ito}, {Liang}, {Schaerer}, \& {Stark}}]{Jiang2020}
{Jiang}, L., {Kashikawa}, N., {Wang}, S., {et~al.} 2021, Nature Astronomy, 5,
  256, \dodoi{10.1038/s41550-020-01275-y}

\bibitem[{{Kamieneski} {et~al.}(2023){Kamieneski}, {Frye}, {Pascale}, {Cohen},
  {Windhorst}, {Jansen}, {Yun}, {Cheng}, {Summers}, {Carleton}, {Harrington},
  {Diego}, {Yan}, {Koekemoer}, {Willmer}, {Petric}, {Furtak}, {Foo},
  {Conselice}, {Coe}, {Driver}, {Grogin}, {Marshall}, {Nonino}, {Pirzkal},
  {Robotham}, {Ryan}, \& {Tompkins}}]{Kamienski2023}
{Kamieneski}, P.~S., {Frye}, B.~L., {Pascale}, M., {et~al.} 2023, \apj, 955,
  91, \dodoi{10.3847/1538-4357/aceb4a}

\bibitem[{{Kannan} {et~al.}(2022){Kannan}, {Garaldi}, {Smith}, {Pakmor},
  {Springel}, {Vogelsberger}, \& {Hernquist}}]{Kannan2022}
{Kannan}, R., {Garaldi}, E., {Smith}, A., {et~al.} 2022, \mnras, 511, 4005,
  \dodoi{10.1093/mnras/stab3710}

\bibitem[{{Kocevski} {et~al.}(2023){Kocevski}, {Onoue}, {Inayoshi}, {Trump},
  {Arrabal Haro}, {Grazian}, {Dickinson}, {Finkelstein}, {Kartaltepe},
  {Hirschmann}, {Aird}, {Holwerda}, {Fujimoto}, {Juneau}, {Amor{\'\i}n},
  {Backhaus}, {Bagley}, {Barro}, {Bell}, {Bisigello}, {Calabr{\`o}}, {Cleri},
  {Cooper}, {Ding}, {Grogin}, {Ho}, {Hutchison}, {Inoue}, {Jiang}, {Jones},
  {Koekemoer}, {Li}, {Li}, {McGrath}, {Molina}, {Papovich},
  {P{\'e}rez-Gonz{\'a}lez}, {Pirzkal}, {Wilkins}, {Yang}, \&
  {Yung}}]{Kocevski2023}
{Kocevski}, D.~D., {Onoue}, M., {Inayoshi}, K., {et~al.} 2023, \apjl, 954, L4,
  \dodoi{10.3847/2041-8213/ace5a0}

\bibitem[{{Koekemoer} {et~al.}(2011){Koekemoer}, {Faber}, {Ferguson}, {Grogin},
  {Kocevski}, {Koo}, {Lai}, {Lotz}, {Lucas}, {McGrath}, {Ogaz}, {Rajan},
  {Riess}, {Rodney}, {Strolger}, {Casertano}, {Castellano}, {Dahlen},
  {Dickinson}, {Dolch}, {Fontana}, {Giavalisco}, {Grazian}, {Guo}, {Hathi},
  {Huang}, {van der Wel}, {Yan}, {Acquaviva}, {Alexander}, {Almaini}, {Ashby},
  {Barden}, {Bell}, {Bournaud}, {Brown}, {Caputi}, {Cassata}, {Challis},
  {Chary}, {Cheung}, {Cirasuolo}, {Conselice}, {Roshan Cooray}, {Croton},
  {Daddi}, {Dav{\'e}}, {de Mello}, {de Ravel}, {Dekel}, {Donley}, {Dunlop},
  {Dutton}, {Elbaz}, {Fazio}, {Filippenko}, {Finkelstein}, {Frazer}, {Gardner},
  {Garnavich}, {Gawiser}, {Gruetzbauch}, {Hartley}, {H{\"a}ussler},
  {Herrington}, {Hopkins}, {Huang}, {Jha}, {Johnson}, {Kartaltepe},
  {Khostovan}, {Kirshner}, {Lani}, {Lee}, {Li}, {Madau}, {McCarthy},
  {McIntosh}, {McLure}, {McPartland}, {Mobasher}, {Moreira}, {Mortlock},
  {Moustakas}, {Mozena}, {Nandra}, {Newman}, {Nielsen}, {Niemi}, {Noeske},
  {Papovich}, {Pentericci}, {Pope}, {Primack}, {Ravindranath}, {Reddy},
  {Renzini}, {Rix}, {Robaina}, {Rosario}, {Rosati}, {Salimbeni}, {Scarlata},
  {Siana}, {Simard}, {Smidt}, {Snyder}, {Somerville}, {Spinrad}, {Straughn},
  {Telford}, {Teplitz}, {Trump}, {Vargas}, {Villforth}, {Wagner}, {Wandro},
  {Wechsler}, {Weiner}, {Wiklind}, {Wild}, {Wilson}, {Wuyts}, \&
  {Yun}}]{Koekemoer2011}
{Koekemoer}, A.~M., {Faber}, S.~M., {Ferguson}, H.~C., {et~al.} 2011, \apjs,
  197, 36, \dodoi{10.1088/0067-0049/197/2/36}

\bibitem[{{Kriek} {et~al.}(2015){Kriek}, {Shapley}, {Reddy}, {Siana}, {Coil},
  {Mobasher}, {Freeman}, {de Groot}, {Price}, {Sanders}, {Shivaei}, {Brammer},
  {Momcheva}, {Skelton}, {van Dokkum}, {Whitaker}, {Aird}, {Azadi}, {Kassis},
  {Bullock}, {Conroy}, {Dav{\'e}}, {Kere{\v{s}}}, \& {Krumholz}}]{Kriek2015}
{Kriek}, M., {Shapley}, A.~E., {Reddy}, N.~A., {et~al.} 2015, \apjs, 218, 15,
  \dodoi{10.1088/0067-0049/218/2/15}

\bibitem[{{Labb{\'e}} {et~al.}(2023){Labb{\'e}}, {van Dokkum}, {Nelson},
  {Bezanson}, {Suess}, {Leja}, {Brammer}, {Whitaker}, {Mathews}, {Stefanon}, \&
  {Wang}}]{Labbe2022}
{Labb{\'e}}, I., {van Dokkum}, P., {Nelson}, E., {et~al.} 2023, \nat, 616, 266,
  \dodoi{10.1038/s41586-023-05786-2}

\bibitem[{{Lagos} {et~al.}(2023){Lagos}, {Bravo}, {Tobar}, {Obreschkow},
  {Power}, {Robotham}, {Proctor}, {Hansen}, {Chandro-Gomez}, \&
  {Carrivick}}]{Lagos2023}
{Lagos}, C. D.~P., {Bravo}, M., {Tobar}, R., {et~al.} 2023, arXiv e-prints,
  arXiv:2309.02310, \dodoi{10.48550/arXiv.2309.02310}

\bibitem[{{Larson} {et~al.}(2023{\natexlab{a}}){Larson}, {Hutchison}, {Bagley},
  {Finkelstein}, {Yung}, {Somerville}, {Hirschmann}, {Brammer}, {Holwerda},
  {Papovich}, {Morales}, \& {Wilkins}}]{Larson2022}
{Larson}, R.~L., {Hutchison}, T.~A., {Bagley}, M., {et~al.} 2023{\natexlab{a}},
  \apj, 958, 141, \dodoi{10.3847/1538-4357/acfed4}

\bibitem[{{Larson} {et~al.}(2023{\natexlab{b}}){Larson}, {Finkelstein},
  {Kocevski}, {Hutchison}, {Trump}, {Arrabal Haro}, {Bromm}, {Cleri},
  {Dickinson}, {Fujimoto}, {Kartaltepe}, {Koekemoer}, {Papovich}, {Pirzkal},
  {Tacchella}, {Zavala}, {Bagley}, {Behroozi}, {Champagne}, {Cole}, {Jung},
  {Morales}, {Yang}, {Zhang}, {Zitrin}, {Amor{\'\i}n}, {Burgarella}, {Casey},
  {Ch{\'a}vez Ortiz}, {Cox}, {Chworowsky}, {Fontana}, {Gawiser}, {Grazian},
  {Grogin}, {Harish}, {Hathi}, {Hirschmann}, {Holwerda}, {Juneau}, {Leung},
  {Lucas}, {McGrath}, {P{\'e}rez-Gonz{\'a}lez}, {Rigby}, {Seill{\'e}},
  {Simons}, {de La Vega}, {Weiner}, {Wilkins}, {Yung}, \& {Ceers
  Team}}]{Larson2023}
{Larson}, R.~L., {Finkelstein}, S.~L., {Kocevski}, D.~D., {et~al.}
  2023{\natexlab{b}}, \apjl, 953, L29, \dodoi{10.3847/2041-8213/ace619}

\bibitem[{{Leung} {et~al.}(2023){Leung}, {Bagley}, {Finkelstein}, {Ferguson},
  {Koekemoer}, {P{\'e}rez-Gonz{\'a}lez}, {Morales}, {Kocevski}, {Yang},
  {Somerville}, {Wilkins}, {Yung}, {Fujimoto}, {Larson}, {Papovich}, {Pirzkal},
  {Berg}, {Lotz}, {Castellano}, {Ch{\'a}vez Ortiz}, {Cheng}, {Dickinson},
  {Giavalisco}, {Hathi}, {Hutchison}, {Jung}, {Kartaltepe}, {Natarajan}, \&
  {Rothberg}}]{Leung2023}
{Leung}, G. C.~K., {Bagley}, M.~B., {Finkelstein}, S.~L., {et~al.} 2023, \apjl,
  954, L46, \dodoi{10.3847/2041-8213/acf365}

\bibitem[{{Livermore} {et~al.}(2017){Livermore}, {Finkelstein}, \&
  {Lotz}}]{Livermore2017}
{Livermore}, R.~C., {Finkelstein}, S.~L., \& {Lotz}, J.~M. 2017, \apj, 835,
  113, \dodoi{10.3847/1538-4357/835/2/113}

\bibitem[{{Lovell} {et~al.}(2021){Lovell}, {Vijayan}, {Thomas}, {Wilkins},
  {Barnes}, {Irodotou}, \& {Roper}}]{Lovell2020}
{Lovell}, C.~C., {Vijayan}, A.~P., {Thomas}, P.~A., {et~al.} 2021, \mnras, 500,
  2127, \dodoi{10.1093/mnras/staa3360}

\bibitem[{{Lynden-Bell}(1971)}]{Lynden-Bell1971}
{Lynden-Bell}, D. 1971, \mnras, 155, 95, \dodoi{10.1093/mnras/155.1.95}

\bibitem[{{Madau} \& {Dickinson}(2014)}]{Madau2014}
{Madau}, P., \& {Dickinson}, M. 2014, \araa, 52, 415,
  \dodoi{10.1146/annurev-astro-081811-125615}

\bibitem[{{Madau} \& {Haardt}(2015)}]{Madau2015}
{Madau}, P., \& {Haardt}, F. 2015, \apjl, 813, L8,
  \dodoi{10.1088/2041-8205/813/1/L8}

\bibitem[{{Madau} {et~al.}(1999){Madau}, {Haardt}, \& {Rees}}]{Madau1999}
{Madau}, P., {Haardt}, F., \& {Rees}, M.~J. 1999, \apj, 514, 648,
  \dodoi{10.1086/306975}

\bibitem[{{Marley} {et~al.}(2021){Marley}, {Saumon}, {Visscher}, {Lupu},
  {Freedman}, {Morley}, {Fortney}, {Seay}, {Smith}, {Teal}, \&
  {Wang}}]{Marley2022}
{Marley}, M.~S., {Saumon}, D., {Visscher}, C., {et~al.} 2021, \apj, 920, 85,
  \dodoi{10.3847/1538-4357/ac141d}

\bibitem[{{Marshall} {et~al.}(2022){Marshall}, {Watts}, {Wilkins}, {Di Matteo},
  {Kuusisto}, {Roper}, {Vijayan}, {Ni}, {Feng}, \& {Croft}}]{Marshall2022}
{Marshall}, M.~A., {Watts}, K., {Wilkins}, S., {et~al.} 2022, \mnras, 516,
  1047, \dodoi{10.1093/mnras/stac2111}

\bibitem[{{Matsuoka} {et~al.}(2019){Matsuoka}, {Iwasawa}, {Onoue}, {Kashikawa},
  {Strauss}, {Lee}, {Imanishi}, {Nagao}, {Akiyama}, {Asami}, {Bosch},
  {Furusawa}, {Goto}, {Gunn}, {Harikane}, {Ikeda}, {Izumi}, {Kawaguchi},
  {Kato}, {Kikuta}, {Kohno}, {Komiyama}, {Koyama}, {Lupton}, {Minezaki},
  {Miyazaki}, {Murayama}, {Niida}, {Nishizawa}, {Noboriguchi}, {Oguri}, {Ono},
  {Ouchi}, {Price}, {Sameshima}, {Schulze}, {Silverman}, {Sugiyama}, {Tait},
  {Takada}, {Takata}, {Tanaka}, {Tang}, {Toba}, {Utsumi}, {Wang}, \&
  {Yamashita}}]{Matsuoka2018c}
{Matsuoka}, Y., {Iwasawa}, K., {Onoue}, M., {et~al.} 2019, \apj, 883, 183,
  \dodoi{10.3847/1538-4357/ab3c60}

\bibitem[{{McLeod} {et~al.}(2016){McLeod}, {McLure}, \& {Dunlop}}]{McLeod2016}
{McLeod}, D.~J., {McLure}, R.~J., \& {Dunlop}, J.~S. 2016, \mnras, 459, 3812,
  \dodoi{10.1093/mnras/stw904}

\bibitem[{{McLeod} {et~al.}(2023){McLeod}, {Donnan}, {McLure}, {Dunlop},
  {Magee}, {Begley}, {Carnall}, {Cullen}, {Ellis}, {Hamadouche}, \&
  {Stanton}}]{McLeod2023}
{McLeod}, D.~J., {Donnan}, C.~T., {McLure}, R.~J., {et~al.} 2023, \mnras,
  \dodoi{10.1093/mnras/stad3471}

\bibitem[{McLure {et~al.}(2013)McLure, Dunlop, Bowler, Curtis-Lake, Schenker,
  Ellis, Robertson, Koekemoer, Rogers, Ono, Ouchi, Charlot, Wild, Stark,
  Furlanetto, Cirasuolo, \& Targett}]{McLure2013}
McLure, R.~J., Dunlop, J.~S., Bowler, R.~A., {et~al.} 2013, \mnras, 432, 2696,
  \dodoi{10.1093/mnras/stt627}

\bibitem[{Momcheva {et~al.}(2016)Momcheva, Brammer, van Dokkum, Skelton,
  Whitaker, Nelson, Fumagalli, Maseda, Leja, Franx, Rix, Bezanson, Cunha,
  Dickey, Schreiber, Illingworth, Kriek, Labbé, Lange, Lundgren, Magee,
  Marchesini, Oesch, Pacifici, Patel, Price, Tal, Wake, van~der Wel, \&
  Wuyts}]{Momcheva2016}
Momcheva, I.~G., Brammer, G.~B., van Dokkum, P.~G., {et~al.} 2016, \apjs, 225,
  27, \dodoi{10.3847/0067-0049/225/2/27}

\bibitem[{{Morishita} {et~al.}(2018){Morishita}, {Trenti}, {Stiavelli},
  {Bradley}, {Coe}, {Oesch}, {Mason}, {Bridge}, {Holwerda}, {Livermore},
  {Salmon}, {Schmidt}, {Shull}, \& {Treu}}]{Morishita2018}
{Morishita}, T., {Trenti}, M., {Stiavelli}, M., {et~al.} 2018, \apj, 867, 150,
  \dodoi{10.3847/1538-4357/aae68c}

\bibitem[{{Moster} {et~al.}(2011){Moster}, {Somerville}, {Newman}, \&
  {Rix}}]{Moster2011}
{Moster}, B.~P., {Somerville}, R.~S., {Newman}, J.~A., \& {Rix}, H.-W. 2011,
  \apj, 731, 113, \dodoi{10.1088/0004-637X/731/2/113}

\bibitem[{{Naidu} {et~al.}(2020){Naidu}, {Tacchella}, {Mason}, {Bose}, {Oesch},
  \& {Conroy}}]{Naidu2020}
{Naidu}, R.~P., {Tacchella}, S., {Mason}, C.~A., {et~al.} 2020, \apj, 892, 109,
  \dodoi{10.3847/1538-4357/ab7cc9}

\bibitem[{{Naidu} {et~al.}(2022){Naidu}, {Oesch}, {van Dokkum}, {Nelson},
  {Suess}, {Brammer}, {Whitaker}, {Illingworth}, {Bouwens}, {Tacchella},
  {Matthee}, {Allen}, {Bezanson}, {Conroy}, {Labbe}, {Leja}, {Leonova},
  {Magee}, {Price}, {Setton}, {Strait}, {Stefanon}, {Toft}, {Weaver}, \&
  {Weibel}}]{Naidu2022}
{Naidu}, R.~P., {Oesch}, P.~A., {van Dokkum}, P., {et~al.} 2022, \apjl, 940,
  L14, \dodoi{10.3847/2041-8213/ac9b22}

\bibitem[{{Newman} {et~al.}(2013){Newman}, {Cooper}, {Davis}, {Faber}, {Coil},
  {Guhathakurta}, {Koo}, {Phillips}, {Conroy}, {Dutton}, {Finkbeiner}, {Gerke},
  {Rosario}, {Weiner}, {Willmer}, {Yan}, {Harker}, {Kassin}, {Konidaris},
  {Lai}, {Madgwick}, {Noeske}, {Wirth}, {Connolly}, {Kaiser}, {Kirby},
  {Lemaux}, {Lin}, {Lotz}, {Luppino}, {Marinoni}, {Matthews}, {Metevier}, \&
  {Schiavon}}]{Newman2013}
{Newman}, J.~A., {Cooper}, M.~C., {Davis}, M., {et~al.} 2013, \apjs, 208, 5,
  \dodoi{10.1088/0067-0049/208/1/5}

\bibitem[{{Niida} {et~al.}(2020){Niida}, {Nagao}, {Ikeda}, {Akiyama},
  {Matsuoka}, {He}, {Matsuoka}, {Toba}, {Onoue}, {Kobayashi}, {Taniguchi},
  {Furusawa}, {Harikane}, {Imanishi}, {Kashikawa}, {Kawaguchi}, {Komiyama},
  {Shirakata}, {Terashima}, \& {Ueda}}]{Niida2020}
{Niida}, M., {Nagao}, T., {Ikeda}, H., {et~al.} 2020, \apj, 904, 89,
  \dodoi{10.3847/1538-4357/abbe11}

\bibitem[{{Oesch} {et~al.}(2018){Oesch}, {Bouwens}, {Illingworth}, {Labb{\'e}},
  \& {Stefanon}}]{Oesch2018}
{Oesch}, P.~A., {Bouwens}, R.~J., {Illingworth}, G.~D., {Labb{\'e}}, I., \&
  {Stefanon}, M. 2018, \apj, 855, 105, \dodoi{10.3847/1538-4357/aab03f}

\bibitem[{{Oesch} {et~al.}(2015){Oesch}, {van Dokkum}, {Illingworth},
  {Bouwens}, {Momcheva}, {Holden}, {Roberts-Borsani}, {Smit}, {Franx},
  {Labb{\'e}}, {Gonz{\'a}lez}, \& {Magee}}]{Oesch2015}
{Oesch}, P.~A., {van Dokkum}, P.~G., {Illingworth}, G.~D., {et~al.} 2015,
  \apjl, 804, L30, \dodoi{10.1088/2041-8205/804/2/L30}

\bibitem[{{Oesch} {et~al.}(2016){Oesch}, {Brammer}, {van Dokkum},
  {Illingworth}, {Bouwens}, {Labb{\'e}}, {Franx}, {Momcheva}, {Ashby}, {Fazio},
  {Gonzalez}, {Holden}, {Magee}, {Skelton}, {Smit}, {Spitler}, {Trenti}, \&
  {Willner}}]{Oesch2016}
{Oesch}, P.~A., {Brammer}, G., {van Dokkum}, P.~G., {et~al.} 2016, \apj, 819,
  129, \dodoi{10.3847/0004-637X/819/2/129}

\bibitem[{{Oke}(1974)}]{Oke1974}
{Oke}, J.~B. 1974, \apjs, 27, 21, \dodoi{10.1086/190287}

\bibitem[{{Oke} \& {Gunn}(1983)}]{Oke1983}
{Oke}, J.~B., \& {Gunn}, J.~E. 1983, \apj, 266, 713, \dodoi{10.1086/160817}

\bibitem[{{Paardekooper} {et~al.}(2013){Paardekooper}, {Khochfar}, \&
  {Dalla}}]{Paardekooper2013}
{Paardekooper}, J.~P., {Khochfar}, S., \& {Dalla}, C.~V. 2013, \mnras, 429,
  L94, \dodoi{10.1093/mnrasl/sls032}

\bibitem[{{Parsa} {et~al.}(2018){Parsa}, {Dunlop}, \& {McLure}}]{Parsa2018}
{Parsa}, S., {Dunlop}, J.~S., \& {McLure}, R.~J. 2018, \mnras, 474, 2904,
  \dodoi{10.1093/mnras/stx2887}

\bibitem[{{P{\'e}rez-Gonz{\'a}lez} {et~al.}(2023){P{\'e}rez-Gonz{\'a}lez},
  {Costantin}, {Langeroodi}, {Rinaldi}, {Annunziatella}, {Ilbert}, {Colina},
  {N{\o}rgaard-Nielsen}, {Greve}, {{\"O}stlin}, {Wright}, {Alonso-Herrero},
  {{\'A}lvarez-M{\'a}rquez}, {Caputi}, {Eckart}, {Le F{\`e}vre}, {Labiano},
  {Garc{\'\i}a-Mar{\'\i}n}, {Hjorth}, {Kendrew}, {Pye}, {Tikkanen}, {van der
  Werf}, {Walter}, {Ward}, {Bik}, {Boogaard}, {Bosman}, {G{\'o}mez}, {Gillman},
  {Iani}, {Jermann}, {Melinder}, {Meyer}, {Moutard}, {van Dishoek}, {Henning},
  {Lagage}, {Guedel}, {Peissker}, {Ray}, {Vandenbussche},
  {Garc{\'\i}a-Argum{\'a}nez}, \& {Mar{\'\i}a M{\'e}rida}}]{Gonzalez2023}
{P{\'e}rez-Gonz{\'a}lez}, P.~G., {Costantin}, L., {Langeroodi}, D., {et~al.}
  2023, \apjl, 951, L1, \dodoi{10.3847/2041-8213/acd9d0}

\bibitem[{{Perrin} {et~al.}(2014){Perrin}, {Sivaramakrishnan}, {Lajoie},
  {Elliott}, {Pueyo}, {Ravindranath}, \& {Albert}}]{Perrin2014}
{Perrin}, M.~D., {Sivaramakrishnan}, A., {Lajoie}, C.-P., {et~al.} 2014, in
  Society of Photo-Optical Instrumentation Engineers (SPIE) Conference Series,
  Vol. 9143, Space Telescopes and Instrumentation 2014: Optical, Infrared, and
  Millimeter Wave, ed. J.~{Oschmann}, Jacobus~M., M.~{Clampin}, G.~G. {Fazio},
  \& H.~A. {MacEwen}, 91433X, \dodoi{10.1117/12.2056689}

\bibitem[{{Perrin} {et~al.}(2012){Perrin}, {Soummer}, {Elliott}, {Lallo}, \&
  {Sivaramakrishnan}}]{Perrin2012}
{Perrin}, M.~D., {Soummer}, R., {Elliott}, E.~M., {Lallo}, M.~D., \&
  {Sivaramakrishnan}, A. 2012, in Society of Photo-Optical Instrumentation
  Engineers (SPIE) Conference Series, Vol. 8442, Space Telescopes and
  Instrumentation 2012: Optical, Infrared, and Millimeter Wave, ed. M.~C.
  {Clampin}, G.~G. {Fazio}, H.~A. {MacEwen}, \& J.~{Oschmann}, Jacobus~M.,
  84423D, \dodoi{10.1117/12.925230}

\bibitem[{{Pontoppidan} {et~al.}(2022){Pontoppidan}, {Barrientes}, {Blome},
  {Braun}, {Brown}, {Carruthers}, {Coe}, {DePasquale}, {Espinoza}, {Marin},
  {Gordon}, {Henry}, {Hustak}, {James}, {Jenkins}, {Koekemoer}, {LaMassa},
  {Law}, {Lockwood}, {Moro-Martin}, {Mullally}, {Pagan}, {Player}, {Proffitt},
  {Pulliam}, {Ramsay}, {Ravindranath}, {Reid}, {Robberto}, {Sabbi}, {Ubeda},
  {Balogh}, {Flanagan}, {Gardner}, {Hasan}, {Meinke}, \&
  {Nota}}]{Pontoppidan2022}
{Pontoppidan}, K.~M., {Barrientes}, J., {Blome}, C., {et~al.} 2022, \apjl, 936,
  L14, \dodoi{10.3847/2041-8213/ac8a4e}

\bibitem[{{Rieke} {et~al.}(2015){Rieke}, {Wright}, {B{\"o}ker}, {Bouwman},
  {Colina}, {Glasse}, {Gordon}, {Greene}, {G{\"u}del}, {Henning}, {Justtanont},
  {Lagage}, {Meixner}, {N{\o}rgaard-Nielsen}, {Ray}, {Ressler}, {van Dishoeck},
  \& {Waelkens}}]{rieke15}
{Rieke}, G.~H., {Wright}, G.~S., {B{\"o}ker}, T., {et~al.} 2015, \pasp, 127,
  584, \dodoi{10.1086/682252}

\bibitem[{{Rieke} {et~al.}(2005){Rieke}, {Kelly}, {Horner}, \& {NIRCam
  Team}}]{rieke05}
{Rieke}, M., {Kelly}, D., {Horner}, S., \& {NIRCam Team}. 2005, in American
  Astronomical Society Meeting Abstracts, Vol. 207, American Astronomical
  Society Meeting Abstracts, 115.09

\bibitem[{{Rieke} {et~al.}(2008){Rieke}, {Eisenstein}, {Engelbracht}, {Kelly},
  {McCarthy}, {Meyer}, {Misselt}, {Rieke}, {Stansberry}, {Willmer}, {Young},
  {Baum}, {Beichman}, {Trauger}, {Doyon}, {Dressler}, {Ferrarese}, {Johnstone},
  {Greene}, {Roellig}, {Hall}, {Hodapp}, {Horner}, {Lilly}, {Martin}, \&
  {Stauffer}}]{rieke08}
{Rieke}, M.~J., {Eisenstein}, D., {Engelbracht}, C.~W., {et~al.} 2008, in
  American Astronomical Society Meeting Abstracts, Vol. 212, American
  Astronomical Society Meeting Abstracts \#212, 79.01

\bibitem[{{Rieke} {et~al.}(2023{\natexlab{a}}){Rieke}, {Kelly}, {Misselt},
  {Stansberry}, {Boyer}, {Beatty}, {Egami}, {Florian}, {Greene}, {Hainline},
  {Leisenring}, {Roellig}, {Schlawin}, {Sun}, {Tinnin}, {Williams}, {Willmer},
  {Wilson}, {Clark}, {Rohrbach}, {Brooks}, {Canipe}, {Correnti}, {DiFelice},
  {Gennaro}, {Girard}, {Hartig}, {Hilbert}, {Koekemoer}, {Nikolov}, {Pirzkal},
  {Rest}, {Robberto}, {Sunnquist}, {Telfer}, {Wu}, {Ferry}, {Lewis}, {Baum},
  {Beichman}, {Doyon}, {Dressler}, {Eisenstein}, {Ferrarese}, {Hodapp},
  {Horner}, {Jaffe}, {Johnstone}, {Krist}, {Martin}, {McCarthy}, {Meyer},
  {Rieke}, {Trauger}, \& {Young}}]{Rieke2022}
{Rieke}, M.~J., {Kelly}, D.~M., {Misselt}, K., {et~al.} 2023{\natexlab{a}},
  \pasp, 135, 028001, \dodoi{10.1088/1538-3873/acac53}

\bibitem[{{Rieke} {et~al.}(2023{\natexlab{b}}){Rieke}, {Robertson},
  {Tacchella}, {Hainline}, {Johnson}, {Hausen}, {Ji}, {Willmer}, {Eisenstein},
  {Pusk{\'a}s}, {Alberts}, {Arribas}, {Baker}, {Baum}, {Bhatawdekar},
  {Bonaventura}, {Boyett}, {Bunker}, {Cameron}, {Carniani}, {Charlot},
  {Chevallard}, {Chen}, {Curti}, {Curtis-Lake}, {Danhaive}, {DeCoursey},
  {Dressler}, {Egami}, {Endsley}, {Helton}, {Hviding}, {Kumari}, {Looser},
  {Lyu}, {Maiolino}, {Maseda}, {Nelson}, {Rieke}, {Rix}, {Sandles}, {Saxena},
  {Sharpe}, {Shivaei}, {Skarbinski}, {Smit}, {Stark}, {Stone}, {Suess}, {Sun},
  {Topping}, {{\"U}bler}, {Villanueva}, {Wallace}, {Williams}, {Willott},
  {Whitler}, {Witstok}, \& {Woodrum}}]{Reike2023}
{Rieke}, M.~J., {Robertson}, B., {Tacchella}, S., {et~al.} 2023{\natexlab{b}},
  \apjs, 269, 16, \dodoi{10.3847/1538-4365/acf44d}

\bibitem[{{Rigby} {et~al.}(2023){Rigby}, {Perrin}, {McElwain}, {Kimble},
  {Friedman}, {Lallo}, {Doyon}, {Feinberg}, {Ferruit}, {Glasse}, {Rieke},
  {Rieke}, {Wright}, {Willott}, {Colon}, {Milam}, {Neff}, {Stark}, {Valenti},
  {Abell}, {Abney}, {Abul-Huda}, {Acton}, {Adams}, {Adler}, {Aguilar}, {Ahmed},
  {Albert}, {Alberts}, {Aldridge}, {Allen}, {Altenburg},
  {{\'A}lvarez-M{\'a}rquez}, {Alves de Oliveira}, {Andersen}, {Anderson},
  {Anderson}, {Argyriou}, {Armstrong}, {Arribas}, {Artigau}, {Arvai},
  {Atkinson}, {Bacon}, {Bair}, {Banks}, {Barrientes}, {Barringer}, {Bartosik},
  {Bast}, {Baudoz}, {Beatty}, {Bechtold}, {Beck}, {Bergeron}, {Bergkoetter},
  {Bhatawdekar}, {Birkmann}, {Blazek}, {Blome}, {Boccaletti}, {B{\"o}ker},
  {Boia}, {Bonaventura}, {Bond}, {Bosley}, {Boucarut}, {Bourque}, {Bouwman},
  {Bower}, {Bowers}, {Boyer}, {Bradley}, {Brady}, {Braun}, {Breda},
  {Bresnahan}, {Bright}, {Britt}, {Bromenschenkel}, {Brooks}, {Brooks},
  {Brown}, {Brown}, {Brown}, {Bunker}, {Burger}, {Bushouse}, {Cale}, {Cameron},
  {Cameron}, {Canipe}, {Caplinger}, {Caputo}, {Cara}, {Carey}, {Carniani},
  {Carrasquilla}, {Carruthers}, {Case}, {Catherine}, {Chance}, {Chapman},
  {Charlot}, {Charlow}, {Chayer}, {Chen}, {Cherinka}, {Chichester}, {Chilton},
  {Chonis}, {Clampin}, {Clark}, {Clark}, {Coe}, {Coleman}, {Comber}, {Comeau},
  {Connolly}, {Cooper}, {Cooper}, {Coppock}, {Correnti}, {Cossou}, {Coulais},
  {Coyle}, {Cracraft}, {Curti}, {Cuturic}, {Davis}, {Davis}, {Dean}, {DeLisa},
  {deMeester}, {Dencheva}, {Dencheva}, {DePasquale}, {Deschenes}, {Hunor
  Detre}, {Diaz}, {Dicken}, {DiFelice}, {Dillman}, {Dixon}, {Doggett},
  {Donaldson}, {Douglas}, {DuPrie}, {Dupuis}, {Durning}, {Easmin}, {Eck},
  {Edeani}, {Egami}, {Ehrenwinkler}, {Eisenhamer}, {Eisenhower}, {Elie},
  {Elliott}, {Elliott}, {Ellis}, {Engesser}, {Espinoza}, {Etienne}, {Etxaluze},
  {Falini}, {Feeney}, {Ferry}, {Filippazzo}, {Fincham}, {Fix}, {Flagey},
  {Florian}, {Flynn}, {Fontanella}, {Ford}, {Forshay}, {Fox}, {Franz}, {Fu},
  {Fullerton}, {Galkin}, {Galyer}, {Garc{\'\i}a Mar{\'\i}n}, {Gardner},
  {Gardner}, {Garland}, {Garrett}, {Gasman}, {Gaspar}, {Gaudreau}, {Gauthier},
  {Geers}, {Geithner}, {Gennaro}, {Giardino}, {Girard}, {Giuliano},
  {Glassmire}, {Glauser}, {Glazer}, {Godfrey}, {Golimowski}, {Gollnitz},
  {Gong}, {Gonzaga}, {Gordon}, {Gordon}, {Goudfrooij}, {Greene}, {Greenhouse},
  {Grimaldi}, {Groebner}, {Grundy}, {Guillard}, {Gutman}, {Ha}, {Haderlein},
  {Hagedorn}, {Hainline}, {Haley}, {Hami}, {Hamilton}, {Hammel}, {Hansen},
  {Harkins}, {Harr}, {Hart}, {Hart}, {Hartig}, {Hashimoto}, {Haskins},
  {Hathaway}, {Havey}, {Hayden}, {Hecht}, {Heller-Boyer}, {Henriques}, {Henry},
  {Hermann}, {Hernandez}, {Hesman}, {Hicks}, {Hilbert}, {Hines}, {Hoffman},
  {Holfeltz}, {Holler}, {Hoppa}, {Hott}, {Howard}, {Howard}, {Hunter},
  {Hunter}, {Hurst}, {Husemann}, {Hustak}, {Ilinca Ignat}, {Illingworth},
  {Irish}, {Jackson}, {Jahromi}, {Jakobsen}, {James}, {James}, {Januszewski},
  {Jenkins}, {Jirdeh}, {Johnson}, {Johnson}, {Jones}, {Jones}, {Jones},
  {Jones}, {Jordan}, {Jordan}, {Jurczyk}, {Jurling}, {Kaleida}, {Kalmanson},
  {Kammerer}, {Kang}, {Kao}, {Karakla}, {Kavanagh}, {Kelly}, {Kendrew},
  {Kennedy}, {Kenny}, {Keski-kuha}, {Keyes}, {Kidwell}, {Kinzel}, {Kirk},
  {Kirkpatrick}, {Kirshenblat}, {Klaassen}, {Knapp}, {Knight}, {Knollenberg},
  {Koehler}, {Koekemoer}, {Kovacs}, {Kulp}, {Kumari}, {Kyprianou}, {La Massa},
  {Labador}, {Labiano}, {Lagage}, {Lajoie}, {Lallo}, {Lam}, {Lamb}, {Lambros},
  {Lampenfield}, {Langston}, {Larson}, {Law}, {Lawrence}, {Lee}, {Leisenring},
  {Lepo}, {Leveille}, {Levenson}, {Levine}, {Levy}, {Lewis}, {Lewis},
  {Libralato}, {Lightsey}, {Link}, {Liu}, {Lo}, {Lockwood}, {Logue}, {Long},
  {Long}, {Loomis}, {Lopez-Caniego}, {Lorenzo Alvarez}, {Love-Pruitt}, {Lucy},
  {Luetzgendorf}, {Maghami}, {Maiolino}, {Major}, {Malla}, {Malumuth},
  {Manjavacas}, {Mannfolk}, {Marrione}, {Marston}, {Martel}, {Maschmann},
  {Masci}, {Masciarelli}, {Maszkiewicz}, {Mather}, {McKenzie}, {McLean},
  {McMaster}, {Melbourne}, {Mel{\'e}ndez}, {Menzel}, {Merz}, {Meyett}, {Meza},
  {Miskey}, {Misselt}, {Moller}, {Morrison}, {Morse}, {Moseley}, {Mosier},
  {Mountain}, {Mueckay}, {Mueller}, {Mullally}, {Murphy}, {Murray}, {Murray},
  {Mustelier}, {Muzerolle}, {Mycroft}, {Myers}, {Myrick}, {Nanavati}, {Nance},
  {Nayak}, {Naylor}, {Nelan}, {Nickson}, {Nielson}, {Nieto-Santisteban},
  {Nikolov}, {Noriega-Crespo}, {O'Shaughnessy}, {O'Sullivan}, {Ochs}, {Ogle},
  {Oleszczuk}, {Olmsted}, {Osborne}, {Ottens}, {Owens}, {Pacifici}, {Pagan},
  {Page}, {Park}, {Parrish}, {Patapis}, {Paul}, {Pauly}, {Pavlovsky}, {Pedder},
  {Peek}, {Pena-Guerrero}, {Penanen}, {Perez}, {Perna}, {Perriello},
  {Phillips}, {Pietraszkiewicz}, {Pinaud}, {Pirzkal}, {Pitman}, {Piwowar},
  {Platais}, {Player}, {Plesha}, {Pollizi}, {Polster}, {Pontoppidan},
  {Porterfield}, {Proffitt}, {Pueyo}, {Pulliam}, {Quirt}, {Quispe Neira},
  {Ramos Alarcon}, {Ramsay}, {Rapp}, {Rapp}, {Rauscher}, {Ravindranath},
  {Rawle}, {Regan}, {Reichard}, {Reis}, {Ressler}, {Rest}, {Reynolds}, {Rhue},
  {Richon}, {Rickman}, {Ridgaway}, {Ritchie}, {Rix}, {Robberto}, {Robinson},
  {Robinson}, {Robinson}, {Rock}, {Rodriguez}, {Rodriguez Del Pino}, {Roellig},
  {Rohrbach}, {Roman}, {Romelfanger}, {Rose}, {Roteliuk}, {Roth}, {Rothwell},
  {Rowlands}, {Roy}, {Royer}, {Royle}, {Rui}, {Rumler}, {Runnels}, {Russ},
  {Rustamkulov}, {Ryden}, {Ryer}, {Sabata}, {Sabatke}, {Sabbi}, {Samuelson},
  {Sapp}, {Sappington}, {Sargent}, {Sauer}, {Scheithauer}, {Schlawin},
  {Schlitz}, {Schmitz}, {Schneider}, {Schreiber}, {Schulze}, {Schwab}, {Scott},
  {Sembach}, {Shanahan}, {Shaughnessy}, {Shaw}, {Shawger}, {Shay}, {Sheehan},
  {Shen}, {Sherman}, {Shiao}, {Shih}, {Shivaei}, {Sienkiewicz}, {Sing},
  {Sirianni}, {Sivaramakrishnan}, {Skipper}, {Sloan}, {Slocum}, {Slowinski},
  {Smith}, {Smith}, {Smith}, {Smith}, {Snyder}, {Soh}, {Sohn}, {Soto},
  {Spencer}, {Stallcup}, {Stansberry}, {Starr}, {Starr}, {Stewart},
  {Stiavelli}, {Straughn}, {Strickland}, {Stys}, {Summers}, {Sun}, {Sunnquist},
  {Swade}, {Swam}, {Swaters}, {Swoish}, {Taylor}, {Taylor}, {Te Plate}, {Tea},
  {Teague}, {Telfer}, {Temim}, {Thatte}, {Thompson}, {Thompson}, {Thomson},
  {Tikkanen}, {Tippet}, {Todd}, {Toolan}, {Tran}, {Trejo}, {Truong},
  {Tsukamoto}, {Tustain}, {Tyra}, {Ubeda}, {Underwood}, {Uzzo}, {Van Campen},
  {Vandal}, {Vandenbussche}, {Vila}, {Volk}, {Wahlgren}, {Waldman}, {Walker},
  {Wander}, {Warfield}, {Warner}, {Wasiak}, {Watkins}, {Weaver}, {Weilert},
  {Weiser}, {Weiss}, {Weissman}, {Welty}, {West}, {Wheate}, {Wheatley},
  {Wheeler}, {White}, {Whiteaker}, {Whitehouse}, {Whiteleather}, {Whitman},
  {Williams}, {Willmer}, {Willoughby}, {Wilson}, {Wirth}, {Wislowski}, {Wolf},
  {Wolfe}, {Wolff}, {Workman}, {Wright}, {Wu}, {Wu}, {Wymer}, {Yates},
  {Yeager}, {Yeates}, {Yerger}, {Yoon}, {Young}, {Yu}, {Zak}, {Zeidler},
  {Zhou}, {Zielinski}, {Zincke}, \& {Zonak}}]{Rigby2022}
{Rigby}, J., {Perrin}, M., {McElwain}, M., {et~al.} 2023, \pasp, 135, 048001,
  \dodoi{10.1088/1538-3873/acb293}

\bibitem[{{Roberts-Borsani} {et~al.}(2016){Roberts-Borsani}, {Bouwens},
  {Oesch}, {Labbe}, {Smit}, {Illingworth}, {van Dokkum}, {Holden}, {Gonzalez},
  {Stefanon}, {Holwerda}, \& {Wilkins}}]{RobertsBorsani2015}
{Roberts-Borsani}, G.~W., {Bouwens}, R.~J., {Oesch}, P.~A., {et~al.} 2016,
  \apj, 823, 143, \dodoi{10.3847/0004-637X/823/2/143}

\bibitem[{{Robertson} {et~al.}(2015){Robertson}, {Ellis}, {Furlanetto}, \&
  {Dunlop}}]{Robertson2015}
{Robertson}, B.~E., {Ellis}, R.~S., {Furlanetto}, S.~R., \& {Dunlop}, J.~S.
  2015, \apjl, 802, L19, \dodoi{10.1088/2041-8205/802/2/L19}

\bibitem[{{Robertson} {et~al.}(2023){Robertson}, {Tacchella}, {Johnson},
  {Hainline}, {Whitler}, {Eisenstein}, {Endsley}, {Rieke}, {Stark}, {Alberts},
  {Dressler}, {Egami}, {Hausen}, {Rieke}, {Shivaei}, {Williams}, {Willmer},
  {Arribas}, {Bonaventura}, {Bunker}, {Cameron}, {Carniani}, {Charlot},
  {Chevallard}, {Curti}, {Curtis-Lake}, {D'Eugenio}, {Jakobsen}, {Looser},
  {L{\"u}tzgendorf}, {Maiolino}, {Maseda}, {Rawle}, {Rix}, {Smit}, {{\"U}bler},
  {Willott}, {Witstok}, {Baum}, {Bhatawdekar}, {Boyett}, {Chen}, {de Graaff},
  {Florian}, {Helton}, {Hviding}, {Ji}, {Kumari}, {Lyu}, {Nelson}, {Sandles},
  {Saxena}, {Suess}, {Sun}, {Topping}, \& {Wallace}}]{Robertson2022}
{Robertson}, B.~E., {Tacchella}, S., {Johnson}, B.~D., {et~al.} 2023, Nature
  Astronomy, 7, 611, \dodoi{10.1038/s41550-023-01921-1}

\bibitem[{{Robotham} {et~al.}(2018){Robotham}, {Davies}, {Driver}, {Koushan},
  {Taranu}, {Casura}, \& {Liske}}]{Robotham2018}
{Robotham}, A.~S.~G., {Davies}, L.~J.~M., {Driver}, S.~P., {et~al.} 2018,
  \mnras, 476, 3137, \dodoi{10.1093/mnras/sty440}

\bibitem[{{Robotham} {et~al.}(2023){Robotham}, {D'Silva}, {Windhorst},
  {Jansen}, {Summers}, {Driver}, {Wilmer}, \& {Bellstedt}}]{Robotham2023}
{Robotham}, A.~S.~G., {D'Silva}, J.~C.~J., {Windhorst}, R.~A., {et~al.} 2023,
  \pasp, 135, 085003, \dodoi{10.1088/1538-3873/acea42}

\bibitem[{{Rowan-Robinson}(1968)}]{rowanrobinson1968}
{Rowan-Robinson}, M. 1968, \mnras, 138, 445, \dodoi{10.1093/mnras/138.4.445}

\bibitem[{{Salmon} {et~al.}(2018){Salmon}, {Coe}, {Bradley}, {Brada{\v{c}}},
  {Strait}, {Paterno-Mahler}, {Huang}, {Oesch}, {Zitrin}, {Acebron}, {Cibirka},
  {Kikuchihara}, {Oguri}, {Brammer}, {Sharon}, {Trenti}, {Avila}, {Ogaz},
  {Andrade-Santos}, {Carrasco}, {Cerny}, {Dawson}, {Frye}, {Hoag}, {Jones},
  {Mainali}, {Ouchi}, {Rodney}, {Stark}, \& {Umetsu}}]{Salmon2018}
{Salmon}, B., {Coe}, D., {Bradley}, L., {et~al.} 2018, \apjl, 864, L22,
  \dodoi{10.3847/2041-8213/aadc10}

\bibitem[{{Schechter}(1976)}]{Schechter1976}
{Schechter}, P. 1976, \apj, 203, 297, \dodoi{10.1086/154079}

\bibitem[{{Schmidt}(1968)}]{Schmidt1968}
{Schmidt}, M. 1968, \apj, 151, 393, \dodoi{10.1086/149446}

\bibitem[{{Shull} {et~al.}(2012){Shull}, {Stevans}, \& {Danforth}}]{Shull2012}
{Shull}, J.~M., {Stevans}, M., \& {Danforth}, C.~W. 2012, \apj, 752, 162,
  \dodoi{10.1088/0004-637X/752/2/162}

\bibitem[{{Stefanon} {et~al.}(2019){Stefanon}, {Labb{\'e}}, {Bouwens}, {Oesch},
  {Ashby}, {Caputi}, {Franx}, {Fynbo}, {Illingworth}, {Le F{\`e}vre},
  {Marchesini}, {McCracken}, {Milvang-Jensen}, {Muzzin}, \& {van
  Dokkum}}]{Stefanon2019}
{Stefanon}, M., {Labb{\'e}}, I., {Bouwens}, R.~J., {et~al.} 2019, \apj, 883,
  99, \dodoi{10.3847/1538-4357/ab3792}

\bibitem[{{Steidel} {et~al.}(1996){Steidel}, {Giavalisco}, {Pettini},
  {Dickinson}, \& {Adelberger}}]{Steidel1996}
{Steidel}, C.~C., {Giavalisco}, M., {Pettini}, M., {Dickinson}, M., \&
  {Adelberger}, K.~L. 1996, \apjl, 462, L17, \dodoi{10.1086/310029}

\bibitem[{{Steidel} \& {Hamilton}(1992)}]{Steidel1992}
{Steidel}, C.~C., \& {Hamilton}, D. 1992, \aj, 104, 941, \dodoi{10.1086/116287}

\bibitem[{{Tamura} {et~al.}(2019){Tamura}, {Mawatari}, {Hashimoto}, {Inoue},
  {Zackrisson}, {Christensen}, {Binggeli}, {Matsuda}, {Matsuo}, {Takeuchi},
  {Asano}, {Sunaga}, {Shimizu}, {Okamoto}, {Yoshida}, {Lee}, {Shibuya},
  {Taniguchi}, {Umehata}, {Hatsukade}, {Kohno}, \& {Ota}}]{Tamura2019}
{Tamura}, Y., {Mawatari}, K., {Hashimoto}, T., {et~al.} 2019, \apj, 874, 27,
  \dodoi{10.3847/1538-4357/ab0374}

\bibitem[{{Tang} {et~al.}(2023){Tang}, {Stark}, {Chen}, {Mason}, {Topping},
  {Endsley}, {Senchyna}, {Plat}, {Lu}, {Whitler}, {Robertson}, \&
  {Charlot}}]{Tang2023}
{Tang}, M., {Stark}, D.~P., {Chen}, Z., {et~al.} 2023, \mnras, 526, 1657,
  \dodoi{10.1093/mnras/stad2763}

\bibitem[{{Treu} {et~al.}(2022){Treu}, {Roberts-Borsani}, {Bradac}, {Brammer},
  {Fontana}, {Henry}, {Mason}, {Morishita}, {Pentericci}, {Wang}, {Acebron},
  {Bagley}, {Bergamini}, {Belfiori}, {Bonchi}, {Boyett}, {Boutsia},
  {Calabr{\'o}}, {Caminha}, {Castellano}, {Dressler}, {Glazebrook}, {Grillo},
  {Jacobs}, {Jones}, {Kelly}, {Leethochawalit}, {Malkan}, {Marchesini},
  {Mascia}, {Mercurio}, {Merlin}, {Nanayakkara}, {Nonino}, {Paris},
  {Poggianti}, {Rosati}, {Santini}, {Scarlata}, {Shipley}, {Strait}, {Trenti},
  {Tubthong}, {Vanzella}, {Vulcani}, \& {Yang}}]{Treu2022}
{Treu}, T., {Roberts-Borsani}, G., {Bradac}, M., {et~al.} 2022, \apj, 935, 110,
  \dodoi{10.3847/1538-4357/ac8158}

\bibitem[{{Trussler} {et~al.}(2023){Trussler}, {Adams}, {Conselice},
  {Ferreira}, {Austin}, {Bhatawdekar}, {Caruana}, {Frye}, {Harvey}, {Lovell},
  {Pascale}, {Roper}, {Verma}, {Vijayan}, \& {Wilkins}}]{Trussler2022}
{Trussler}, J. A.~A., {Adams}, N.~J., {Conselice}, C.~J., {et~al.} 2023,
  \mnras, 523, 3423, \dodoi{10.1093/mnras/stad1629}

\bibitem[{Ulm(1990)}]{Ulm1990}
Ulm, K. 1990, American Journal of Epidemiology, 131, 373,
  \dodoi{10.1093/oxfordjournals.aje.a115507}

\bibitem[{{Vijayan} {et~al.}(2021){Vijayan}, {Lovell}, {Wilkins}, {Thomas},
  {Barnes}, {Irodotou}, {Kuusisto}, \& {Roper}}]{Vijayan2021}
{Vijayan}, A.~P., {Lovell}, C.~C., {Wilkins}, S.~M., {et~al.} 2021, \mnras,
  501, 3289, \dodoi{10.1093/mnras/staa3715}

\bibitem[{Virtanen {et~al.}(2020)Virtanen, Gommers, Oliphant, Haberland, Reddy,
  Cournapeau, Burovski, Peterson, Weckesser, Bright, {van der Walt}, Brett,
  Wilson, Millman, Mayorov, Nelson, Jones, Kern, Larson, Carey, Polat, Feng,
  Moore, {VanderPlas}, Laxalde, Perktold, Cimrman, Henriksen, Quintero, Harris,
  Archibald, Ribeiro, Pedregosa, {van Mulbregt}, \& {SciPy 1.0
  Contributors}}]{2020SciPy-NMeth}
Virtanen, P., Gommers, R., Oliphant, T.~E., {et~al.} 2020, Nature Methods, 17,
  261, \dodoi{10.1038/s41592-019-0686-2}

\bibitem[{{Wang} {et~al.}(2023){Wang}, {Fujimoto}, {Labb{\'e}}, {Furtak},
  {Miller}, {Setton}, {Zitrin}, {Atek}, {Bezanson}, {Brammer}, {Leja}, {Oesch},
  {Price}, {Chemerynska}, {Cutler}, {Dayaghoffl}, {van Dokkum}, {Goulding},
  {Greene}, {Fudamoto}, {Khullar}, {Kokorev}, {Marchesini}, {Pan}, {Weaver},
  {Whitaker}, \& {Williams}}]{Wang2023}
{Wang}, B., {Fujimoto}, S., {Labb{\'e}}, I., {et~al.} 2023, \apjl, 957, L34,
  \dodoi{10.3847/2041-8213/acfe07}

\bibitem[{Wang {et~al.}(1986)Wang, Jewell, \& Tsai}]{Wang1986}
Wang, M.-C., Jewell, N.~P., \& Tsai, W.-Y. 1986, The Annals of Statistics, 14,
  1597.
\newblock \url{http://www.jstor.org/stable/2241492}

\bibitem[{{Whitaker} {et~al.}(2019){Whitaker}, {Ashas}, {Illingworth}, {Magee},
  {Leja}, {Oesch}, {van Dokkum}, {Mowla}, {Bouwens}, {Franx}, {Holden},
  {Labb{\'e}}, {Rafelski}, {Teplitz}, \& {Gonzalez}}]{Whitaker2019}
{Whitaker}, K.~E., {Ashas}, M., {Illingworth}, G., {et~al.} 2019, \apjs, 244,
  16, \dodoi{10.3847/1538-4365/ab3853}

\bibitem[{{Wilkins} {et~al.}(2017{\natexlab{a}}){Wilkins}, {Feng}, {Di Matteo},
  {Croft}, {Lovell}, \& {Waters}}]{Bluetides-II}
{Wilkins}, S.~M., {Feng}, Y., {Di Matteo}, T., {et~al.} 2017{\natexlab{a}},
  \mnras, 469, 2517, \dodoi{10.1093/mnras/stx841}

\bibitem[{{Wilkins} {et~al.}(2017{\natexlab{b}}){Wilkins}, {Feng}, {Di Matteo},
  {Croft}, {Lovell}, \& {Waters}}]{Wilkins2017}
---. 2017{\natexlab{b}}, \mnras, 469, 2517, \dodoi{10.1093/mnras/stx841}

\bibitem[{{Wilkins} {et~al.}(2023){Wilkins}, {Vijayan}, {Lovell}, {Roper},
  {Irodotou}, {Caruana}, {Seeyave}, {Kuusisto}, {Thomas}, \&
  {Parris}}]{Wilkins2023}
{Wilkins}, S.~M., {Vijayan}, A.~P., {Lovell}, C.~C., {et~al.} 2023, \mnras,
  519, 3118, \dodoi{10.1093/mnras/stac3280}

\bibitem[{{Williams} {et~al.}(2018){Williams}, {Curtis-Lake}, {Hainline},
  {Chevallard}, {Robertson}, {Charlot}, {Endsley}, {Stark}, {Willmer},
  {Alberts}, {Amorin}, {Arribas}, {Baum}, {Bunker}, {Carniani}, {Crandall},
  {Egami}, {Eisenstein}, {Ferruit}, {Husemann}, {Maseda}, {Maiolino}, {Rawle},
  {Rieke}, {Smit}, {Tacchella}, \& {Willott}}]{Williams18}
{Williams}, C.~C., {Curtis-Lake}, E., {Hainline}, K.~N., {et~al.} 2018, \apjs,
  236, 33, \dodoi{10.3847/1538-4365/aabcbb}

\bibitem[{{Willott} {et~al.}(2023){Willott}, {Desprez}, {Asada}, {Sarrouh},
  {Abraham}, {Brada{\v{c}}}, {Brammer}, {Estrada-Carpenter}, {Iyer}, {Martis},
  {Matharu}, {Mowla}, {Muzzin}, {Noirot}, {Sawicki}, {Strait},
  {Rihtar{\v{s}}i{\v{c}}}, \& {Withers}}]{Willott2023}
{Willott}, C.~J., {Desprez}, G., {Asada}, Y., {et~al.} 2023, arXiv e-prints,
  arXiv:2311.12234, \dodoi{10.48550/arXiv.2311.12234}

\bibitem[{{Windhorst} {et~al.}(2023){Windhorst}, {Cohen}, {Jansen}, {Summers},
  {Tompkins}, {Conselice}, {Driver}, {Yan}, {Coe}, {Frye}, {Grogin},
  {Koekemoer}, {Marshall}, {O'Brien}, {Pirzkal}, {Robotham}, {Ryan}, {Willmer},
  {Carleton}, {Diego}, {Keel}, {Porto}, {Redshaw}, {Scheller}, {Wilkins},
  {Willner}, {Zitrin}, {Adams}, {Austin}, {Arendt}, {Beacom}, {Bhatawdekar},
  {Bradley}, {Broadhurst}, {Cheng}, {Civano}, {Dai}, {Dole}, {D'Silva},
  {Duncan}, {Fazio}, {Ferrami}, {Ferreira}, {Finkelstein}, {Furtak}, {Gim},
  {Griffiths}, {Hammel}, {Harrington}, {Hathi}, {Holwerda}, {Honor}, {Huang},
  {Hyun}, {Im}, {Joshi}, {Kamieneski}, {Kelly}, {Larson}, {Li}, {Lim}, {Ma},
  {Maksym}, {Manzoni}, {Meena}, {Milam}, {Nonino}, {Pascale}, {Petric},
  {Pierel}, {del Carmen Polletta}, {R{\"o}ttgering}, {Rutkowski}, {Smail},
  {Straughn}, {Strolger}, {Swirbul}, {Trussler}, {Wang}, {Welch}, {B. Wyithe},
  {Yun}, {Zackrisson}, {Zhang}, \& {Zhao}}]{Windhorst2023}
{Windhorst}, R.~A., {Cohen}, S.~H., {Jansen}, R.~A., {et~al.} 2023, \aj, 165,
  13, \dodoi{10.3847/1538-3881/aca163}

\bibitem[{Woodroofe(1985)}]{Woodroofe1985}
Woodroofe, M. 1985, The Annals of Statistics, 13, 163 ,
  \dodoi{10.1214/aos/1176346584}

\bibitem[{{Yan} {et~al.}(2023{\natexlab{a}}){Yan}, {Ma}, {Ling}, {Cheng}, \&
  {Huang}}]{Yan2022}
{Yan}, H., {Ma}, Z., {Ling}, C., {Cheng}, C., \& {Huang}, J.-S.
  2023{\natexlab{a}}, \apjl, 942, L9, \dodoi{10.3847/2041-8213/aca80c}

\bibitem[{{Yan} {et~al.}(2023{\natexlab{b}}){Yan}, {Cohen}, {Windhorst},
  {Jansen}, {Ma}, {Beacom}, {Ling}, {Cheng}, {Huang}, {Grogin}, {Willner},
  {Yun}, {Hammel}, {Milam}, {Conselice}, {Driver}, {Frye}, {Marshall},
  {Koekemoer}, {Willmer}, {Robotham}, {D'Silva}, {Summers}, {Lim},
  {Harrington}, {Ferreira}, {Diego}, {Pirzkal}, {Wilkins}, {Wang}, {Hathi},
  {Zitrin}, {Bhatawdekar}, {Adams}, {Furtak}, {Maksym}, {Rutkowski}, \&
  {Fazio}}]{Yan2023}
{Yan}, H., {Cohen}, S.~H., {Windhorst}, R.~A., {et~al.} 2023{\natexlab{b}},
  \apjl, 942, L8, \dodoi{10.3847/2041-8213/aca974}

\bibitem[{{Yoshiura} {et~al.}(2017){Yoshiura}, {Hasegawa}, {Ichiki}, {Tashiro},
  {Shimabukuro}, \& {Takahashi}}]{Yoshiura2017}
{Yoshiura}, S., {Hasegawa}, K., {Ichiki}, K., {et~al.} 2017, \mnras, 471, 3713,
  \dodoi{10.1093/mnras/stx1754}

\bibitem[{{Yung} {et~al.}(2024){Yung}, {Somerville}, {Finkelstein}, {Wilkins},
  \& {Gardner}}]{Yung2023}
{Yung}, L.~Y.~A., {Somerville}, R.~S., {Finkelstein}, S.~L., {Wilkins}, S.~M.,
  \& {Gardner}, J.~P. 2024, \mnras, 527, 5929, \dodoi{10.1093/mnras/stad3484}

\bibitem[{{Zavala} {et~al.}(2023){Zavala}, {Buat}, {Casey}, {Finkelstein},
  {Burgarella}, {Bagley}, {Ciesla}, {Daddi}, {Dickinson}, {Ferguson}, {Franco},
  {Jim{\'e}nez-Andrade}, {Kartaltepe}, {Koekemoer}, {Bail}, {Murphy},
  {Papovich}, {Tacchella}, {Wilkins}, {Aretxaga}, {Behroozi}, {Champagne},
  {Fontana}, {Giavalisco}, {Grazian}, {Grogin}, {Kewley}, {Kocevski},
  {Kirkpatrick}, {Lotz}, {Pentericci}, {P{\'e}rez-Gonz{\'a}lez}, {Pirzkal},
  {Ravindranath}, {Somerville}, {Trump}, {Yang}, {Yung}, {Almaini},
  {Amor{\'\i}n}, {Annunziatella}, {Haro}, {Backhaus}, {Barro}, {Bell},
  {Bhatawdekar}, {Bisigello}, {Buitrago}, {Calabr{\`o}}, {Castellano},
  {Ch{\'a}vez Ortiz}, {Chworowsky}, {Cleri}, {Cohen}, {Cole}, {Cooke},
  {Cooper}, {Cooray}, {Costantin}, {Cox}, {Croton}, {Dav{\'e}}, {de La Vega},
  {Dekel}, {Elbaz}, {Estrada-Carpenter}, {Fern{\'a}ndez}, {Finkelstein},
  {Freundlich}, {Fujimoto}, {Garc{\'\i}a-Argum{\'a}nez}, {Gardner}, {Gawiser},
  {G{\'o}mez-Guijarro}, {Guo}, {Hamilton}, {Hathi}, {Holwerda}, {Hirschmann},
  {Huertas-Company}, {Hutchison}, {Iyer}, {Jaskot}, {Jha}, {Jogee}, {Juneau},
  {Jung}, {Kassin}, {Kurczynski}, {Larson}, {Leung}, {Long}, {Lucas},
  {Magnelli}, {Mantha}, {Matharu}, {McGrath}, {McIntosh}, {Medrano}, {Merlin},
  {Mobasher}, {Morales}, {Newman}, {Nicholls}, {Pandya}, {Rafelski}, {Ronayne},
  {Rose}, {Ryan}, {Santini}, {Seill{\'e}}, {Shah}, {Shen}, {Simons}, {Snyder},
  {Stanway}, {Straughn}, {Teplitz}, {Vanderhoof}, {Vega-Ferrero}, {Wang},
  {Weiner}, {Willmer}, {Wuyts}, \& {(The Ceers Team)}}]{Zavala2022}
{Zavala}, J.~A., {Buat}, V., {Casey}, C.~M., {et~al.} 2023, \apjl, 943, L9,
  \dodoi{10.3847/2041-8213/acacfe}

\bibitem[{{Zhang} {et~al.}(2022){Zhang}, {Shan}, {Gu}, {Zheng}, {Xu}, {Yue},
  {Liu}, {Zhu}, \& {Guo}}]{Zhang2022}
{Zhang}, Z., {Shan}, H., {Gu}, J., {et~al.} 2022, \mnras, 516, 1573,
  \dodoi{10.1093/mnras/stac2208}

\bibitem[{{Zitrin} {et~al.}(2015){Zitrin}, {Labb{\'e}}, {Belli}, {Bouwens},
  {Ellis}, {Roberts-Borsani}, {Stark}, {Oesch}, \& {Smit}}]{Zitrin2015}
{Zitrin}, A., {Labb{\'e}}, I., {Belli}, S., {et~al.} 2015, \apjl, 810, L12,
  \dodoi{10.1088/2041-8205/810/1/L12}

\end{thebibliography}
\bibliographystyle{aasjournal}

\appendix
\section{Detailed breakdown of comparisons between studies.}
Within the SMACS-0723 field, we compare our findings with \citet{Atek2022}. We find cross matches to all 15 of their original sources, however only two of these meet our selection criteria. This is the source listed within that work as SMACS-z10a and z10b.There are a total of 5 sources where we have matching photometric redshift solutions. 10 sources are found to have dominant low-z solutions ($z<3$). Of the three unselected high-z contenders, two have low-z solutions with $\chi^2$ values comparable to the high-z solution and one has a very broad redshift PDF.

Also conducted in the SMACS-0723 field, the work of \citet{Yan2022} identified 88 objects exhibiting a spectral break within the JWST photometric bands. Cross-matching our catalogues to this sample, we find 78 sources are present in our photometric catalogues. Of these, we find that 0 sources pass all of our selection criteria and 11 sources have primary photo-z solutions which agree with the estimations from \citet{Yan2022}. Of these 11, 2 sources are masked, 7 are fainter than our rest-frame UV detection limits and 2 have low-z solutions with $\chi^2$ values too close the primary solution. Following their original naming, the 11 sources with agreeing redshift solutions are: F150DA-019, F150DA-031, F150DA-050, F150DB-004, F150DB-050, F150DB-056, F150DB-069, F200DA-033, F200DA-098, F200DB-086, F200DB-159. Low-z solutions appears to be a mixture of dusty galaxies at $2<z<5$ (22 sources) and redshifted 1.6 micron bump features of $z<1$ galaxies (34 sources).

Within the GLASS field, we recover both of the initially reported high-z objects that overlap between the \citet{Naidu2022} and \citet{Castellano2022} studies. In our work, these two high redshift candidates have solutions of $z=10.2-10.9$ and $z=12.3-13.6$. The study by \citet{Castellano2022} reports an additional 5 candidates, of which only one (GHZ4) has a matching redshift solution in our work and this meets all of our selection criteria. The remaining sources have visible breaks, but obtain low-z solutions of $2.4<z<2.8$ in our catalogues, indicating these breaks are potentially Balmer breaks instead of Lyman breaks.

In the first four pointings of CEERS that were taken in mid-2022 (P1,P2,P3,P6), we select 12 out of the full sample of 26 sources found by \citet{Finkelstein2022c}. An additional 6 sources have photo-z solutions which agree with the CEERS team but are not selected in our final robust sample. For these 6, 4 are fainter than our detection limits, 1 is masked and 1 has a comparable quality low-z solution. Recently, \citet{Finkelstein2023} published an updated catalogue of CEERS targets. This more comprehensive list contains 88 objects, of which we find 87 are present in our catalogues. A total of 29 sources are selected as robust candidates in our work and 62 sources have agreeing high-z solutions. For the 33 unselected sources, 10 are masked, 22 are fainter than our selection limits and 1 source has a comparable quality low-z solution. The CEERS team utilise a smaller aperture size of diameter 0.2 arcseconds in their work, resulting in slightly increased depth at the cost of increased dependence on fewer pixels and larger PSF corrections. As a result, we find there are many sources that may be reasonable candidates, but lie at the $4\sigma$ level in our work.

The work by \citet{Harikane2023} presented a sample of 21 galaxies across the GLASS, SMACS-0723 and CEERS fields. We find matches to all objects and find only 4 enter our final sample, with an additional two agreeing on high-z solutions. There are 6 galaxies where we find better fits with a Balmer break and 7 galaxies where a 1.6 micron bump at $z<1$ can be fit. Within the GLASS field, only the tho luminous objects at $z=10$ and $z=12$ and the additional source GHZ4 in \citet{Castellano2022} overlap between our catalogues. Following their original naming ssytem, the two high-z galaxies which we do not consider robust enough to enter our sample are GL-z9-2 and GL-z9-9. GL-z9-2 is rejected for a very broad redshift PDF in our work. GL-z9-9 is found to fall below our detection limits.

The work by \citet{Donnan2022} also covered the fields of GLASS, SMACS-0723 and CEERS. Of the 43 sources cross matched to our sample, we find 28 sources have agreeing high-z solutions and 16 of these reach our final sample. There are 3 sources with 1.6 micron bump solutions at $z<1$ and 6 sources with Balmer Break solutions at $2<z<5$. Of the high-z sources that were later rejected as unreliable, there were 3 that were masked, 2 had a very broad redshift PDF, 6 were fainter than our detection limits and 1 had a comparable quality low-z solution.

Comparing to \citet{Bouwens2022c}'s independent analysis of SMACS-0723, GLASS and CEERS, we find 20/33 objects have photo-z solutions that agree but only 6/33 objects enter our final sample. A total of 4 objects are better fit with a 1.6 micron bump at $z<1$ and 9 objects are fit by a Balmer break at $1.5<z<4$. Of the objects with agreeing primary photo-z solutions, we find 2 are masked, 10 sources are $z\sim6.5$ with the Lyman break not completely exiting the bluest (F090W) band we utilise in SMACS-0723 and GLASS, 1 object is fainter than our limits and 2 objects have comparable quality low-z solutions.

Finally, we compare to a similar work conducted in \citet{McLeod2023}, which also compiled together a variety of JWST survey programmes. Of the 32 objects which were cross matched between our samples, we find 28 have matching photo-z solutions and 24 reach our final sample. Since \citet{McLeod2023} implements conservative $7\sigma$ detection limits on their sample, we find that the rate of objects scattering below our own $5\sigma$ limits its much lower than other studies. This ultimately means the sample presented in \citet{McLeod2023} is largely recovered in our own sample. We find one object is better fit by a 1.6 micron bump at $z<1$ (their GLASS-Z11-1481) and two objects better fit by a Balmer break at $2<z<4$ (their CEERS-NE-z10-3069854 and CEERS-2-3-z12-8536). The remaining object we disagree upon is CEERS-SW-z11-26592, which we identify at $z=9.5$ instead of $z=11.4$. For the four rejected objects where we agree on primary photo-z solutions, we do not select them because all four fall within our masked regions. Of these four, only one would be rejected for another reason (GLASS-Z14-33570 has a comparable low-z solution), indicating all other three sources may be valid.

\section{SED's of Highlight Sources}

{Within this section we show in Figure \ref{fig:highlightseds} the best fitting SED's for the UV luminous objects described in \S 3.3.1. Alongside these we also present the SED of the highest redshift, spectroscopically confirmed source in the sample from the JADES field \citep{CurtisLake2022}. In addition, we also present a selection of two random candidate galaxies from each redshift bin used in measuring the UV LF in Figure \ref{fig:exampleseds}. The full list of sources, including their coordinates, fluxes and a selection of other properties will be presented in full as a downloadable catalogue within Conselice et al In Prep.

\begin{figure*}
     \begin{subfigure}
         \centering
         \includegraphics[width=0.49\textwidth]{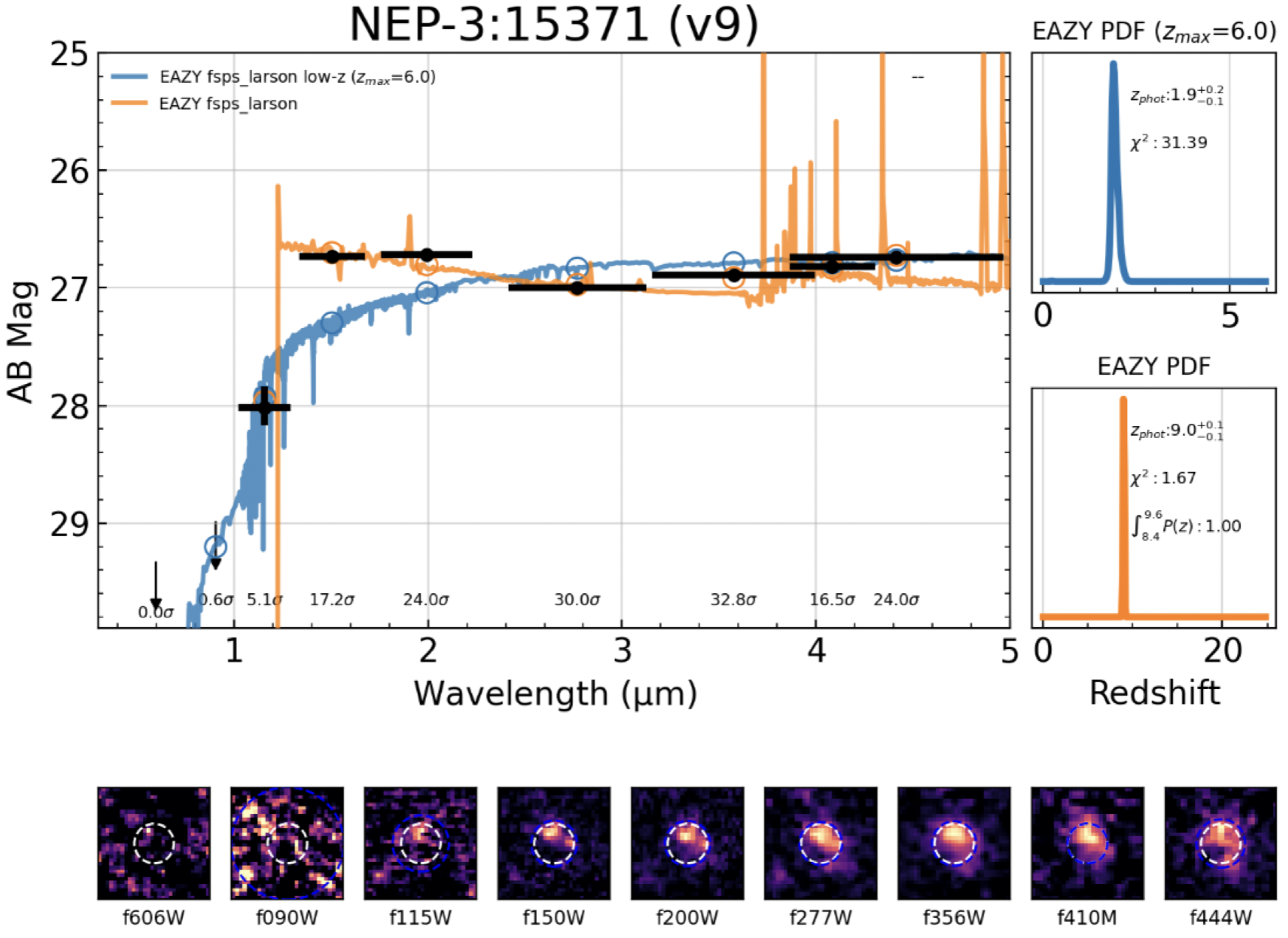}
     \end{subfigure}
     \hfill
     \begin{subfigure}
         \centering
         \includegraphics[width=0.49\textwidth]{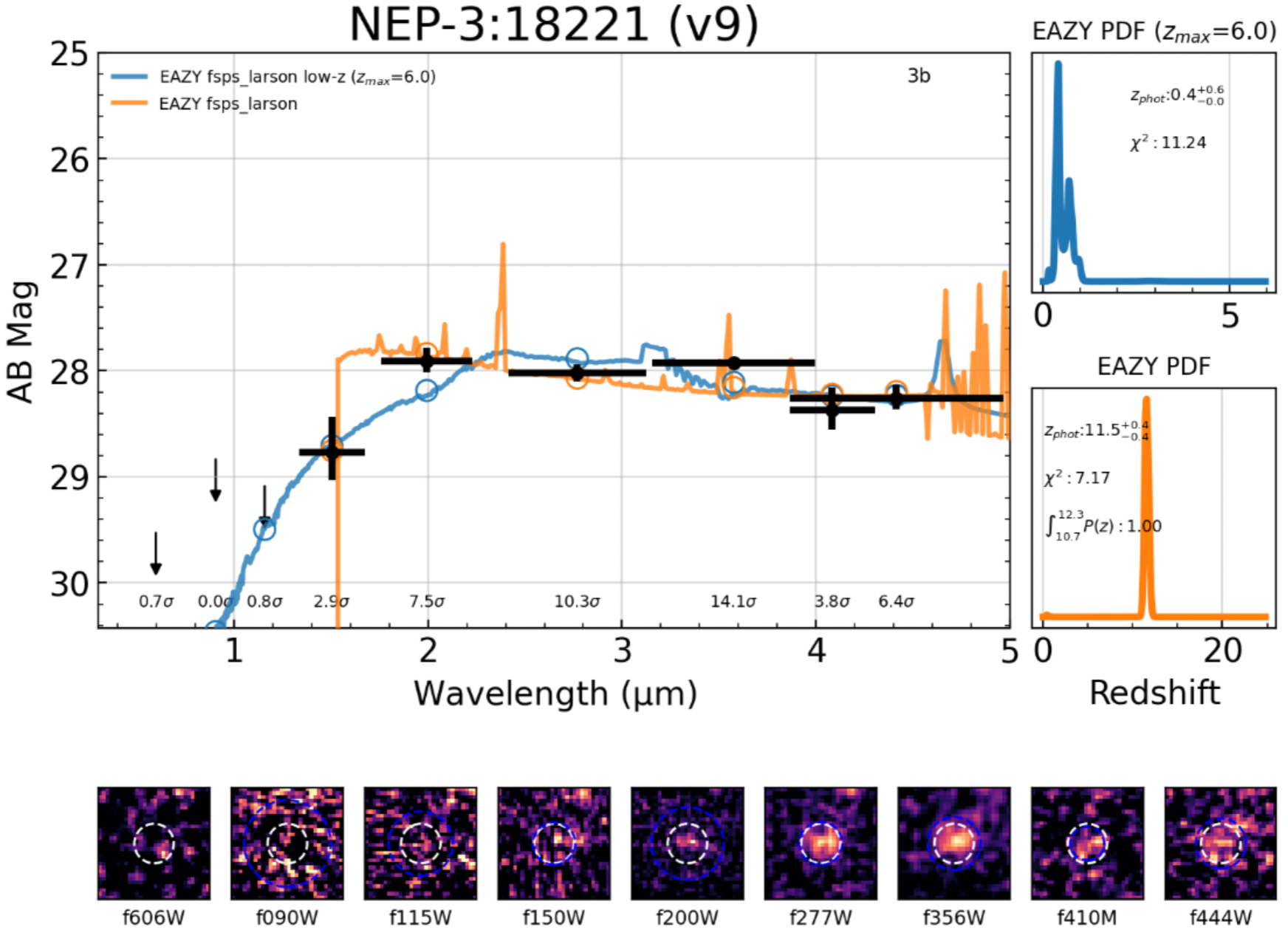}
     \end{subfigure}
     \hfill
     \begin{subfigure}
         \centering
         \includegraphics[width=0.49\textwidth]{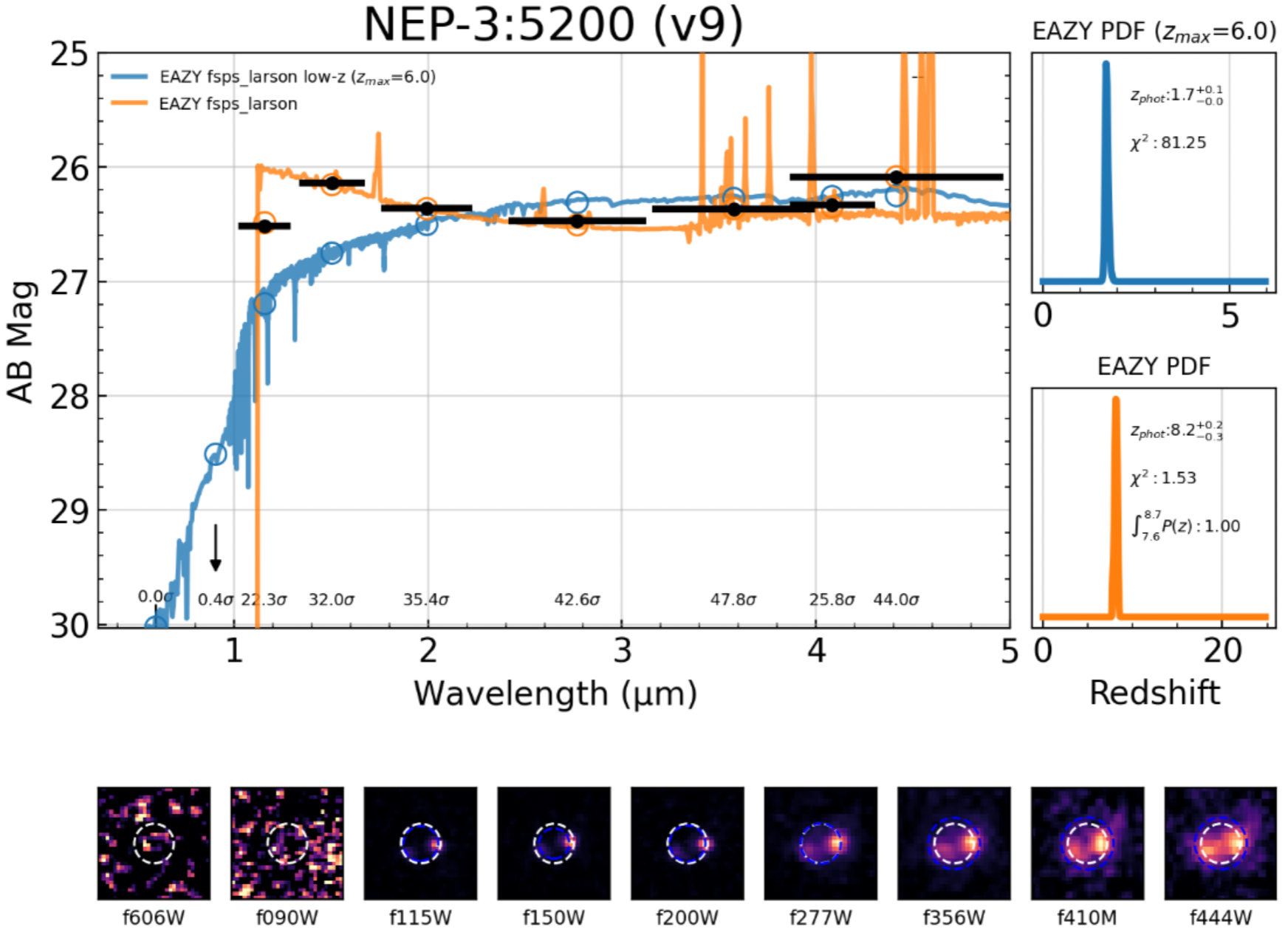}
     \end{subfigure}
    \hfill
    \begin{subfigure}
         \centering
         \includegraphics[width=0.49\textwidth]{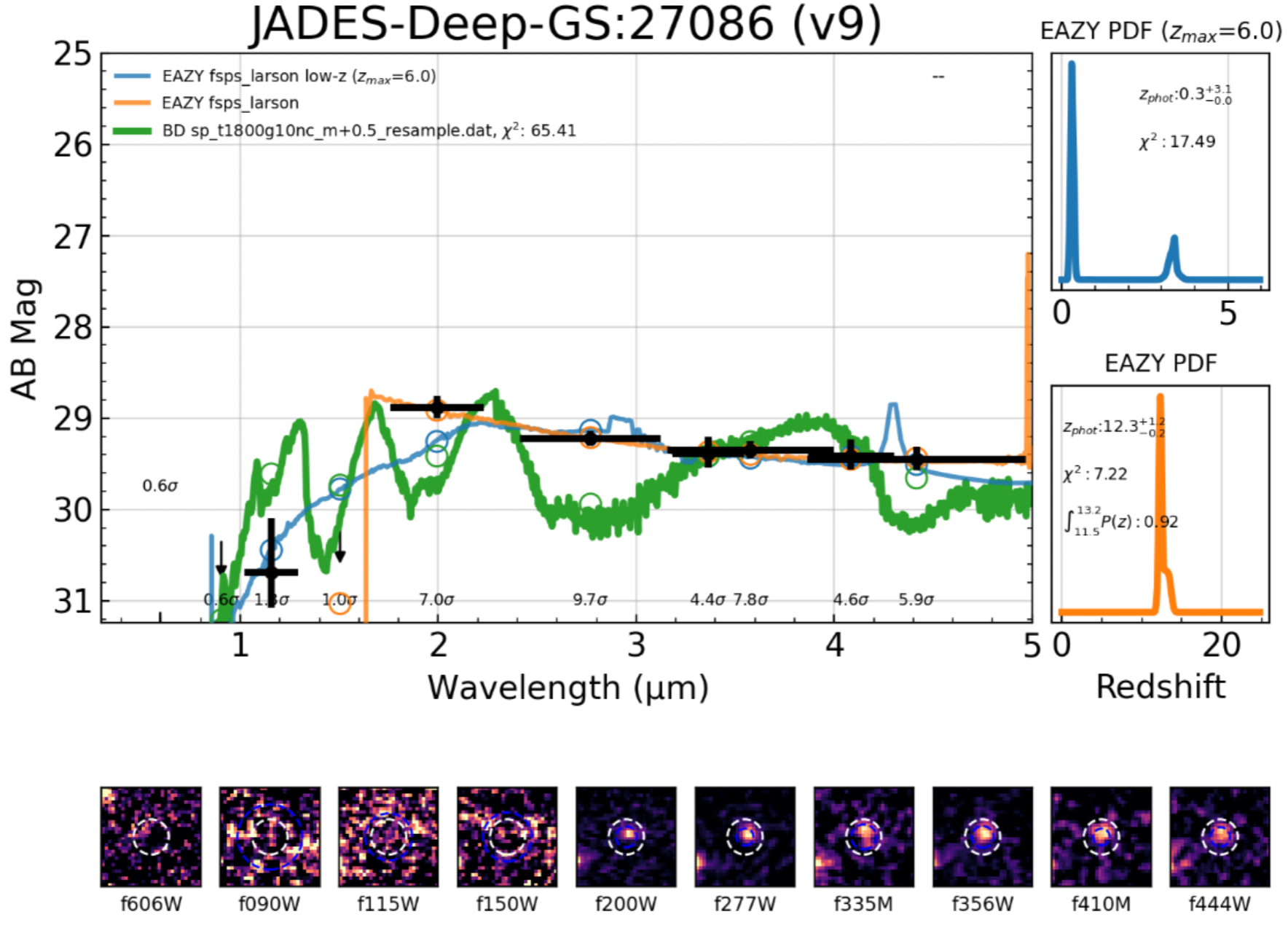}
     \end{subfigure}
     \centering
     \caption{SEDs for a selection of sources noted in this work. Naming of NEP-TDF sources identifies which spoke number (epoch) the object was identified in. In this case, all NEP sources presented are from the third spoke (NEP-3) taken in February 2023 from PID 2738. \emph{Top left:} The source with the second brightest intrinsic $M_{\rm UV}$ in our sample, after the AGN identified in \citet{Larson2023}, with $M_{\rm UV} = -21.34$. \emph{Top right:} The candidate source at $z>10$ with the greatest $M_{\rm UV}$. This is at $z=11.5$ and has $M_{\rm UV} = -21.03$. \emph{Bottom left:} An example of one of the five bright sources ($M_{\rm UV} < -21$) at $8.1<z<8.3$ in the NEP-TDF field. \emph{Bottom right:} The SED of the highest redshift spectroscopic source in the sample from the JADES team \citep{CurtisLake2022}. All SED's show in orange the best fit template, in blue the best fit low-z template ($z<6$), the significance of the detection just above the primary x-axis, $1\sigma$ upper limits if measurement is less, 1 arcsec cutouts in each filter used and accompanying photo-z PDF/statistics. In green, we show an example of an attempt to fit a brown dwarf to the JADES galaxy.}
     
\label{fig:highlightseds}
\end{figure*}  

\begin{figure*}
     \vspace{-0.3cm}
     \begin{subfigure}
         \centering
         \vspace{-0.25cm}\includegraphics[width=0.45\textwidth]{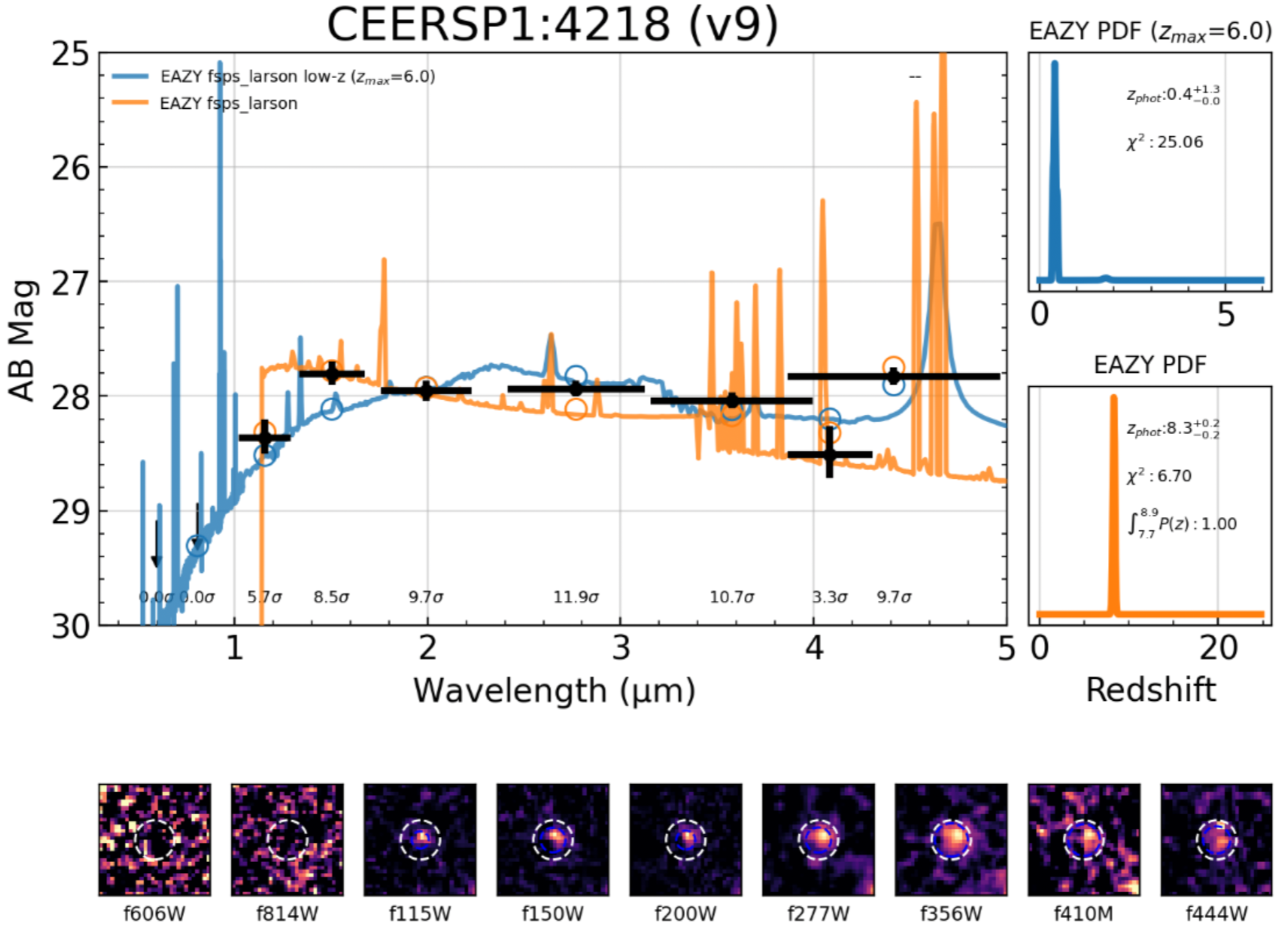}
     \end{subfigure}
     \hfill
     \begin{subfigure}
         \centering
         \includegraphics[width=0.45\textwidth]{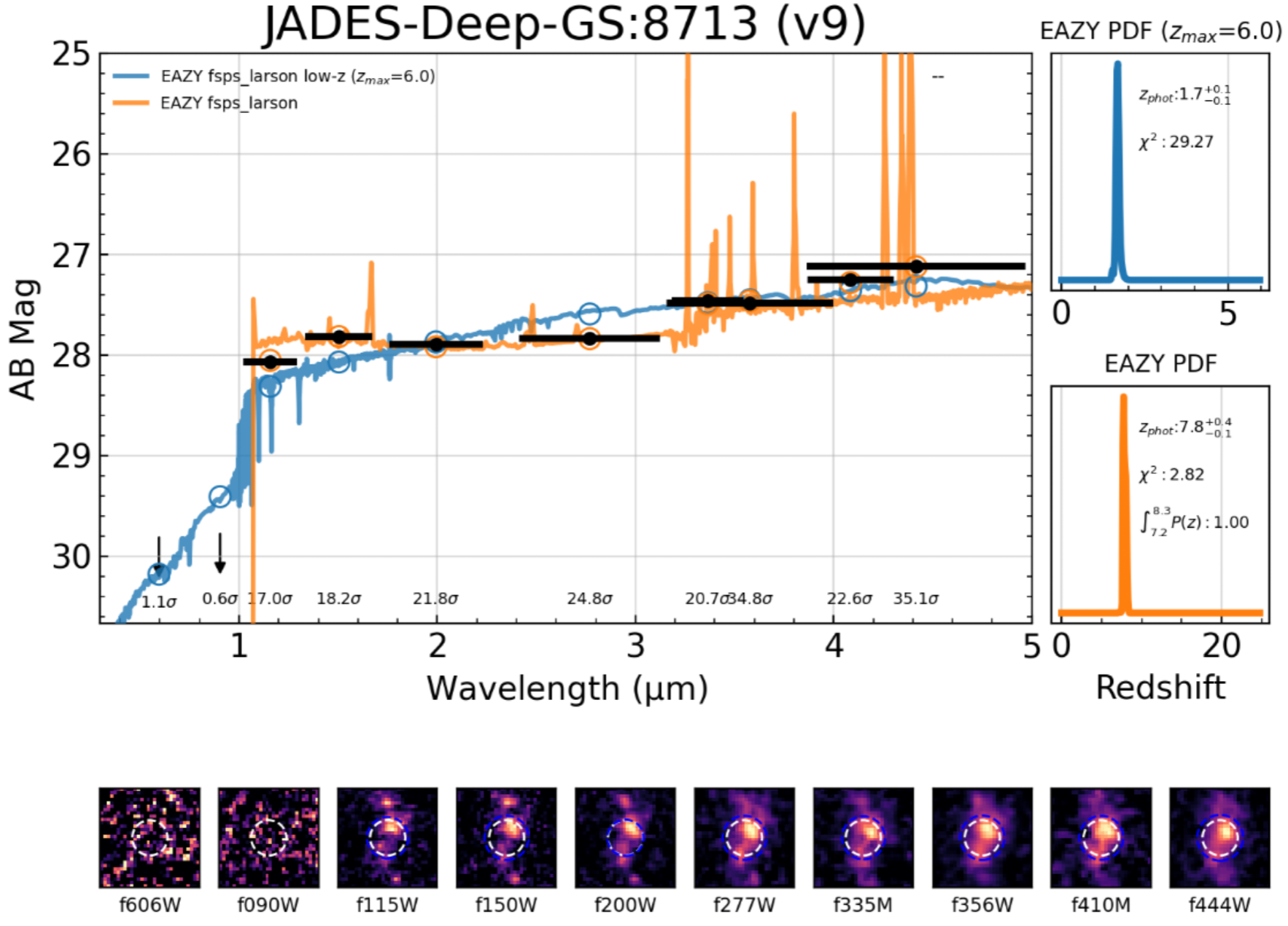}
     \end{subfigure}
     \hfill
     \begin{subfigure}
         \centering
         \vspace{-0.25cm}\includegraphics[width=0.45\textwidth]{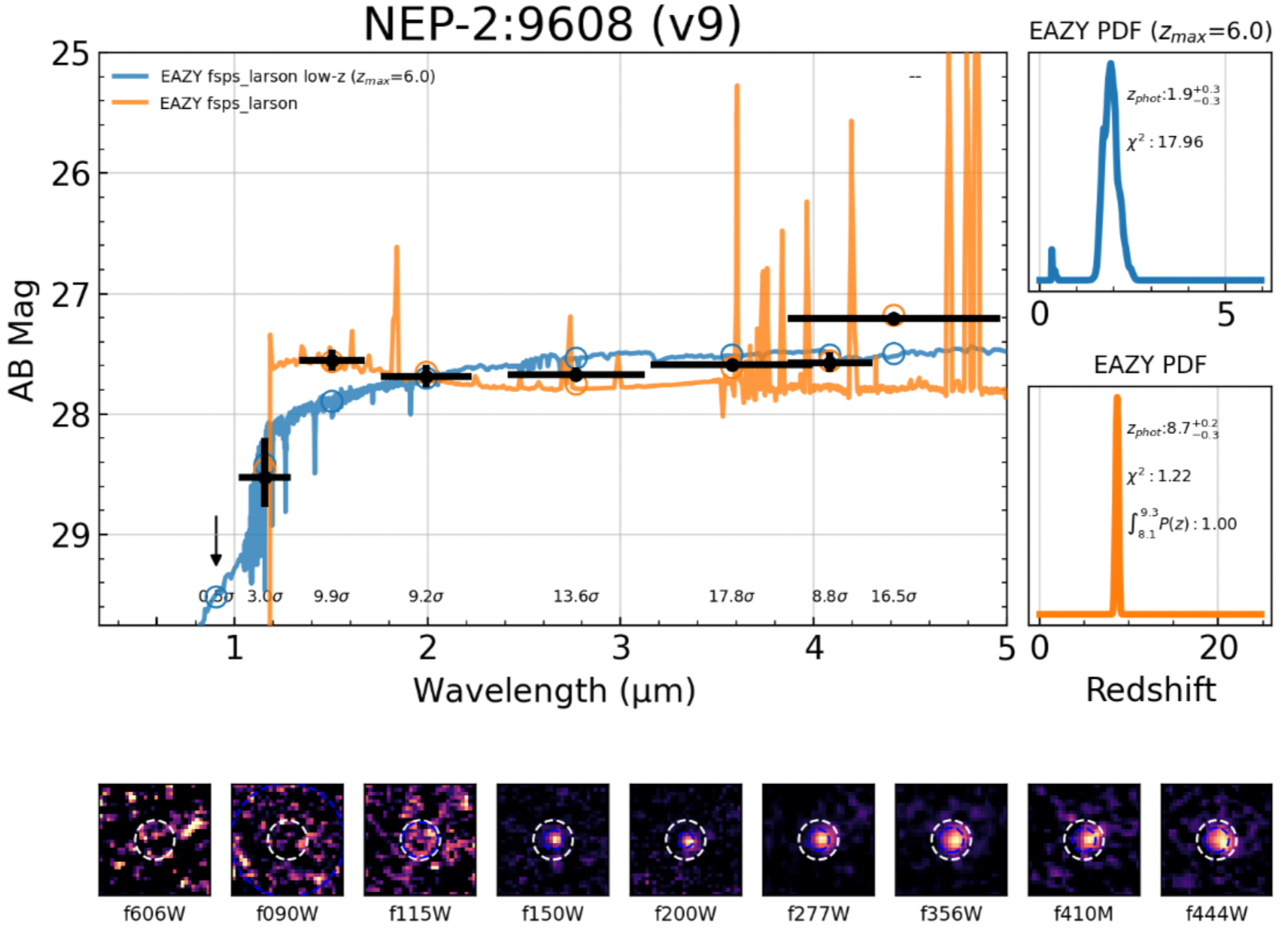}
     \end{subfigure}
    \hfill
    \begin{subfigure}
         \centering
         \includegraphics[width=0.45\textwidth]{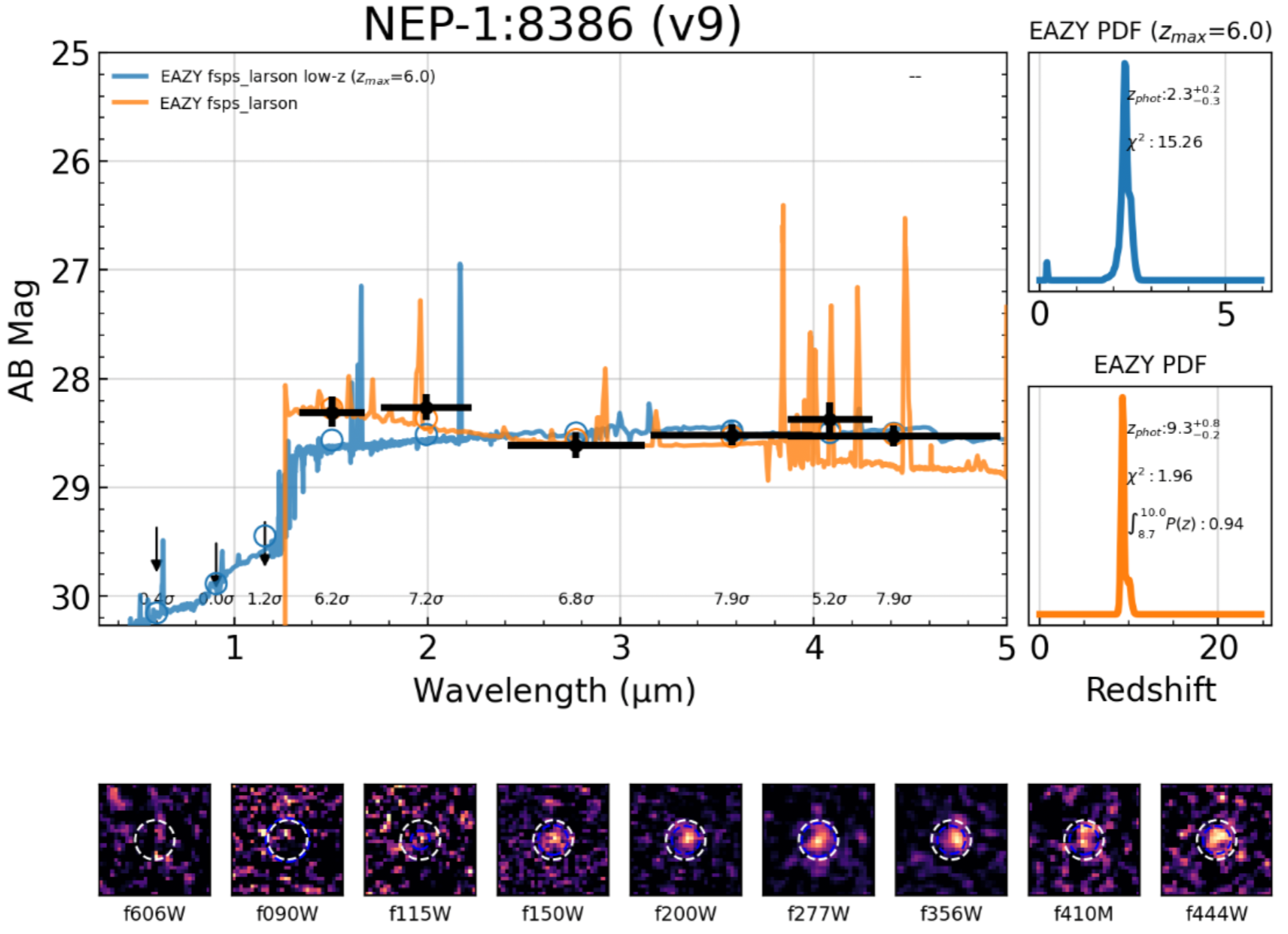}
     \end{subfigure}
          \begin{subfigure}
         \centering
         \vspace{-0.25cm}\includegraphics[width=0.45\textwidth]{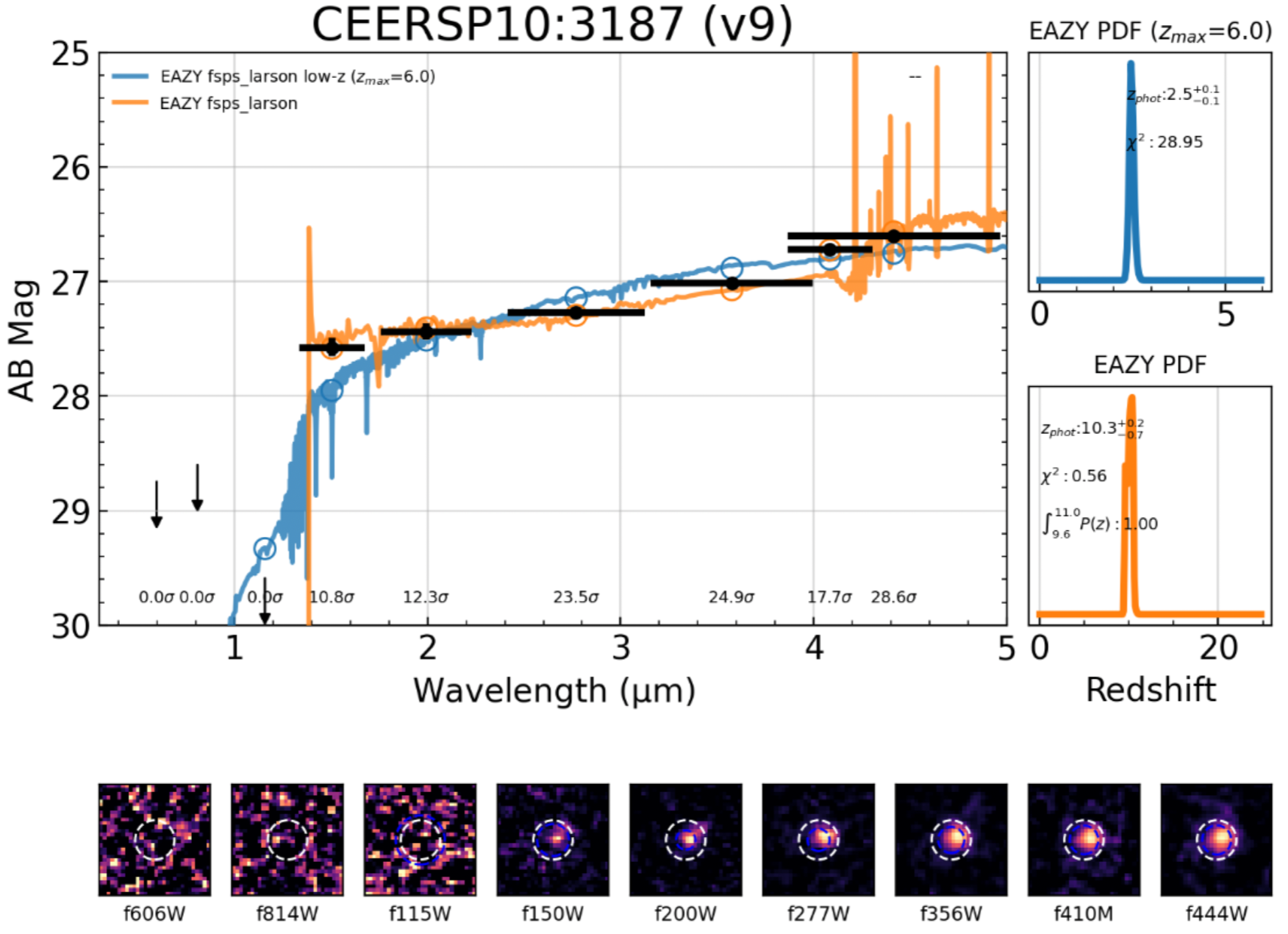}
     \end{subfigure}
     \hfill
     \begin{subfigure}
         \centering
         \includegraphics[width=0.45\textwidth]{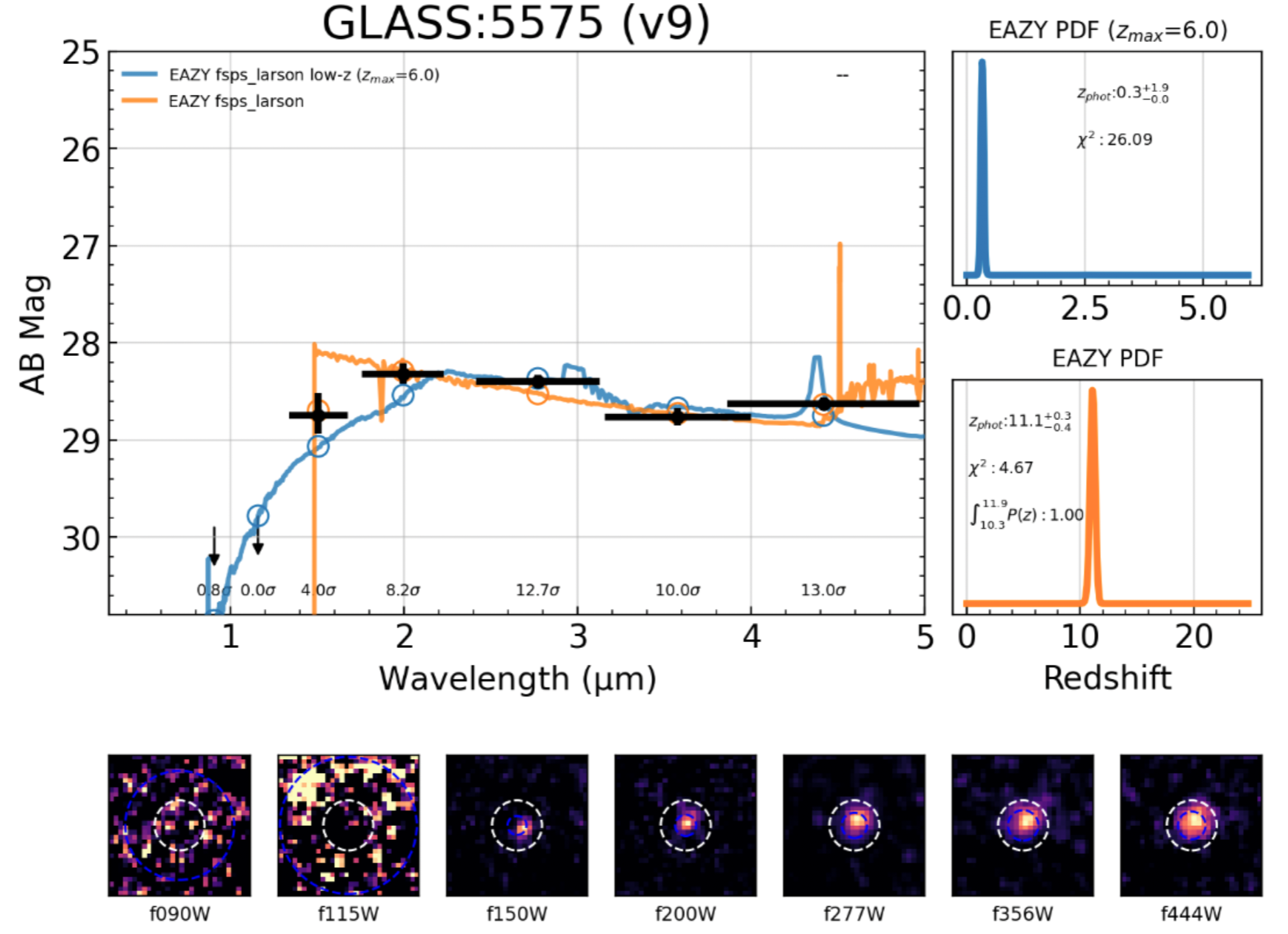}
     \end{subfigure}
     \hfill
     \begin{subfigure}
         \centering
         \includegraphics[width=0.45\textwidth]{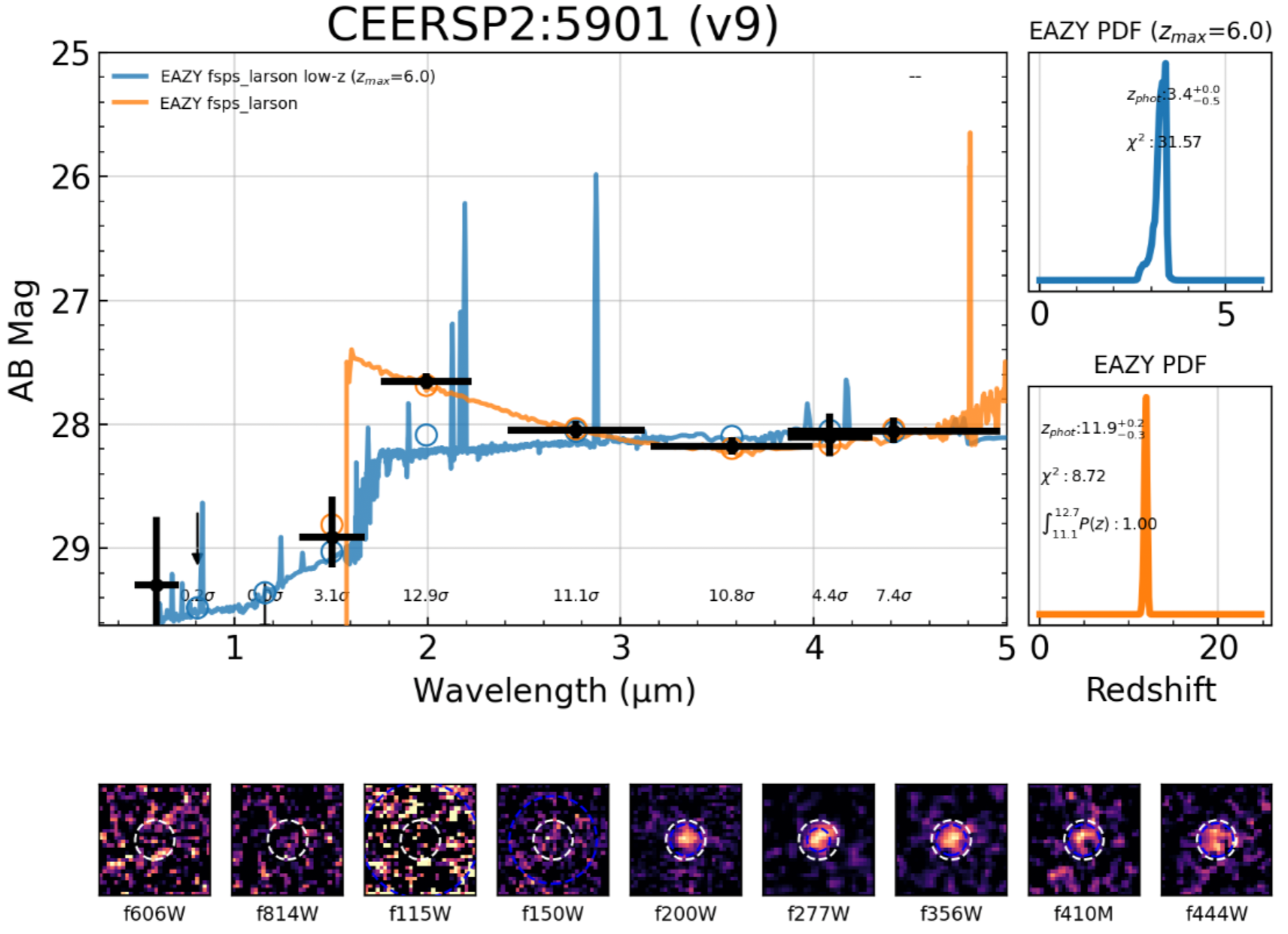}
     \end{subfigure}
    \hfill
    \begin{subfigure}
         \centering
         \includegraphics[width=0.45\textwidth]{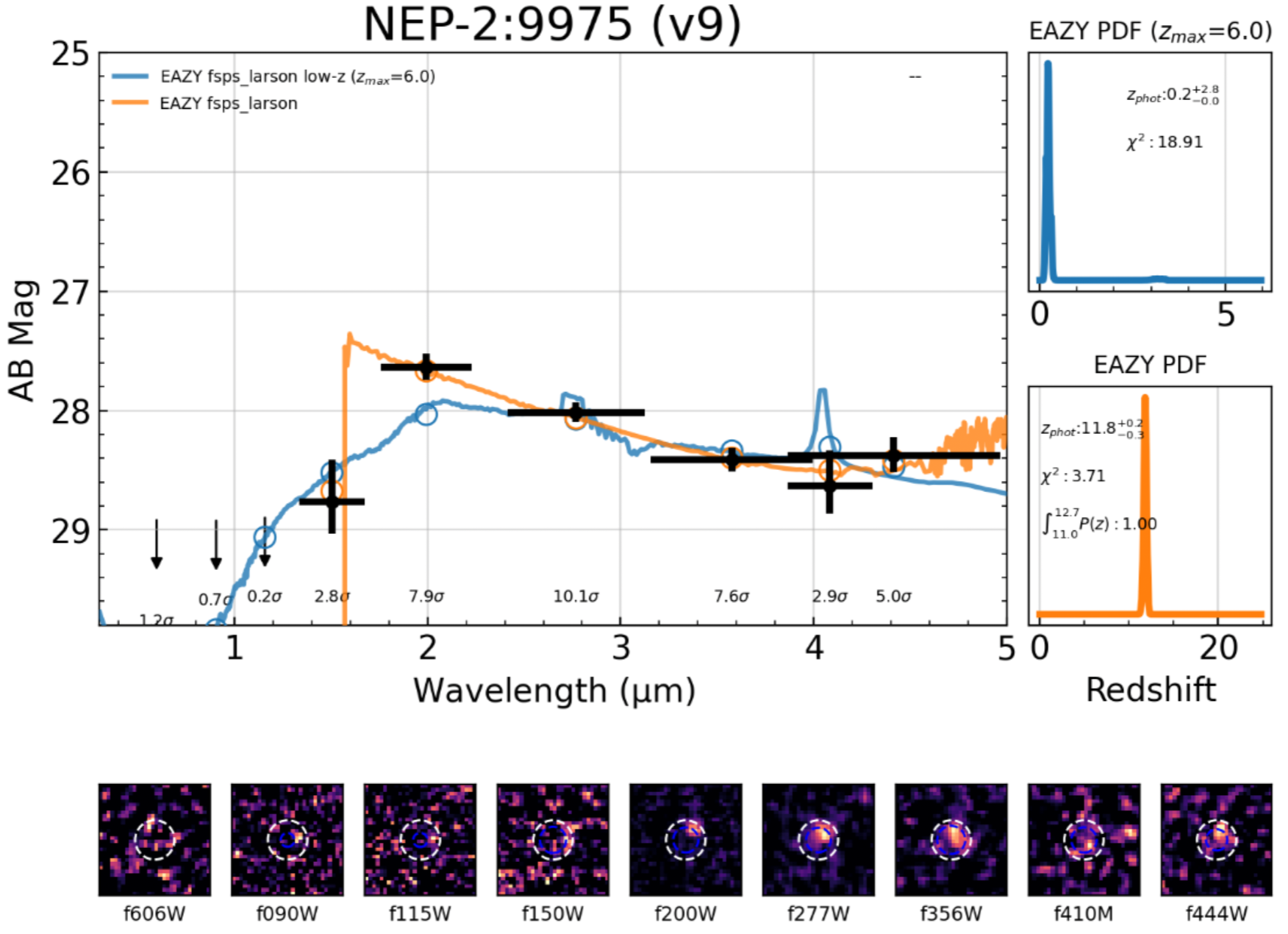}
     \end{subfigure}
     \centering
     \caption{As with Figure \ref{fig:highlightseds}, but for two random galaxies from each of our four redshift bins used for measuring the UV LF. The top row is $z\sim8$, and the following rows are $z\sim9$, $z\sim10.5$ and $z\sim12.5$.}
\label{fig:exampleseds}
\end{figure*}  

\end{document}